\RequirePackage{snapshot}
\documentclass[5p,times,12pt,authoryear,twocolumn]{elsarticle}
\usepackage{aas_macros}
\usepackage{booktabs}
\usepackage{tabularray}
\UseTblrLibrary{booktabs}

\usepackage{hyperref}
\hypersetup{colorlinks=true,
    linkcolor=blue}
\usepackage{xcolor}

\usepackage{amssymb}

\newcommand{\alphaSS}{\alpha_{\rm{SS}}}
\newcommand{\St}{\mathrm{St}}

\journal{New Astronomy Reviews}

\begin{document}

\begin{frontmatter}



\title{Empirical constraints on turbulence in proto-planetary discs}


\author[Milano,Leicester,Leiden]{Giovanni P. Rosotti\corref{corl}}

\cortext[corl]{Corresponding author.}

\ead{giovanni.rosotti@unimi.it}

\affiliation[Milano]{organization={Dipartimento di Fisica ``Aldo Pontremoli'', Universita degli Studi di Milano},
            addressline={via Celoria 16}, 
            city={Milano},
            postcode={20133}, 
            country={Italy}}
            
     \affiliation[Leicester]{organization={School of Physics \& Astronomy, University of Leicester},
            addressline={University Road}, 
            city={Leicester},
            postcode={LE1 7RH}, 
            country={UK}}        
            
\affiliation[Leiden]{organization={Leiden Observatory, Leiden University},
            addressline={PO Box 9513}, 
            city={Leiden},
            postcode={2300 RA}, 
            country={the Netherlands}}            

\begin{abstract}
Proto-planetary discs, the birth environment of planets, are an example of a structure commonly found in astrophysics, accretion discs. Identifying the mechanism responsible for accretion is a long-standing problem, dating back several decades. The common picture is that accretion is a consequence of turbulence, with several instabilities proposed for its origin. While traditionally this field used to be a purely theoretical endeavour, the landscape is now changing thanks mainly to new observational facilities such as the ALMA radio interferometer. Thanks to large improvements in spatial and spectral resolution and sensitivity (which have enabled the study of disc substructure, kinematics and surveys of large disc populations), multiple techniques have been devised to observationally measure the amount of turbulence in discs. This review summarises these techniques, ranging from attempts at direct detection of turbulence from line broadening, to more indirect approaches that rely on properties of the dust or consider the evolution of global disc properties (such as masses, radii and accretion rates) for large samples, and what their findings are. Multiple lines of evidence suggest that discs are in fact not as turbulent as thought one decade ago. On the other hand, direct detection of turbulence in some discs and the finite radial extent of dust substructures and in some cases the finite vertical extent strongly indicate that turbulence must be present at some level in proto-planetary discs. It is still an open question whether this amount of turbulence is enough to power accretion or if this is instead driven by other mechanisms, such as MHD winds.
\end{abstract}




\end{frontmatter}


\section{Introduction}
\label{sec:intro}
Accretion discs are a very common structure in the universe and explaining how and why accretion takes place is a classical problem in astrophysics. The process of accretion is intimately linked to that of angular momentum transport; since the specific angular momentum of a particle in Keplerian orbit increases with radius as $l \propto \sqrt{r}$, accretion requires a mechanism to lose angular momentum and move inwards. Much has been written on this subject and it is not the point of this review to discuss it here; the interested reader can consult the seminal review by \citet{Pringle1981} and the textbook by \citet{AccretionBook}.

For the purpose of this review, it is sufficient to remember that the classical solution to the problem of accretion is to invoke the presence of turbulence in the disc. On the macroscopic scale, turbulence acts as an effective viscosity and redistributes angular momentum in the disc, leading mass to move inwards while the angular momentum is transported outwards. Because of the strong link between turbulence and viscosity, for convenience in this review we will mostly use the two terms as synonyms\footnote{Though of course they are not; viscosity, as intended in accretion disc literature, is the manifestation of turbulence on scales much larger than those of the turbulent eddies. But we will ignore this distinction for simplicity's sake}. The origin of turbulence is still an open problem, though it is believed that the magneto-rotational instability (MRI) \citep{BalbusHawley} can be in many astrophysical cases a viable explanation.

This review focuses on a specific astrophysical case, proto-planetary discs, i.e. accretion discs around young (a few Myr) stars. Compared to other accretion discs found in nature, proto-planetary discs are cold (with typical temperatures in the range of tens of K) and have weak magnetic fields\footnote{If any - no magnetic field has been detected in proto-planetary disc at the time of writing (see \citealt{Vlemmings2019,Harrison2021} for the current upper limits). That being said, the current upper limits are not stringent enough to exclude that MRI can operate.} - this is a very challenging situation for the MRI to operate, since a minimum degree of ionization is required for coupling the gas to the magnetic field. For this reason, the theoretical community started doubting relatively early on that the MRI could be effective across the whole disc. For example, \citet{Gammie1996} proposed that the MRI could operate only in the very inner region ($<$0.1 au), where ionisation is provided by collisions, and at large radii, where the whole vertical column is ionised by cosmic rays. Sandwiched in between, he identified a so-called \textit{dead zone} where the bulk of the disc would be essentially laminar, except for a thin layer in the disc atmosphere ionized by cosmic rays. While the work of \citet{Gammie1996} focused only on Ohmic resistivity as non-ideal effect quenching the MRI, subsequent works \citep[see][]{LesurPPVII} found that the concept of a dead zone survives and is in fact common to many theoretical studies studying in which regions of the disc the MRI can operate. There are therefore reasonable objections to the idea that the MRI can drive the level of turbulence required to explain accretion.

Nevertheless, there is another long list of purely hydrodynamical processes that could give rise to turbulence in discs (see \citealt{LesurPPVII} for a review) and it is therefore reasonable to think that, however our ignorance, the disc \textit{must} be turbulent to explain the observational fact that accretion does take place. In light of this, for many years the field operated essentially hiding our ignorance about the nature of accretion. In practise, most works just assumed some efficiency of angular momentum transport, usually quantified by the $\alphaSS$ parameter popularised by \citet{ShakuraSunyaev1973}, in most cases neglecting its possible dependence with radius and vertical coordinate.

Putting aside for a second the question about the physical origin of turbulence, it is natural to ask if observations can provide independent constraints by assessing the level of turbulence in proto-planetary discs. Unfortunately, this proves to be a surprisingly difficult task. It is worth pausing a second to reflect on why this is the case. After all, the presence of turbulence in the interstellar medium is widely accepted and turbulence measurements are done routinely (e.g. \citealt{ZuckermanEvans1974,Larson1981}, or \citealt{Chevance2022} for a modern review). The fundamental issue is that turbulence in accretion discs is expected to be sub-sonic. This can be seen from the fact that the amplitude of velocity fluctuations is roughly $\sqrt{\alphaSS} c_s$, and in accretion theory $\alphaSS<1$. It follows that the easiest tool in our arsenal, line broadening, is of difficult application in discs, as it requires to have a very precise knowledge of the temperature to disentangle thermal broadening from turbulent broadening. Nevertheless, attempts at measuring turbulence using line broadening are of clear scientific value, and we will discuss them in this review - it is worth anticipating that, however difficult, in a few cases these measurements have been successful.

To our rescue, in the last few years the field of proto-planetary discs has been completely revolutionised from the observational point of view. This is primarily thanks to the Atacama Large Millimeter/submillimeter Array (ALMA), the most expensive ground based telescope ever built, and a vast improvement over the previous generation of interferometers (a factor $\sim$10x in spatial resolution and a factor $\sim$100x in sensitivity). In addition to improving direct measurements of turbulence, ALMA has also opened many ways in which turbulence can be measured indirectly. Summarising all the available constraints is the purpose of this review.

This review is structured as follows. We first briefly remind the reader about the definitions commonly used in the field in section \ref{sec:definitions} and we then discuss the direct constraints on turbulence in section \ref{sec:direct}. We then move to discuss the numerous indirect constraints, first on individual discs (section \ref{sec:indirect_individual}) and then on disc populations (section \ref{sec:populations}). We then discuss the implications of these measurements in section \ref{sec:discussion} and summarise our conclusions in section \ref{sec:conclusions}.

\section{Definitions}
\label{sec:definitions}

Before discussing the available observational constraints, we will briefly remind the reader of common definitions. On macroscopic scales turbulence provides an effective viscosity, which can be characterised by a kinematic viscosity coefficient $\nu$. It is common to parametrise this, following \citet{ShakuraSunyaev1973}, as:
\begin{equation}
\nu=\alphaSS c_s H,
\end{equation}
where $\alphaSS$ is a dimensionelss parameter quantifying the magnitude of turbulence, $c_s$ is the sound speed, and $H=c_s/\Omega$ is the disc scale-height, with $\Omega$ being the disc angular velocity. With this parametrization the velocity fluctuations induced by turbulence are of order $\sqrt{\alphaSS} c_s$, i.e. $\sqrt{\alphaSS}$ quantifies their amplitude in units of the sound speed. In practise in the rest of the review, when discussing the constraints on turbulence, we intend them as constraints on $\alphaSS$.

In terms of the disc secular evolution (see especially section \ref{sec:populations}), it is useful also to remember the definition of the viscous time, which gives the timescale for disc evolution:
\begin{equation}
t_\nu = \frac{R^2}{\nu} \simeq 5 \times 10^4 \left(\frac{\alphaSS}{10^{-2}}\right) \ \mathrm{yr},
\label{eq:tnu}
\end{equation}
where $R$ is the relevant length scale (typically the disc outer radius), which we have set to 10 au for the purpose of giving a numerical value; in addition, we have assumed a disc aspect ratio of 0.033 at 1 au and assumed a stellar mass of 1 solar mass. This timescale can be found through simple dimensional arguments, but analytical solutions of the disc evolution equation with viscosity \citep{LyndenBellPringle1974} show this is indeed the correct timescale on which discs evolve.

Although we will not use it directly in this review, we also briefly remind the reader that the accretion radial velocity induced by viscosity is of order
\begin{equation}
v_r \simeq \alphaSS \left(\frac{H}{R}\right)^2 v_K = \alphaSS \frac{H}{R} c_s
\end{equation} 
where $v_K$ is the Keplerian velocity. Remembering that both $\alphaSS$ and $H/R$ are (much) smaller than 1, the accretion velocity is thus expected to be significantly smaller than the Keplerian velocity, and even much smaller than the sound speed. For this reason attempting to measure radial velocity is not viable from the observational point, and we will not discuss it further.

\section{Direct detection}
\label{sec:direct}

\subsection{Infrared lines}
\label{sec:IR_lines}

\begin{figure}
\includegraphics[angle=-90,width=\columnwidth]{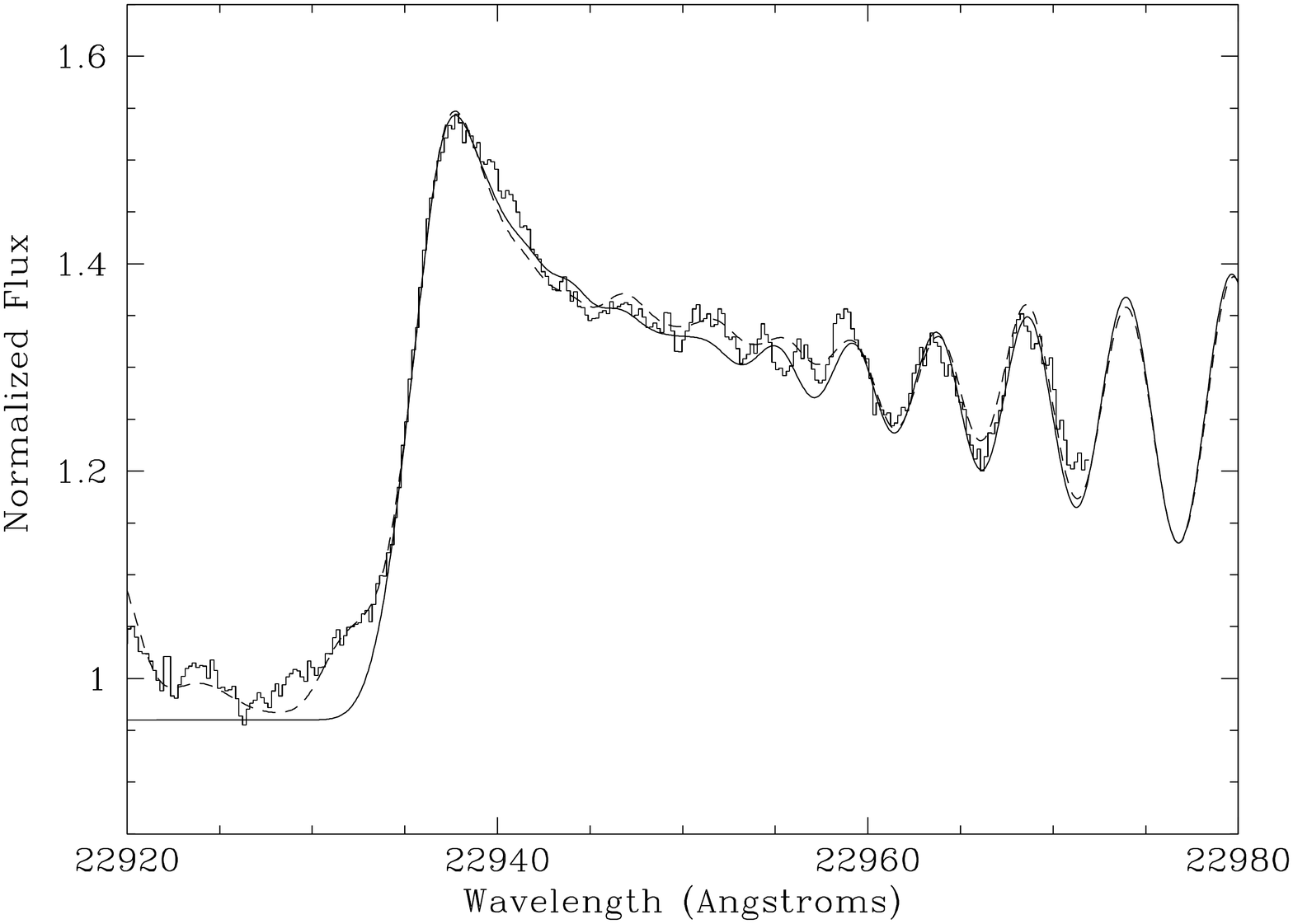}
\caption{CO overtone ($\nu=2-0$) spectrum of SVS13 \citep{Carr2004}. Visible in the right part of the plot are individual lines, corresponding to different rotational levels. The levels pile up on the left so that individual lines are no longer recognizable, determining the big jump at 22940 Angstroms called bandhead. The presence of multiple lines and the morphology of the bandhead allow one to at least partially disentangle temperature from turbulent broadening.}
\label{fig:CO_overtone_spectrum}
\end{figure}

As stated in the introduction, the fundamental issue in detecting turbulence via line broadening is that one needs to have an accurate estimate of the temperature given the small amplitude expected. Fortunately in some situations the temperature can be estimated independently. This is the case for the $\nu=2-0$ CO overtone emission at $\sim 2.3\mu m$, where one gets multiple rovibrational transitions which pile up at a specific J level (see \autoref{fig:CO_overtone_spectrum}). In this case the temperature affects the relative intensity of the individual rotational transitions because of the variation in excitation conditions, while turbulence affects the width of each single transition. It follows that in this case measurements of the line turbulent width are possible. This has been pioneered first by \citet{Najita1996} who showed that thermal broadening alone could not account for the observed line profiles, and subsequently by \citet{Carr2004} who reported supersonic turbulence in the source SVS 13. Other studies \citep[e.g.,][]{Hartmann2004,Najita2009,Ilee2014} have found similar evidence for turbulence at least comparable to the sound speed. While this review focuses on the case of solar mass stars, we note that similar results have been obtained also for the case of discs around massive stars \citep[e.g.,][]{Blum2004,Wheelwright2010,Ilee2013}.

Supersonic turbulence may be sound surprising given what we stated in the introduction. It should be kept in mind, however, that these lines come from a specific region of the disc, namely at small radii ($\sim$ 0.1 au, e.g. \citealt{Carr2004}) and from relatively high up in the disc atmosphere. In these locations the disc is very warm (thousands of K) and these upper layers are also ionized by cosmic rays (see discussion of the \textit{dead zone} in the introduction), making it plausible that the MRI can operate efficiently. MHD simulations (e.g., \citealt{FromangNelson2006a,Simon2015}) show that indeed one should expect to reach high values of the $\alphaSS$ parameter, possibly even higher than unity (i.e. supersonic turbulence), in the disc upper layers.

Caveats to keep in mind are that these observations are based on pure spectroscopy and so spatially unresolved. In addition, the vast majority of the studies that have reported such high values of the turbulent linewidth focus on the CO overtone emission, while neither the fundamental CO transition nor transitions from other molecules (e.g. water) show a similar evidence of high turbulence. Considering that these lines are optically thick and therefore trace only a thin layer of disc material, a possible explanation could be that at the deeper and colder layers of the disc, where the other lines/transitions originate, turbulence is less vigorous. Finally, the last caveat to keep in mind for these studies is that the sample is highly biased, and a large survey that can account for the entirety of the disc population is lacking. Looking at the future, it is expected that the upgraded capabilities of the instrument CRIRES+ at the Very Large Telescope (VLT) should significantly improve the size of the observational sample. The combination of CRIRES+ with optical interferometry with the instrument GRAVITY (e.g. \citealt{GRAVITYCO2021}) should also help to alleviate the problem of lack of spatial resolution.

\subsection{Sub-mm}

\begin{figure*}
\includegraphics[width=\textwidth]{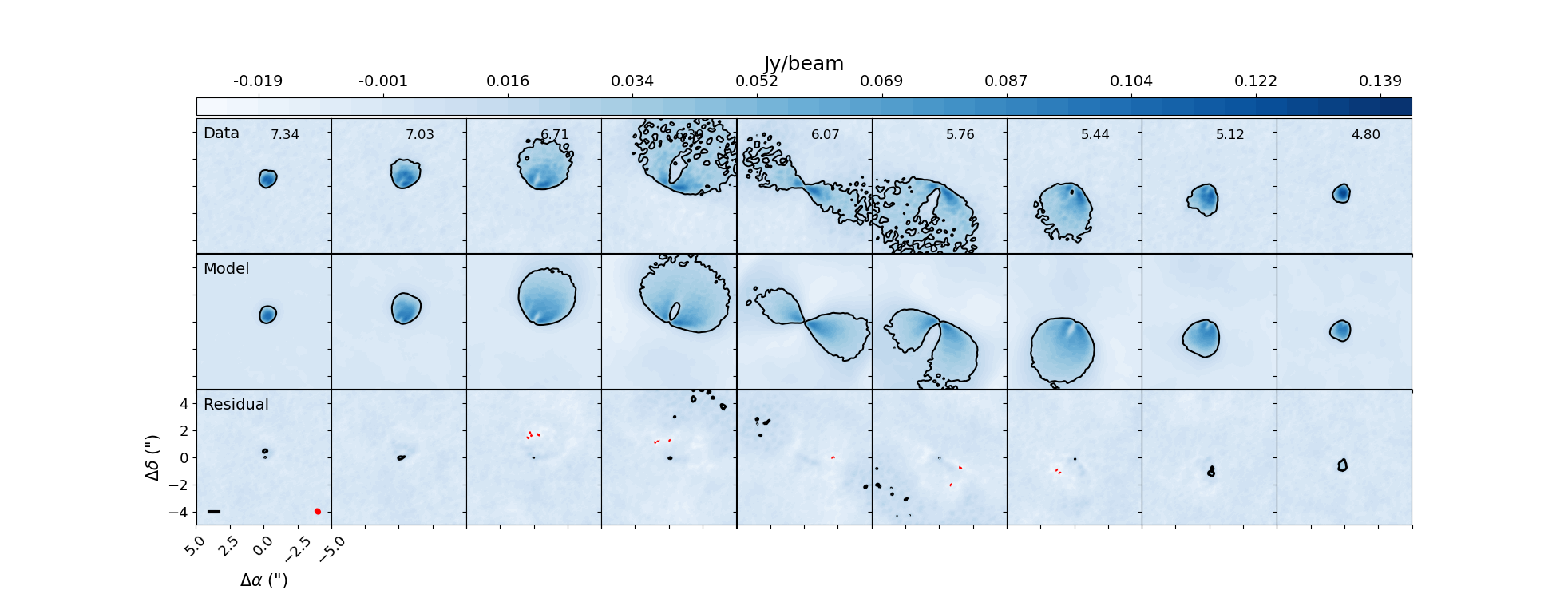}
\caption{Comparison between the channel maps of the data and those of the best fitting model in DM Tau \citep{Flaherty2020}. In this disc the channel maps are particularly spatially extended; as confirmed by the model, this is a signature of turbulence that can help disentanglign it from thermal effects.}
\label{fig:turb_DMTau}
\end{figure*}

\begin{table*}
\centering
\begin{tabular}{ccccc}
\toprule
Object & Tracer & $\delta v_\mathrm{turb}/c_s$ & $\alphaSS$ & Reference\\
\midrule
HD163296 & CO and DCO$^{+}$ & $<$0.05 & $<2.5 \times 10^{-3}$ & (1,2) \\
TW Hya & CO and CS& $<$0.08 & $<6 \times 10^{-3}$ &(3,4) \\
MWC480 & CO & $<$0.12 & $<$0.01 &(5)\\
V4046 Sgr & CO & $<$0.15 & $<$0.02 &(5)\\
DM Tau & CO & 0.25 & 0.0625 & (5)\\
IM Lup & CO & 0.18-0.30 & 0.03-0.09 & (6)\\
\bottomrule
\end{tabular}
\caption{Existing constraints on direct detection of turbulence from sub-mm lines. We have converted the measurements into a constraint on $\alphaSS$ using $\delta v_\mathrm{turb} = \sqrt{\alphaSS} c_s$. References: (1) \citet{Flaherty2015} (2) \citet{Flaherty2017} (3) \citet{Flaherty2018} (4) \citet{Teague2018CS} (5) \citet{Flaherty2020} (6) \citet{FlahertyPrep} }
\label{table:COlines}
\end{table*}

This topic has been recently reviewed also by \citet{PintePPVII}, but for the sake of completeness on the existing constraints on turbulence we will briefly review here once more the existing studies, as well as updating them with the paper of \citet{FlahertyPrep}.

The sub-mm is arguably the best wavelength at which to study discs due to the possibility of getting good spatial and spectral resolution, as well as tracing the cold material (tens of K) that comprises the bulk of the disc at tens or hundreds of au from the star. Considering the high scientific importance of this topic, direct detection of turbulence has been attempted already in the pre-ALMA era \citep[e.g.,][]{Qi2004,Pietu2007,Hughes2011,Chapillon2012}, but these attempts were marred by the low sensitivity and/or spectral resolution available. Already at this stage however it was clear that turbulence in the cold outer disc must be subsonic. Only in DM Tau \citep{GuilloteauDutrey1998,Dartois2003,Guilloteau2012} there were hints of turbulence, which became stronger with time, with an amplitude of 0.07-0.15 km/s. As we shall see, these early results were then vindicated by the later studies with ALMA.

Although the fundamental problem of disentangling thermal and turbulent line broadening will always remain, it is worth highlighting that the degeneracy can at least partially be reduced. If the chosen emission line is optically thick, its brightness temperature gives an accurate measurement of the physical temperature; ultimately the limiting factor is the calibration uncertainty. Other ways to reduce the degeneracy is having good spatial resolution since the line width affects the peak-to-through ratio of a spatially integrated spectrum \citep{Flaherty2015} and the spatial extent of channel maps. We show in \autoref{fig:turb_DMTau} an example in DM Tau of spatially extended channel maps, which are suggestive of high turbulence as confirmed by model fitting.

In the literature there are two different ways in which these studies have been conducted. One is the method introduced by \citet{Flaherty2015}, which uses a parametric model for the disc physical and thermal structure. The turbulence is one of the parameters of the model. The main limitation of this model is that one needs to make assumptions on the disc structure; notably, the method currently assumes a smooth disc neglecting gaps and rings. The method of \citet{Teague2016} uses a different, non-parametric method, which requires high spatial resolution to measure well the emission profile of the chosen tracers. The method uses multiple transitions to disentangle temperature from turbulent broadening, but it needs to assume that they come from the same layer. Reassuringly, for TW Hya the two methods yield similar results \citep{Flaherty2018,Teague2018CS}. While these studies use a single value of the turbulence across the whole disc, observations in HD135344B \citep{Casassus2021} find broadened line widths along spiral arms, suggesting that turbulence may vary spatially as expected from theoretical considerations.

We summarise the existing measurements in \autoref{table:COlines}. It should be highlighted that the current sample is very limited in size. Currently turbulence has been detected in two discs out of a total of six, with only upper limits on the other four discs. The limiting factor to conduct these studies on a larger sample is that even for ALMA there are stringent requirements on the data quality requested, since both high spatial and spectral resolution are needed, pushing the integration time to long values. Future observations such as the large programme exoALMA, which will survey 15 discs at high spatial and spectral resolutions, should significantly expand the sample size in the next years.

\subsection{Summary}
The direct searches for turbulence paint a rather mixed picture. On one side, it is reassuring that turbulence is detected in the disc upper layers. On the other side, the midplane seems to be a much quieter environment; if present, turbulence must be subsonic, and in many cases turbulence is not detected at all. The most straightforward interpretation is that turbulence must exhibit a strong gradient with the vertical coordinate. In addition, at the time of writing turbulence is detected in the midplane in 2 discs out of 6. For the non detections the upper limits are much lower than the values of the detections, suggesting that there could be significant variation from disc to disc rather than an universal value of $\alphaSS$. That being said, the current sample size is obviously limited and it is difficult to gauge how representative the constraints we discussed are for the disc population as a whole. This motivates us to discuss other tracers of turbulence in the next sections.

\section{Indirect methods - individual discs}
\label{sec:indirect_individual}

Considering the small sample discussed in the previous section, we wish here to review methods that rely on more indirect tracers of turbulence, i.e. methods that rely on phenomenological consequences of turbulence. As we shall see the field has been quite creative and devised multiple ways to constrain turbulence. The degree of indirectness varies across the methods, since the measurements involve the interaction with other physical processes, which are not always well understood. For this reason we tentatively list the methods below in decreasing order of directness, though we note that this classification is somewhat subjective. Since almost every work in the area of planet formation needs to make assumptions about the magnitude of turbulence, at times it has been difficult to decide if some works should be included in this discussion. We draw the (admittedly subjective) line at including only works that have some constraining power on viscosity, rather than including all works whose results depend on the magnitude of viscosity. We review in this section methods that study individual discs and in the next section methods that measure turbulence averaged over a disc population.

\subsection{Disc vertical extent}
\label{sec:vertical_extent}

The vertical equilibrium of dust grains is a competition between settling, i.e. gravity which makes the grains fall towards the midplane, though at a speed reduced by gas drag, and turbulence, which stirs up the grains in the vertical direction. The more turbulent the disc is, therefore, the thicker it will become, and this provides an indirect way to assess turbulence. The indirectness here is hidden in the way dust couples to the gas, which is parametrised by the Stokes number $\St$, and to turbulence, which is parametrised by the Schmidt number $\mathrm{Sc}$. The former is given by the following relation:
\begin{equation}
\St=\frac{\pi}{2} \frac{\rho_s a}{\Sigma},
\label{eq:St}
\end{equation}
where $\rho_s$ is the bulk density of the dust, $a$ the dust grain size and $\Sigma$ the gas surface density. Since $a$ and $\Sigma$ are typically unknown the Stokes number is also unknown. The latter coefficient, the Schmidt number, is instead not a limitation for these studies. The Schmidt number is defined as the ratio between the gas and dust diffusivities: $\mathrm{Sc}=D_\mathrm{gas}/D_\mathrm{dust}$, where $D_\mathrm{gas}=\nu=\alphaSS c_s H$ is simply the kinematical viscosity. At the typical conditions of proto-planetary discs turbulence acts as an effective diffusion with the same coefficient of diffusion of the gas \citep[e.g.,][]{YoudinLithwick2007}: $\mathrm{Sc}=1+\mathrm{St}^2$, and typically $\St \ll 1$ so that $\mathrm{Sc}=1$. 

With these definitions, it can be shown \citep{Dubrulle1995} that the scale height of the dust is approximately given by
\begin{equation}
H_\mathrm{dust} = H_\mathrm{gas} \left(1+\frac{\St}{\alphaSS}\right)^{-1/2}.
\end{equation}
This provides a quantitative way to measure $\alphaSS$ provided one can measure or estimate the scale-heights of gas and dust. Since the Stokes number is usually unknown, studies relying on dust tracers can only measure the ratio$\alphaSS/St$. In this section and the next we will report the existing constraints in terms of this parameter to make the comparison between different works and objects uniform, since the existing works have assumed different values of $\mathrm{St}$ when quoting their results. When quoting $\alphaSS$, we have assumed a typical value $\mathrm{St}=10^{-2}$, but we stress this is not a measurement and should be considered uncertain.

It is worth noting that the difference between gas and dust scale-heights is maximised for large grains. Therefore, observationally these studies are best done comparing a tracer of large grains (e.g. the mm emission) with a tracer of the gas (e.g. the small grains; otherwise the gas scale-height can be estimated if one has an estimate of the temperature). Natural candidates where to measure the disc vertical extent are discs at high inclination, where projection gives a direct view of the disc vertical extent. Hints that the mm is thinner came already in the pre-ALMA era \citep[e.g.,][]{Duchene2003,Pinte2007,Grafe2013}, but this kind of studies need high spatial resolution to be done properly (\citealt{Boehler2013} showed through radiative transfer that a resolution of at least 0.1" is desirable; ideally, one would like a resolution comparable to the vertical thickness) and so are really best suited for ALMA. Other indications come also from SED modelling, but this is degenerate with many other parameters  \citep[e.g.,][]{MuldersDominik2012,Boneberg2016}, and so attempts at measuring $\alphaSS$ from SEDs \citep[e.g.,][]{Ribas2020} should be taken with care.

\begin{figure}
\includegraphics[width=\columnwidth]{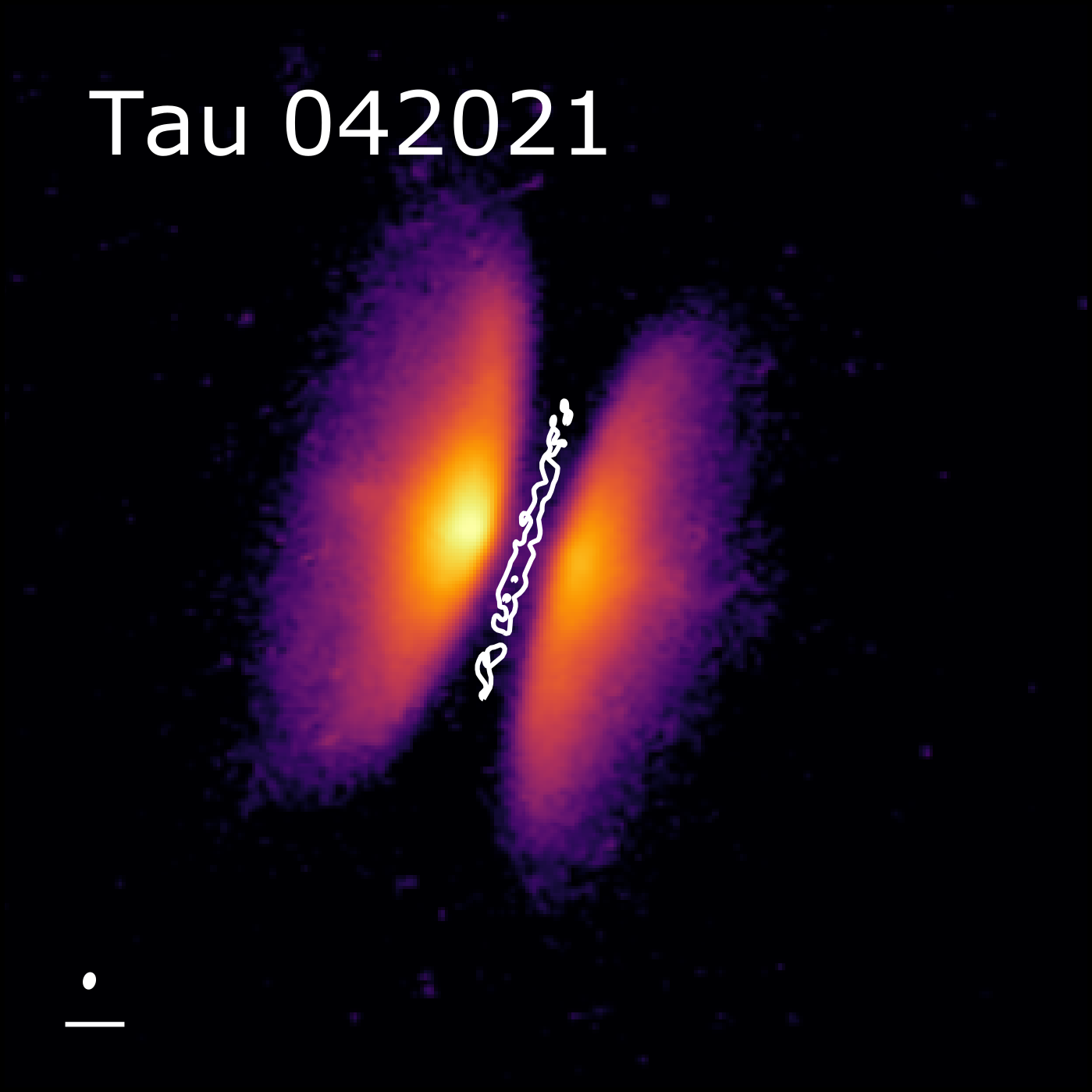}
\caption{Comparison between HST (colour-scale) and ALMA image (contours) for one of the discs in the \citet{Villenave2020} sample, showing that the vertical extent at ALMA wavelengths is much smaller than in scattered light. Considering the very different optical depths, however, only radiative transfer modelling can confirm that mm scale-height is truly smaller.}
\label{fig:villenave_single}
\end{figure}

The most complete study of edge-on discs has been presented in \citet{Villenave2020}, a small survey of 12 discs with high-resolution ALMA imaging ($\sim$0.1") and complementary HST imaging in scattered light. A visual comparison between images show that the disc at optical wavelengths is thicker than in the sub-mm, see \autoref{fig:villenave_single}. Here however it is important to make a strong caveat: since the optical depth for $\mu$m grains is much higher than in the sub-mm, we expect the emitting surface in scattered light to be at much higher heights than in the sub-mm even in absence of settling. The visual comparison is therefore \textit{not} enough to confirm settling and \textit{only} through radiative transfer modelling one can confirm it. The problem is also made complex by the fact that the vertical extent in the images, for the same scale-height, depends sensitively on the disc mass and inclination, making the inverse problem (measuring scale heights from images) highly degenerate. Indeed, \citet{Villenave2020} also present radiative transfer modelling and from this they conclude that only for three discs in the sample there is clear evidence of settling. In these cases, the dust scale height looks more compatible with value of $\sim$au at 100 au rather than 10 au. A more precise measurement of the vertical scale-height is however not possible given the quality of the data. Assuming that the gas scale-height is roughly 10au at this distance, this corresponds to value of $\alphaSS / \mathrm{St} \sim 10^{-2}$, yielding $\alphaSS \sim 10^{-4}$.

\begin{figure*}
\includegraphics[width=\textwidth]{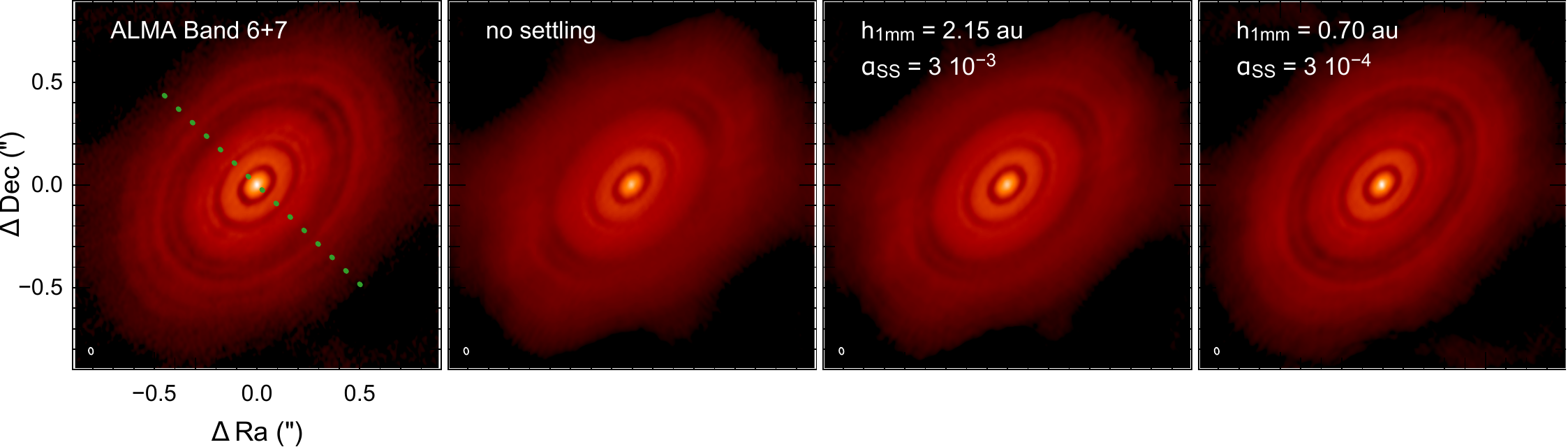}
\caption{Radiative transfer models from \citet{Pinte2016} in comparison with the data. The thicker the disc is, the more the gaps along the minor axis (see green dashed line) get filled and the image becomes blurry. This points to a rather small (H/R$\lesssim 0.01$) disc dust scale height.}
\label{fig:pinte_models_h}
\end{figure*}

Perhaps surprisingly, the study of the disc vertical extent is not however limited to edge-on discs. \citet{Pinte2016} proposed a new method that can be used in discs with gaps and rings. Most discs probably fall in this category \citep{Andrews2018,Long2018} so that this makes the method widely applicable, although so far only 3 discs have published results. The method relies on the fact that due to projection effects gaps get ``filled'' by adjacent rings along the disc minor axis, while the ``filling'' effect is much smaller along the disc major axis. The thicker the disc, the bigger the filling effect and the difference between major and minor axis, as illustrated in \autoref{fig:pinte_models_h}. Through radiative transfer modelling one can therefore estimate the disc vertical extent at the emitting wavelength. \citet{Pinte2016} applied this method to the first high-resolution ALMA image of HL Tau, finding that the disc must be extremely thin: H/R$\simeq 0.01$ at 100au (similar to the constraint of \citealt{Villenave2022}). More recently \citet{Liu2022HD163296} applied this method also to HD163296, finding similarly small values for two of the three gaps in the source, but a significantly larger value for the second gap, where the gas and dust scale-heights are comparable and thus point to $\alphaSS/\mathrm{St} \sim 1$. An ongoing analysis on the DSHARP sample \citep{Pizzatiprep} is less constraining than on these two discs, but when the disc is sufficiently inclined and gaps are sufficiently deep for the method to be applicable finds that we can exclude that the dust scale height is thicker than $H/R \sim 0.04$.
\citet{DoiKataoka2021} devised a similar method to that of \citet{Pinte2016}, which uses instead the azimuthal variation in rings rather than in gaps. The method works only if the emission in the ring is optically thin. They note that the emission from a geometrically thick (thin) ring is azimuthally asymmetric (symmetric). Using this fact and detail radiative transfer modelling of HD163296, the same source studied by \citet{Liu2022HD163296}, they come to similar conclusions regarding the vertical scale-height.

\citet{Villenave2022} presented very high resolution ($\sim$0.02") observations of Oph 163131. This case somehow combines both the study of edge-on discs, since it is very inclined ($\sim$ 84 $\deg$), with the method developed by \citet{Pinte2016} since it shows gaps and rings. The source had already been studied in lower-resolution data by \citet{Wolff2021}, who through combined Bayesian modelling of sub-mm and scattered light imaging found evidence for settling, but could not clearly measure the extent of the sub-mm dust due to the degeneracies in lower resolution data. Thankfully, these degeneracies disappear with enough spatial resolution and confirm that the dust scale-height in the sub-mm must be $< 1$au at 100 au distance from the star, yielding $\alphaSS / \mathrm{St} \lesssim 6 \times 10^{-3}$, or $\alphaSS \lesssim 6 \times 10^{-5}$.

So far we have considered only class II discs (~Myr old), which are the focus of this review. It is worth highlighting there is a potential tension with younger evolutionary phases such as class 0 and I discs, which either lack evidence of dust settling \citep[e.g.,][]{Lin2021HH212,Michel2022VLA1623W,Sheehan2022} or appear to be less settled than class II discs \citep{Villenave2023} - notably the latter work studies a class I object, suggesting that the disc vertical extent may be decreasing with evolutionary stage. It is possible this means turbulence in the early phases of disc formation is much higher than in evolved discs, but on the other hand it is unclear how much the ongoing infall from the envelope may resupply the disc with grains far from the midplane.

\subsection{Radial width of gas structures}
\label{sec:radial_width}

\begin{table*}
\begin{tabular}{cccc}
\toprule
Object & $\alphaSS / \mathrm{St}$ & $\alphaSS$ & Reference\\
\midrule
AS 209 B74 & 3.1e-2 - 5.7e-1 & 3.1e-4 - 5.7e-3 & (1) \\
AS 209 B120 & 4.6e-2 - 1.9e-1 & 4.6e-4 - 1.9 e-3 & (1) \\
Elias 24 & 7.7e-2 - 6.6e-1 & 7.7e-4 - 6.6e-3 & (1) \\
HD163296 B67 & 3.3e-1 - ... & 3.3e-3 - ... & (1) \\
HD163296 B100 & 1.3e-1 - 7.7e-1 & 1.3e-3 - 7.7e-3 & (1) \\
GW Lup & 3.1e-1 - 6.8e-1 & 3.1e-3 - 6.8e-3 & (1) \\
HD143006 B41 & 1.8e-1 - ... & 1.8e-3 - ... & (1) \\
HD143006 B65 & 1.1 - ... & 1.1e-3 - ... & (1) \\
J1610 & 15.3 - ... & 1.5e-2 - ... & (2) \\
LkCa15 & ... - 0.29 & ... - 2.9e-3 & (2) \\
HD163296 B67 & 0.23 & 2.3e-3 & (3) \\
HD163296 B100 & 0.04 & 4e-4 & (3) \\
HD163296 B155 & 0.04 & 4e-4 & (3) \\
AS 209 B74 & 0.18 & 1.8e-3 & (3) \\
AS 209 B120 & 0.13 & 1.3e-3 & (3) \\
HD169142 & 1.35 & 1.35e-2 & (4) \\
\bottomrule
\end{tabular}
\caption{Existing constraints from the radial width of gas substructures, as discussed in section \ref{sec:radial_width}. When sources have multiple rings to which the method can be applied, we use the Bxx nomenclature to distinguish between them, where xx is the distance in au from the star of the ring. Note that AS 209 and HD163296 appear twice because they were studied both by \citet{Dullemond2018} and \citet{Rosotti2020}. We provide ranges rather than values when the gas width is not available (ellipsis denote when the constraint is only a lower/upper limit). In order to convert to $\alphaSS$, we have assumed a typical value of $\mathrm{St}=10^{-2}$ so that the constraints coming from different works in the literature (which often assumed different values of $\mathrm{St}$) can be made uniform. We caution however that these values should be taken with much care since $\St$ is in general unknown. The only exception is HD169142, where \citet{Sierra2019} were able to break the degeneracy because they also had measurements of the dust spectral index. In this case we converted their measured value of $\alphaSS$ to $\alphaSS / \mathrm{St}$ for the sake of comparison with the other works. References: (1) \citet{Dullemond2018} (2) \citet{Facchini2020} (3) \citet{Rosotti2020} (4) \citet{Sierra2019}}
\label{table:radial_alphaSt}
\end{table*}

In the previous section we have focused on the \textit{vertical} extent of the dust. It turns out that, if the gas is \textit{radially} concentrated, the dust will also tend to radially concentrate on a smaller spatial scale. This is the dust trapping scenario first envisaged by \citet{Whipple1972}, where the concentration is operated by the gas pressure gradient affecting the dust radial velocity through drag. In this case drag tends to concentrate the dust, while the effective dust diffusion operated by turbulence opposes the concentration. It can be shown that the relation between the two radial length scales follows exactly the same relation we illustrated for the vertical scale-heights:
\begin{equation}
w_d = w \left( 1 + \frac{\St}{\alphaSS}\right)^{-1/2},
\end{equation}
where $w_d$ and $w$ are the dust and gas radial widths. As before, measurements of the two length scales can therefore provide a measurement of $\alphaSS/\St$, and indirectly a constraint on $\alphaSS$. This method relies on the presence of radial concentrations of gas and dust, and implicitly assumes that the dust structures are pressure traps. As already discussed before, radial concentrations of dust are observed in the majority of discs as bright rings. There is less evidence that the gas is similarly radially concentrated and that the observed rings are created by pressure confinement; for example sub-structures in line emission do not appear to be correlated with those in the dust \citep{Law2021Radial,Jiang2022}. Alternative explanations for the presence of bright rings in protoplanetary discs such as snow lines effects however do not appear to pass simple tests regarding their expected radial locations \citep{Long2018,Huang2018,vanderMarel2019,BaePPVII} and in a limited number of cases the analysis of rotation curves provides evidence that rings are indeed pressure traps \citep{Teague2018,IzquierdoMAPS}, making this explanation the preferred one at the moment of writing\footnote{Note this does \textit{not} imply that these structures are due to planets. Any mechanism, for example due to the interaction with the magnetic field, radially concentrating gas will have the same effect on the dust}. 

In principle, measuring dust radial widths is fairly straightforward as it is can be done directly on the ALMA images, provided one has enough spatial resolution. Nevertheless, we classify this method as more indirect than that discussed in the previous section, because measuring the gas radial width is a hard problem: in most cases we do not even have evidence of radial concentration. \citet{Dullemond2018} side-stepped this problem \textit{assuming} the gas is radially concentrated, and imposing reasonable upper and lower limits on the gas radial width using simple arguments. In this way they were able to derive a range of $\alphaSS/\St$ for the sources in the high-resolution DSHARP survey \citep{Andrews2018} with the most prominent dust rings. \citet{Facchini2020} followed the same procedure to get limits in J1610 and LkCa15. \citet{Rosotti2020} improved on this method by measuring the gas radial width for two discs, HD163296 and AS209, using the kinematics of line emission. They show that the gas width can be inferred from the slope of the deviation from Keplerian rotation, and in this way they were able to derive measurements for $\alphaSS/\St$. Currently the quality of kinematics data is not enough to conduct this analysis for a larger disc samples, but in the near future the large programme exoALMA should significantly expand the sample.

While the methods discussed so far only look at prominent rings, there is value in trying to use the full radial information coming from line and continuum observations. This is what \citet{Sierra2019} attempted to do in HD169142, solving the dust radial transport equation in steady state, i.e. looking for the equilibrium between turbulence and radial drift imposed by gas drag. They assumed it is possible to ``invert'' the dust and gas emission profiles in order to reconstruct the underlying surface densities. Critical steps of this inversion are knowledge of the dust optical depth, which they constrain using multi wavelength continuum data, and dealing with the intricacies of gas emission - namely optical depth and the unknown abundance of chosen tracer. To minimise the former, they select the $^{13}$CO isotopologue, assuming it is optically thin, and for the latter they simply assume a constant CO abundance. Considering $^{13}$CO may still be optically thick and the results of the chemical survey MAPS regarding radial variations of CO abundance \citep{Alarcon2021}, their results are worth revisiting in the future. The additional benefit of having multiwavelength continuum data is that they are able to break the degeneracy between $\alphaSS$ and $\mathrm{St}$. We will discuss this in more detail in section \ref{sec:discussion_degeneracy}.

We summarise the existing constraints in \autoref{table:radial_alphaSt}. There is certainly a range of values in the results obtained - however, we note that in most cases $\alphaSS/\mathrm{St}<1$ (sometimes significantly so), yielding values of $\alphaSS<10^{-2}$ assuming $\mathrm{St}=10^{-2}$. These constraints are similar to those listed in the previous section about the vertical extent.

Finally, while we focused here on \textit{radial} dust trapping, dust can also be trapped \textit{azimuthally} in vortices. \citet{Casassus2019} studied trapping in this case in MWC758, constraining $\log_{10}{\alphaSS} < -2.1$.

\subsection{Polarization}
\label{sec:polarization}

This method ultimately measures the vertical extent of the disc and so in principle it could have been discussed in section \ref{sec:vertical_extent}. However, we list it in a separate section because polarization introduces another layer of indirection with respect to the other methods measuring the vertical extent, which rely on simpler radiative transfer and geometrical considerations. Proto-planetary disc dust emission at mm wavelengths is polarized because the dust grains have a significant scattering opacity, and therefore self-scatter their own thermal emission. \citet{OhashiKataoka2019}, analysing ALMA polarization observations of HD163296, noted that the height of the dust layer affects the polarization pattern: thin dust layers tend to produce a polarization pattern aligned with the disc minor axis while thick dust layers have an azimuthal polarization pattern. By performing radiative transfer modelling, in comparison with the observations, they conclude that they can only explain the morphology in the observations if they assume that the inner part of the disc is thin (1/3 of the gas scale-height) while outside 70 au the disc must be thicker (at least 2/3 of the gas scale-height). These constraints are complementary to the other constraints on the same source discussed in section \ref{sec:vertical_extent} and section \ref{sec:radial_width} since those concern the rings, while this method can best be applied in the gaps.

\citet{Ueda2020} uses instead the fraction of polarized intensity to constrain the turbulence in HL Tau. This is a complex and degenerate problem because the polarization fraction depends sensitively on the maximum grain size and the emission in HL Tau is most likely optically thick. This implies that one cannot see emission from all the dust grains, but only from those above the $\tau=1$ surface; because of differential settling, these are in general smaller than those in the midplane. The emission at longer wavelengths, such as ALMA band 3, is however dominated by these larger grains. Disentangling the relation between the polarization fraction and how this changes with wavelengths is therefore complex, but nevertheless the authors estimate that the grain size must be $\sim$ mm, with low values of turbulence implied ($\alphaSS < 10^{-5}$), in line with the results of \citet{Pinte2016}.

\subsection{Disc-planet interaction}
\label{sec:disc-planet}

This is arguably the most indirect of the methods we discuss in this section since it needs to go through another layer of uncertainty, our understanding of disc-planet interaction (see \citealt{PaardekooperPPVII} for a recent review). Even more fundamentally, this method needs to assume that planets are present in proto-planetary discs and creating the structures (gaps, rings, etc) we observe, which is far from established.

\citet{Liu2018HD163296} applied disc-planet interaction models to the disc around HD163296, finding that they need a radially varying profile of $\alphaSS$ to fit the observations. Their analysis has not been repeated however for the higher resolution data of the DSHARP survey.

\citet{Zhang2018DSHARP} discusses how the morphology of structures created by planets depend on turbulence. With low viscosity ($\alphaSS = 10^{-4}$), the structures tend to become asymmetrical, showing for example vortices triggered by Rossby Wave instability (see e.g. \citealt{GodonLivio1999,Ataiee2013,Regaly2017,Rometsch2021} for dedicated studies) and long-lived dust accumulations at the Lagrangian points \citep{Montesinos2020}. Since most of the discs in the DSHARP survey do not show asymmetries, overall this disfavours low viscosities of $10^{-4}$ and points towards values of $\alphaSS \gtrsim 10^{-3}$.

This conclusion excluding low viscosities however does not apply to all discs - after all, asymmetries \textit{are} observed, although less frequently than azimuthally symmetric discs. Additionally, from the theoretical side, asymmetries are not expected to last forever but eventually dissipate, as vortices decay and dust eventually diffuses out of the Lagrangian points. The lifetime depends on viscosity and so the statistical incidence of discs with asymmetries in an observed population can be used to place constraints on turbulence, though one should bear in mind that the lifetime depends also on other parameters (most importantly here, the mass of the planet creating the structure). \citet{Hammer2021} performed this experiment using the fraction of detected discs with asymmetries, in comparison with the Taurus survey by \citet{Long2018}. They conclude that in this case $\alphaSS \geq 10^{-4}$, but this estimate is highly uncertain since it based on low number statistics. 

In addition to asymmetries, in a few cases observations show multiple, narrow gaps close together: e.g. two of the gaps in HL Tau \citep{HLTau2015}, the gaps in AS 209 \citep{Zhang2018DSHARP}, HD169142 \citep{Perez2019} and LkCa 15 \citep{Facchini2020}. Rather than requiring a planet per gap, it has been noted in the simulations that a single planet can create multiple gaps \citep{Bae2017,Dong2017,Dong2018} provided that the viscosity is low enough ($\alphaSS \leq 10^{-4}$). Many recent developments in the theory are making however difficult to understand the exact implications of multiple gaps; they can also be opened by migrating planets \citep{Meru2019,Nazari2019,Weber2019} and the value of the viscosity at which the gap goes from single to multiple is degenerate with the cooling timescale and planet mass \citep{MirandaRafikov2019,Facchini2020,Zhang2020}; it is even possible that in discs with realistic cooling planets do not open multiple gaps \citep{Ziampras2020}. At the moment, multiple, narrow gaps remain suggestive of low viscosity, but it is hard to be quantitative on the exact value implied.

Finally, \citet{deJuanOvelar2016} analysed the observational morphology of structures created by planets for different values of the viscosity considering also the effect of grain growth - as such this result is more indirect than the others in this subsection. They find that with $\alphaSS=10^{-2}$ dust diffusion results in weak dust trapping outside planetary orbits, which appears incompatible with the strong trapping observed in transition discs. This is a similar conclusion to that of \citet{Zormpas2022} that we will discuss in section \ref{sec:populations}. For $\alphaSS=10^{-4}$, dust growth proceeds very effectively in the traps, depleting the small grains and producing strong rings in scattered light which are normally not observed. For this reason they prefer an intermediate value of the viscosity of $\alphaSS=10^{-3}$.

In synthesis, the brief review of this topic highlights that, despite the high number of studies, it is hard to draw definite conclusions using disc-planet interaction because it is a highly degenerate problem. The more robust conclusion is probably that we can exclude the highest values of viscosity (e.g. \citealt{deJuanOvelar2016}) since explaining the observed structures would require very massive planets to trap dust; on the other hand the scarcity of asymmetries probably excludes low values of the viscosity for the general population, but individual sources highly suggestive of low viscosity do exist. Further progress on this topic may come by coupling these studies with those using planetary kinematics \citep{PintePPVII}, since they estimate the unseen planet mass in a way almost independent of viscosity, removing one of the strongest sources of degeneracy in these studies.

\section{Indirect methods - population level diagnostics}
\label{sec:populations}

In contrast to the previous section, we describe here methods that study whole populations. Turbulence controls angular momentum transport in the disc, which in turn affects the evolution of disc properties (mass, radius, ...) over time; from the observational study of these properties one can therefore constrain the properties of turbulence. Since it is not possible to study the evolution of a single disc over time, surveys offer a substitute by providing samples of discs at different evolutionary stages. From the observational side, the study of disc populations has flourished in the last few years, with hundreds of proto-planetary discs surveyed over multiple star forming regions. Disc surveys have been recently reviewed by \citet{ManaraPPVII}; the review also contains an extensive discussion of the theoretical implications of disc surveys on disc evolution theory, but in the spirit of this review here we will focus only on the constraints on turbulence and we will not enter the debate regarding which process is the main driver of disc evolution. As before, we list the methods in tentative decreasing order of directness.

\subsection{Radius evolution}
\label{sec:radius_evolution}

We consider this as the most direct method since, in absence of other mechanisms truncating the disc, disc expansion is a natural consequence of the angular momentum transfer operated by turbulence. Already in the pre-ALMA era some studies attempted to establish whether the disc size followed a trend with stellar age \citep{Isella2009,Guilloteau2011}. Unfortunately establishing what is the ``disc size'' is surprisingly more nuanced than one might imagine, since it depends on the observational tracer. In the pre-ALMA era, in the vast majority of cases it meant using the dust continuum size. Dust however is expected to drift inwards in proto-planetary discs, as a result of aerodynamic drag with the gas; this does not make it a good tracer of expansion. Although an analysis with dust evolution models \citep{Rosotti2019Paper} finds that overall viscous expansion is expected to overcome the shrinking effect of radial drift, in practise tracing this expansion requires exceptionally high sensitivity even by ALMA standards. At the typical sensitivity of disc surveys, we expect shrinking to dominate, in line with what is observed (see \citealt{Tobin2020,Hendler2020} for the observational data and \citealt{Zagaria2022} for a comparison with evolutionary models).

\begin{figure}
\includegraphics[width=\columnwidth]{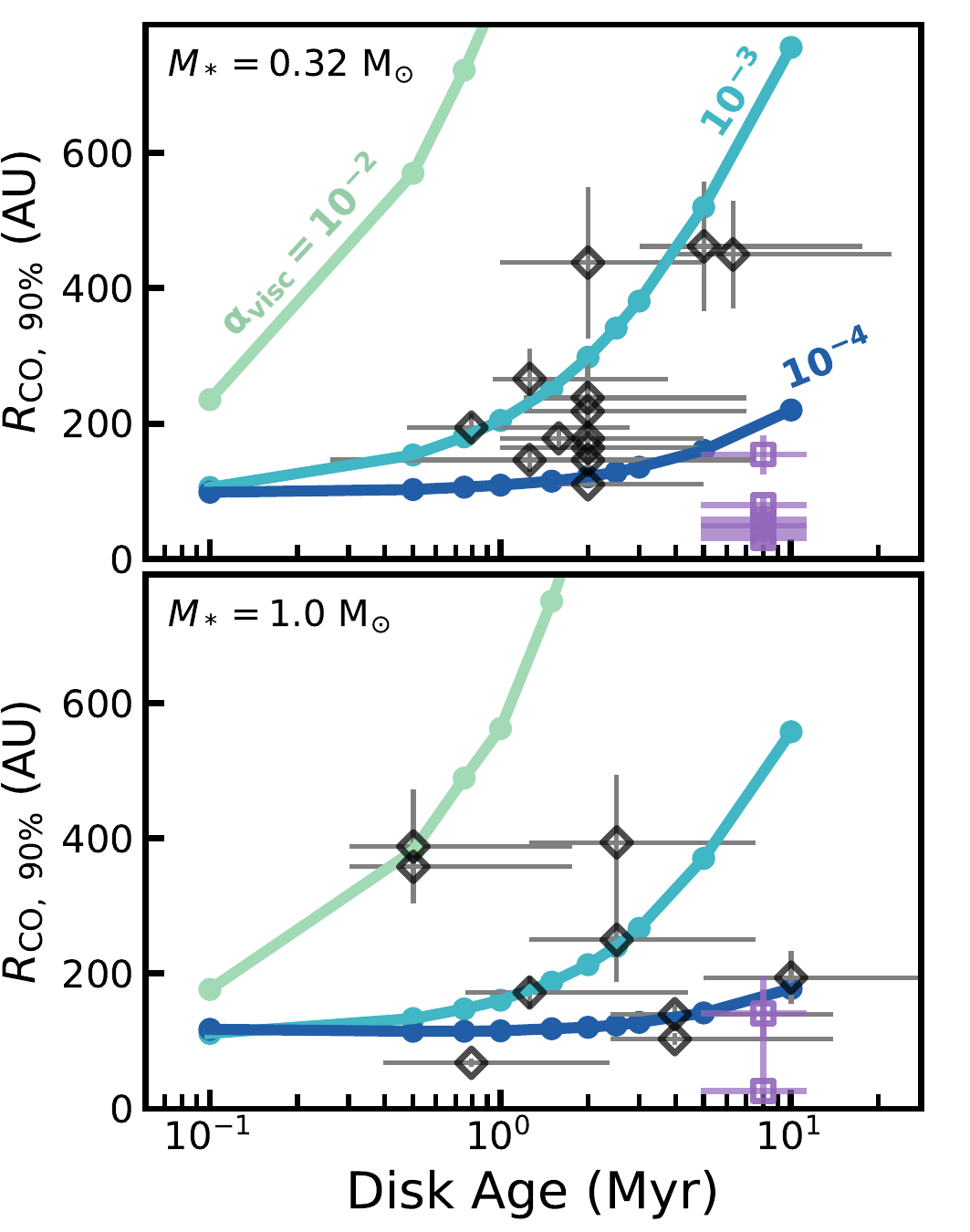}
\caption{Disc size as a function time from models (coloured lines) and observations (datapoints), from \citet{Trapman2020}. The two panels correspond to different stellar masses. While no trend is visible with stellar age, if the viscosity was as high as $\alphaSS=10^{-2}$ we should expect to see much larger discs than observed. We can therefore conclude that the viscosity must be lower than this value.}
\label{fig:disc_expansion}
\end{figure}

Since most of the mass is in the gas, it is natural to employ gas tracers. \citet{NajitaBergin2018} first reported a trend of expanding disc size with stellar age from multiple gas tracers. However, disc sizes in the sample were measured with different tracers, complicating the comparison. Observationally, the gas tracer with the largest available sample is CO \citep{Barenfeld2017,Ansdell2018,Sanchis2021}, although it has the significant drawback of being highly optically thick, as well as subject to abundance variations in the disc due to photo-dissociation and freeze-out. The \textit{observed} CO radius therefore does not coincide with the \textit{physical} radius, and only through modelling we can link the two. \citet{Trapman2020} confirmed through detailed thermo-chemical modelling that, these problems notwithstanding, the CO radius is a good tracer of disc expansion. However, the authors did not find any trend with age in the available data (see \autoref{fig:disc_expansion}), implying that there is no observational evidence of viscous spreading. \citet{Long2022}, with a larger sample, reached a similar conclusion. However, even with their additions, the sample with measured gas radii remains of limited size, masking any real trend. While waiting for larger samples of disc radii (such as those that will be provided by the ALMA large programmes AGE-PRO and DECO), we can however already draw an interim conclusion. High values of the viscosity ($\alphaSS \sim 10^{-2}$) lead to discs that are as large as hundreds of au. This conclusion does not depend on the (unknown) initial conditions because for such high values of the viscosity the initial memory of the initial conditions is quickly lost. Since such large discs are not observed (see also \citealt{Toci2021}), this places an upper limit of $\alphaSS \lesssim 10^{-3}$ on the viscosity.

\subsection{Spread in the $\dot{M}-M_\mathrm{disc}$ correlation}

\begin{figure}
\includegraphics[width=\columnwidth]{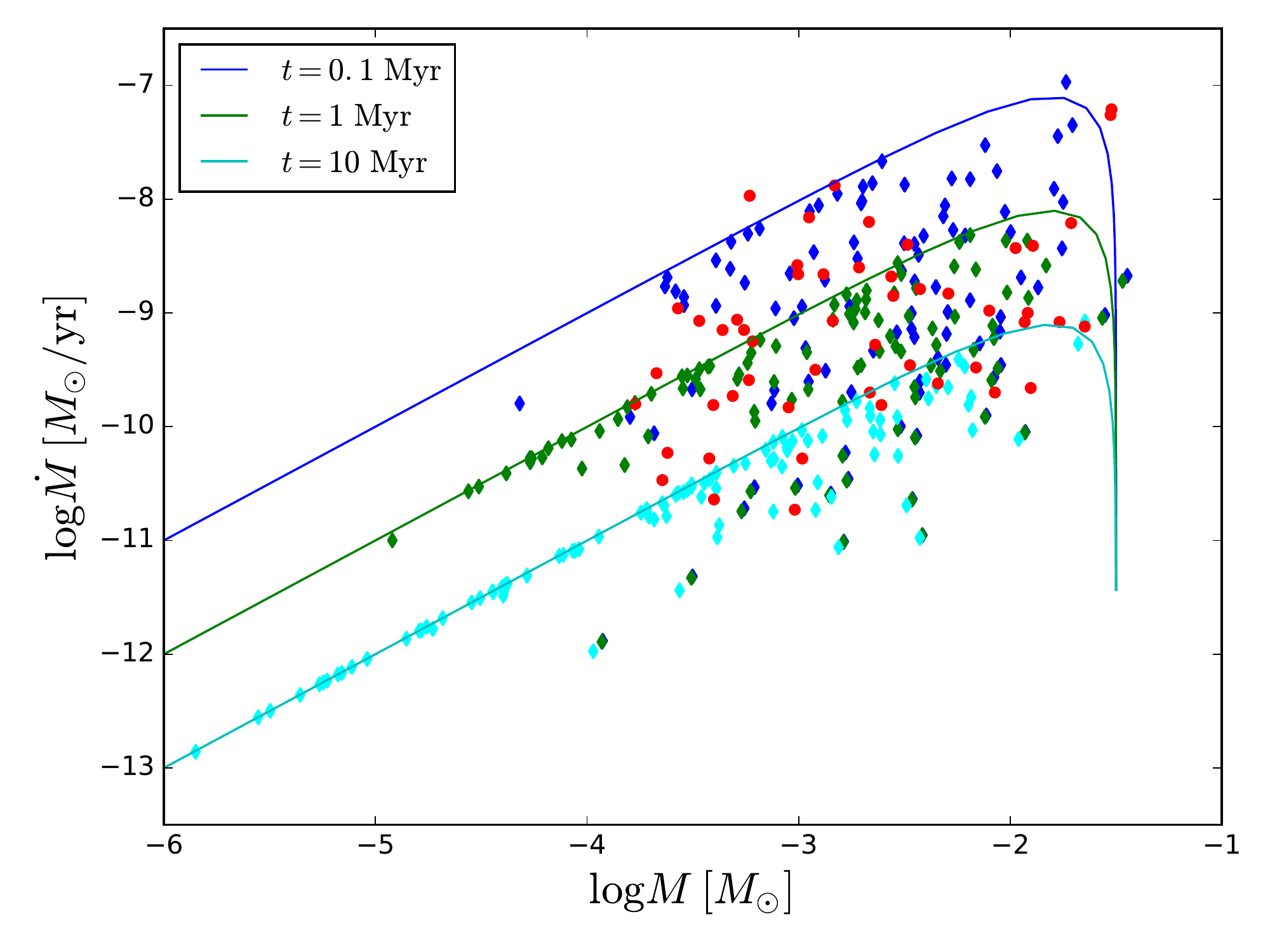}
\caption{Comparison between models and observations in the $\dot{M}-M_\mathrm{disc}$ plane, from \citet{Lodato2017}. The red points are observations in Lupus, whereas the coloured diamonds are a synthetic disc population at different times (see legend). The solid lines represent the theoretical isochrones, that is, the lines along which the points should lie provided that a time much longer than the viscous time has passed. Note the significant scatter in the observations and that the scatter in the models becomes smaller with time.}
\label{fig:spread_mdot_mdisc}
\end{figure}

\citet{Manara2016} first reported the discovery of a correlation between the mass accretion rate $\dot{M}$ and the disc mass $M_\mathrm{disc}$ in the Lupus star forming region. This is expected for viscous evolution \citep{Jones2012,Rosotti2017}, though the correlation is not a proof that accretion is driven by turbulence \citep{Mulders2017,Tabone2022Letter}. \citet{Lodato2017} showed that there is information also in the spread of the correlation. For viscously evolving discs, the spread tends to become smaller with time (see \autoref{fig:spread_mdot_mdisc}), eventually vanishing after enough time. This is because in viscous models one expects eventually to reach the condition $\dot{M}\simeq M_\mathrm{disc}/t$ (\citealt{Lodato2017} discusses the exact analytical expression), regardless of other parameters; the observed datapoints should therefore align on a line with negligible spread. The relevant timescale over which the spread becomes small is, naturally, the viscous time; the observed spread can then be used to constrain the viscous time, provided one has some idea of the spread in the initial conditions. Given the relatively high spread observed in Lupus (0.32 dex), the problem of the initial conditions is somehow bypassed since one can immediately conclude that the viscous timescale must be long compared to the age of the region. Making this analysis quantitative, they conclude that the viscous time must be of order Myr, implying a relatively low $\alphaSS$ of $\sim 5 \times 10^{-4}$ (see \autoref{eq:tnu}). A key prediction is that the spread in the correlation should become smaller with time. Observationally, this is not verified in the older (5-10 Myr vs 1-3 Myr of Lupus) Upper Sco region \citep{Manara2020}, where the spread is comparable to Lupus. It is unclear whether this is due to other evolutionary processes, such as dust evolution, photo-evaporation and unresolved binaries \citep{Sellek2020Correlation,Somigliana2020,Zagaria2022UpperSco}, adding further spread, or whether this is because disc evolution is not driven by turbulence. In this latter case, the constraint on turbulence reported by \citet{Lodato2017} should be regarded as an upper limit since another process must be dominating angular momentum transport.

\subsection{Flux-radius correlation}
We classify this method as more indirect than the previous ones because it relies also on dust growth and transport models. Surveys have established \citep{Tripathi2017,Andrews2018FluxRadius} that the disc continuum size correlates strongly with the sub-mm flux, with an exponent close to a value of 2. There are many interpretations for the origin of this correlation; a straightforward one would be that the disc is optically thick, though given the measured surface brightness temperatures the filling factor must be smaller than one.

\begin{figure}
\includegraphics[width=\columnwidth]{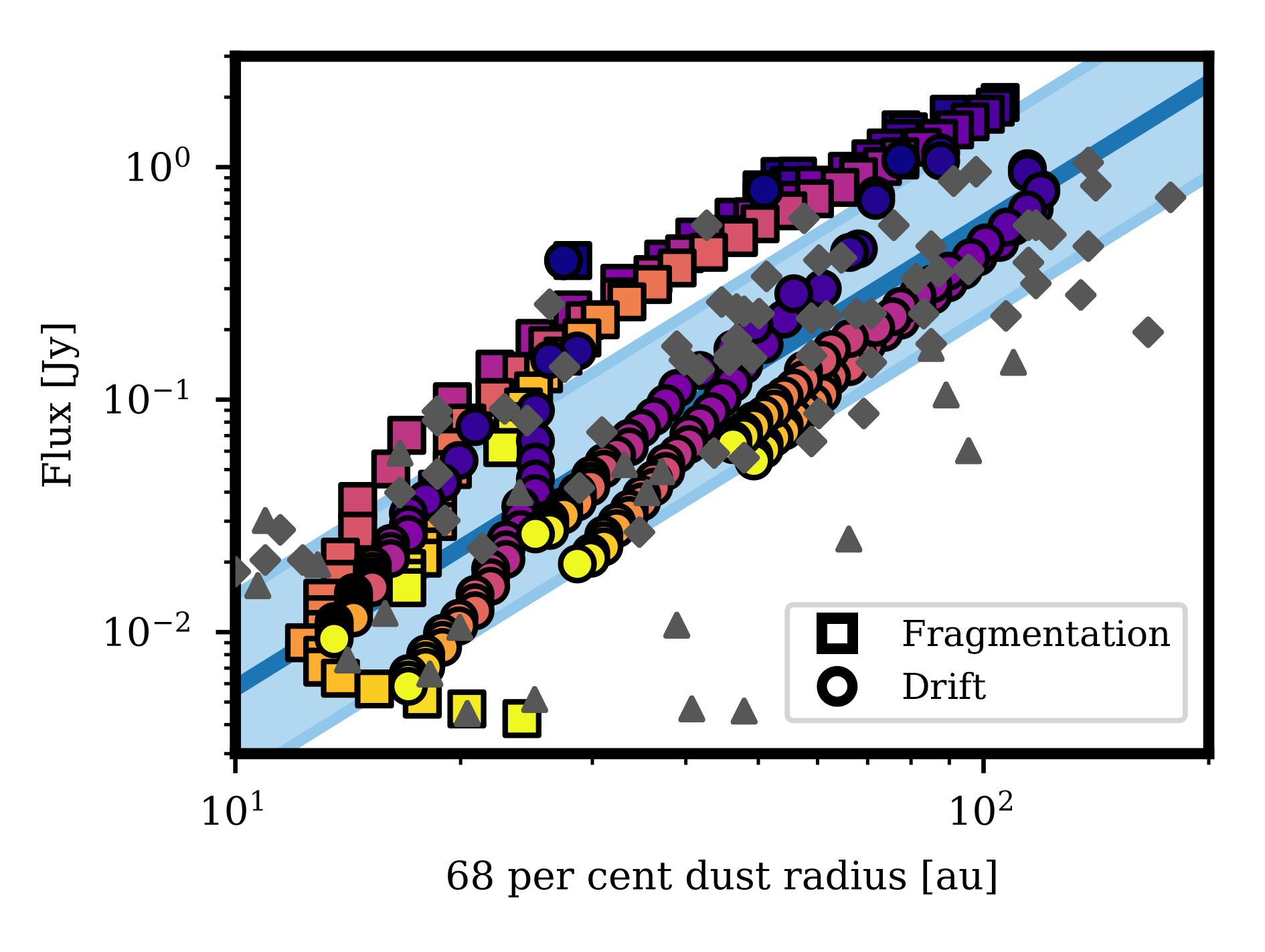}
\caption{Models (coloured symbols) and observations (grey points, from \citealt{Tripathi2017,Andrews2018FluxRadius}) on the sub-mm flux and disc continuum size parameter space, from \citet{Rosotti2019letter}. The blue line and shaded region represent the observed correlation with the associated spread. Models in the drift limited regime ($\alphaSS \lesssim 10^{-3}$) reproduce well the observed correlation, while models in the turbulent regime overpredict the observed fluxes for a given radius. Although these models are smooth, models including substructure \citep{Zormpas2022} reach a similar conclusion regarding the values of $\alphaSS$ that reproduce the observed correlation.}
\label{fig:flux_radius_correlation}
\end{figure}

\citet{Rosotti2019letter} showed that the correlation is a natural result of dust evolutionary processes, if the main processing limiting dust coagulation is radial drift (i.e., the grains grow as fast as they drift towards the star). In dust evolutionary model the other process that can limit grain growth is turbulent fragmentation: dust grains caught in turbulent eddies collide at relatively high velocities, leading to their fragmentation. Since this process is driven by turbulence, it dominates only for high enough values of $\alphaSS$ (in the outer disc, typically this happens for $\alphaSS \gtrsim 10^{-2}$, though the exact value depends also on the unknown fragmentation velocity). If the flux-radius correlation is due to grain growth being in the drift dominated regime, we can place an upper limit on $\alphaSS \lesssim 10^{-3}$, since the correlation is not established if turbulence dominates grain growth (see \autoref{fig:flux_radius_correlation}). We should however make the caveat that the models of \citet{Rosotti2019letter} assume that the disc is smooth, i.e. there are no dust traps. This is contradiction with the observational evidence that many discs show gaps and rings, as well with disc spectral indices \citep{Tazzari2021} and the relative gas and disc sizes \citep{Toci2021}. \citet{Zormpas2022} revisited the problem considering also non smooth discs, under the assumption that dust traps are created by planets. In this case the flux-radius correlation is established not as a result of dust growth processes, but because the dust emission has high optical depth due to dust accumulation in the rings. The correlation in this case has a slightly different slope, suggesting that maintaining the correlation over the observed range could need a combination of discs with substructure (those bright and large) and smooth discs (those small and faint). For the present discussion the most important conclusion is that, similar to the previous case, they find that significant trapping still requires similar values of $\alphaSS \lesssim 10^{-3}$; otherwise, strong diffusion prevents dust trapping and the models do not reproduce the flux - radius correlation. Note that this conclusion is possible because the authors include a physical model for forming traps, namely the presence of a planet; in this case the amplitude of a trap cannot be increased arbitrarily since it would correspond to large planet masses that would be easily detectable.

\subsection{Combining the three fundamental disc diagnostics}
\label{sec:most_empirical}

In principle, the instantaneous value of the viscosity $\alphaSS$ can be estimated from measurements of the disc radius, disc mass $M_d$ and mass accretion rate $\dot{M}$ from the following formula:
\begin{equation}
\alphaSS=\frac{\dot{M}}{M_d} \frac{\mu}{kT} \Omega r_d^2,
\label{eq:alpha_from_radius}
\end{equation}
where quantities are evaluated at the disc radius $r_d$. A few studies \citep{Hartmann1998,Andrews2009,Andrews2010,Rafikov2017,Ansdell2018} have attempted to apply this relation. However, we should make a caveat that, as we discussed in section \ref{sec:radius_evolution}, the observed radius is \textit{not} the physical radius. The existence of radial drift implies that one should not use dust radii \citep{Rosotti2019Paper}, but even for gas radii one should be careful since the difference can easily be as large as a factor ten \citep{Trapman2020}. Inverting the problem to go from to the observed to the physical radius is complex, and, to the best of our knowledge, no study yet has applied \autoref{eq:alpha_from_radius} to a large sample of \textit{physical} radii. The other thing to note is that the formula is at the same time the most empirical estimate we discuss in this review, but also the most indirect. This can be understood remembering that $\alphaSS$ was introduced to explain the observational fact that discs accrete; the formula above quantifies the values of $\alphaSS$ needed to explain accretion. On the other hand, the formula cannot tell us that accretion is indeed driven through turbulence; the values it returns must be understood as an \textit{effective} $\alphaSS$ and any mechanism invoked to explain angular momentum transport in discs would need an equivalent efficiency of angular momentum transport.

Nevertheless, in line with the empirical spirit of this chapter, it is worth to reflect more on the range of viscosities allowed by \autoref{eq:alpha_from_radius}. We will consider here the observational sample of \citet{Ansdell2018} since it is the only one available for gas tracers. In this case, we can make the safe assumption that the observed radii are always larger than the physical radius due to the high CO optical depth (see e.g. the results of \citealt{Trapman2020}). This implies that values obtained by applying \autoref{eq:alpha_from_radius} are overestimates of $\alphaSS$, assuming that the error coming from the determination of mass accretion rate and disc mass are smaller than those coming from the radius\footnote{While this is arguably true for the mass accretion rate, the determination of disc masses is notoriously complex. We will not enter this debate and refer the reader to \citet{MiotelloPPVII} for an extensive discussion.}. \citet{Ansdell2018} obtains values ranging from $\sim 10^{-4}$ to $\sim 10^{-1}$, with a median of roughly $3 \times 10^{-3}$. The results of \citet{Trapman2020} imply that these values may be overestimated of a factor up to 10. Therefore, a median $\alphaSS = 3 \times 10^{-4} - 3 \times 10^{-3}$ is the empirical value required to explain accretion. While this consideration applies to the median, note that it is harder to interpret the large range of variation; this may be genuine (i.e., $\alphaSS$ varies from disc to disc) but could also reflect the observational errors that go into \autoref{eq:alpha_from_radius}, e.g. a variation of the dust-to-gas ratio from disc to disc. The scatter could also be due to the fact that the formula makes the assumption that the large scales at the disc outer radius are linked to the accretion rate onto the star. This needs to be true on secular timescales since most of the mass of the disc is at large radii, but it does not need to be true at any instant; part of the scatter may then reflect time variability and localised variations on $\alphaSS$.

\section{Discussion}
\label{sec:discussion}

\subsection{Breaking the degeneracy between $\alphaSS$ and $\mathrm{St}$}
\label{sec:discussion_degeneracy}

As already discussed in sections \ref{sec:vertical_extent} and \ref{sec:radial_width}, those methods measure $\alphaSS/\St$. It would thus be highly beneficial to have complementary measurements of $\mathrm{St}$, which implies (see \autoref{eq:St}) to measure the gas surface density and the dust grain size, as was done for example by \citet{Sierra2019}. These are both difficult problems; the former requires having multiple line transitions \citep[e.g.][]{Zhang2021MAPS} and is affected by the unknown abundance of chemical tracers, while the latter requires multiwavelength continuum studies and is affected by theoretical uncertainties in the dust opacity (see e.g. the comparison between \citealt{Sierra2021} and \citealt{Guidi2022}), as well as observational uncertainties regarding what is the best way to measure spectral indices from observations (e.g. visibility vs image plane analysis). Nevertheless, significant steps forward are expected since observational studies targeting multiple molecular transitions are now becoming the norm, and for what concerns grain growth the planned ALMA band 1 upgrade should give a significant boost to those studies.

\subsection{Summary of existing constraints}

\begin{table*}
\begin{tblr}{cQ[c,11cm]}
\toprule
Method & Constraints\\
\midrule
IR lines & Supersonic turbulence\\
Sub-mm lines & High ($\alphaSS \sim 10^{-2}$) turbulence in 2 discs out of 6; $\alphaSS < 10^{-2}$ in the other cases\\
Disc vertical extent & In general $\alphaSS \sim 10^{-4}$\\
Radial width & In most cases $\alphaSS \lesssim 5 \times 10^{-3}$ \\
Disc-planet interaction & Excludes $\alphaSS = 10^{-2}$. Weaker evidence for $\alphaSS \gtrsim 10^{-4}$ in most discs\\
Disc radius evolution & $\alphaSS \lesssim 10^{-3}$\\
Spread in the $\dot{M}-M_\mathrm{disc}$ correlation & $\alphaSS \lesssim 10^{-3}$\\
Flux-radius correlation & $\alphaSS \lesssim 10^{-3}$\\
\midrule
Empirical constraint from accretion & $\alphaSS = 3 \times 10^{-4} - 3 \times 10^{-3}$\\
\bottomrule
\end{tblr}
\caption{Summary of the existing constraints that have been discussed in this review.}
\label{table:summary}
\end{table*}

We summarise the existing constraints on turbulence we have discussed in this review in \autoref{table:summary}. We stress that there is inevitably some degree of simplification involved when summarising multiple works into a single line in a table; we have tried here to summarise the general trend for the sake of discussion, but we refer the reader back to the individual sections to have a full overview of the existing constraints.

The fist thing we wish to note is appreciating the variety of methods that the field has devised, especially considering this is such a complex problem to tackle directly. We are fortunate to be supported by instruments that allow us to attack this problem from multiple directions. On the other hand, having multiple indicators could be a curse rather than a blessing if they were to disagree. Fortunately this is not the case - we will discuss now common trends among these studies.

One of the common themes is that most tracers agree in excluding that discs are highly turbulent ($\alphaSS \sim 10^{-2}$). Although individual discs (such as DM Tau and IM Lup, as well as some of thick discs in the vertical direction discussed in section \ref{sec:vertical_extent}) where turbulence is high do exist, the evidence available up to date allows us to exclude that this is true \textit{at the population level}. This also agrees with the most empirical constraints we discussed in section \ref{sec:most_empirical}, which consist in a median $\alphaSS=3 \times 10^{-4} - 3 \times 10^{-3}$. As discussed those should be regarded as constraints on the mechanism driving accretion rather than constraints on turbulence per se; this means that turbulence can be lower than this value if accretion is driven by another mechanism, but finding a higher value would be problematic. This conclusion excluding the case of high turbulence is far from obvious; a value $\alphaSS=10^{-2}$ was regarded until recently (e.g. \citealt{Hartmann1998}) as canonical and marks the progress happened in the last few years. Although we have not discussed them in this review, this also relaxes the constrains on the theoretical mechanisms generating turbulence which do not need to be so efficient and it is perhaps more in line with the theoretical idea, mentioned in the introduction, that a large part of the disc should be a \textit{dead zone}.

The other question we need to ask is whether we have evidence that discs are indeed turbulent. The strongest evidence of turbulence comes from direct detection, but we note that this is available only in a very small number of cases. It is mildly worrying that none of the constraints we presented at the population level in section \ref{sec:populations} implies that discs are turbulent, since they only provide upper limits. In this case, if discs are not turbulent at all, then another mechanism is needed to explain how discs accrete and evolve, with MHD winds currently being the best alternative (see \citealt{LesurPPVII} and \citealt{ManaraPPVII}). Going to the case of individual discs, more circumstantial evidence is that the radial substructures observed in proto-planetary discs (see section \ref{sec:radial_width}) in most cases have a small, but finite and measurable width - this can be said also of the vertical disc extent (see section \ref{sec:vertical_extent} and \ref{sec:polarization}), though in a more limited number of cases because in most cases the disc height is lower than we can measure with the current quality of the data. This implies that at least some finite level of turbulence (however small) must be present. Interpreting the evidence coming from planet-disc interaction is more difficult, as we discussed in section \ref{sec:disc-planet}; the lack of asymmetries seems to exclude very low turbulence, but this is tentative. In any case, even if discs are turbulent, it is still unknown if the turbulence implied by these observations is enough to explain the observed accretion rates; it is also conceivable that we are in a situation in which accretion is driven by another process (e.g. MHD winds) but the disc is characterised by non-negligible turbulence.

If this were the case, we note that the enterprise of characterising turbulence would still be relevant for our understanding of planet formation, even if it would no longer serve the purpose (explaining accretion) for which it was introduced. This is because many processes happening in proto-planetary discs are sensitive to the local value of the turbulence; to make just a few examples, the efficiency of dust accretion onto forming planets \citep{JohansenLambrechtsReview}, the formation of planetesimals through the streaming instability \citep{DrazkowskaPPVII}, the location of the water snow line \citep{MinDullemond2011} and the vertical mixing of molecular species \citep{SemenovWiebe2011,Krijt2020}, affecting the chemical composition of planets.

\section{Conclusions}
\label{sec:conclusions}

In this review we have summarised the existing empirical constraints on turbulence in proto-planetary discs. Turbulence is commonly invoked to explain the observational evidence that discs accrete. The last few years have seen an explosion in this sub-field, no doubt mainly (but not only) due to the advent of the ALMA telescope. The magnitude of turbulence is a very long-standing question which even pre-dates the observational discovery of proto-planetary discs and it is remarkable to see how much progress has been made on this topic in the last few years. There is now direct detection of turbulence both in the IR and in sub-mm lines, though still in a very limited number of cases, and the field has been very creative in inventing multiple methods to constrain turbulence empirically. Aside from exceptional cases, however, we have not proven yet that discs are indeed turbulent and we do not know how the properties of turbulence vary over the disc population. The fact that radial sub-structures in discs have a finite, measurable radial width is a strong indication that turbulence is operating in these rings, but it is not a definite proof. The same argument applies to the disc vertical extent, though it concerns a more limited number of cases since in many cases it is only possible to give upper limits for the disc vertical extent. For these reasons, it is still an open question whether the levels of turbulence in proto-planetary discs are enough to explain accretion, or if another mechanism is needed. The strongest conclusion to date, supported by many of the methods we have described in this review, is that turbulence is weaker than what used to be the commonly used value in the field until a few years ago of $\alphaSS=10^{-2}$. In line with this conclusion, from modern surveys explaining the observed accretion rates requires an efficiency of angular momentum transport $\alphaSS=3 \times 10^{-4} - 3 \times 10^{-3}$.

Looking at the future, it is remarkable that, even if ALMA has been operating for more than a decade, its potential is far from being exhausted; for example large programmes are planned in the short term (exoALMA, AGE-PRO and DECO) which will significantly expand the sample sizes and the precisions of the methods we have discussed. At this moment, it looks like the momentum of the field is still increasing, though of course eventually a successor or an upgrade to the ALMA capabilities (as already being discussed; see \citealt{CarpenterALMAUpgrade2022}) will be needed to make further progress. As highlighted in section \ref{sec:IR_lines}, we expect also CRIRES+ and GRAVITY on the VLT to contribute to direct detection of turbulence in the IR, and looking on the longer term the Extremely Large Telescope (ELT) will surely revolutionise this aspect as well.

\section*{Funding}
GR acknowledges support from the Netherlands Organisation for Scientific Research (NWO, program number 016.Veni.192.233) and from an STFC Ernest Rutherford Fellowship (grant number ST/T003855/1). This project has received funding from the European Research Council (ERC) under the European Union's Horizon Europe Research \& Innovation Programme grant agreement No 101039651 (DiscEvol). Views and opinions expressed are however those of the author(s) only and do not necessarily reflect those of the European Union or the European Research Council. Neither the European Union nor the granting authority can be held responsible for them.

\section*{Acknowledgements}
We thank the Editors for their invitation to write this review, and in particular Cathie Clarke for her patience and ongoing support of this work, and an anonymous reviewer for their insightful and enthusiastically positive comments. We thank John Ilee for discussions on CO emission and excitation and Marion Villenave for discussions on dust settling.



\bibliographystyle{model2-names-astronomy}
\bibliography{turb_biblio}

\begin{thebibliography}{141}
\expandafter\ifx\csname natexlab\endcsname\relax\def\natexlab#1{#1}\fi
\providecommand{\url}[1]{\texttt{#1}}
\providecommand{\href}[2]{#2}
\providecommand{\path}[1]{#1}
\providecommand{\DOIprefix}{doi:}
\providecommand{\ArXivprefix}{arXiv:}
\providecommand{\URLprefix}{URL: }
\providecommand{\Pubmedprefix}{pmid:}
\providecommand{\doi}[1]{\href{http://dx.doi.org/#1}{\path{#1}}}
\providecommand{\Pubmed}[1]{\href{pmid:#1}{\path{#1}}}
\providecommand{\bibinfo}[2]{#2}
\ifx\xfnm\relax \def\xfnm[#1]{\unskip,\space#1}\fi
\bibitem[{{Alarc{\'o}n} et~al.(2021){Alarc{\'o}n}, {Bosman}, {Bergin}, {Zhang},
  {Teague}, {Bae}, {Aikawa}, {Andrews}, {Booth}, {Calahan}, {Cataldi},
  {Czekala}, {Huang}, {Ilee}, {Law}, {Le Gal}, {Liu}, {Long}, {Loomis},
  {M{\'e}nard}, {{\"O}berg}, {Schwarz}, {van't Hoff}, {Walsh} and
  {Wilner}}]{Alarcon2021}
{Alarc{\'o}n}, F. et~al., \bibinfo{year}{2021}.
\newblock \bibinfo{title}{{Molecules with ALMA at Planet-forming Scales (MAPS).
  VIII. CO Gap in AS 209-Gas Depletion or Chemical Processing?}}
\newblock \bibinfo{journal}{\apjs} \bibinfo{volume}{257}, \bibinfo{pages}{8}.
\newblock \DOIprefix\doi{10.3847/1538-4365/ac22ae},
  \href{http://arxiv.org/abs/2109.06263}{{\tt arXiv:2109.06263}}.
\bibitem[{{ALMA Partnership} et~al.(2015){ALMA Partnership}, {Brogan},
  {P{\'e}rez}, {Hunter}, {Dent}, {Hales}, {Hills}, {Corder}, {Fomalont},
  {Vlahakis}, {Asaki}, {Barkats}, {Hirota}, {Hodge}, {Impellizzeri}, {Kneissl},
  {Liuzzo}, {Lucas}, {Marcelino}, {Matsushita}, {Nakanishi}, {Phillips},
  {Richards}, {Toledo}, {Aladro}, {Broguiere}, {Cortes}, {Cortes}, {Espada},
  {Galarza}, {Garcia-Appadoo}, {Guzman-Ramirez}, {Humphreys}, {Jung}, {Kameno},
  {Laing}, {Leon}, {Marconi}, {Mignano}, {Nikolic}, {Nyman}, {Radiszcz},
  {Remijan}, {Rod{\'o}n}, {Sawada}, {Takahashi}, {Tilanus}, {Vila Vilaro},
  {Watson}, {Wiklind}, {Akiyama}, {Chapillon}, {de Gregorio-Monsalvo}, {Di
  Francesco}, {Gueth}, {Kawamura}, {Lee}, {Nguyen Luong}, {Mangum}, {Pietu},
  {Sanhueza}, {Saigo}, {Takakuwa}, {Ubach}, {van Kempen}, {Wootten},
  {Castro-Carrizo}, {Francke}, {Gallardo}, {Garcia}, {Gonzalez}, {Hill},
  {Kaminski}, {Kurono}, {Liu}, {Lopez}, {Morales}, {Plarre}, {Schieven},
  {Testi}, {Videla}, {Villard}, {Andreani}, {Hibbard} and
  {Tatematsu}}]{HLTau2015}
{ALMA Partnership}et~al., \bibinfo{year}{2015}.
\newblock \bibinfo{title}{{The 2014 ALMA Long Baseline Campaign: First Results
  from High Angular Resolution Observations toward the HL Tau Region}}.
\newblock \bibinfo{journal}{\apjl} \bibinfo{volume}{808}, \bibinfo{pages}{L3}.
\newblock \DOIprefix\doi{10.1088/2041-8205/808/1/L3},
  \href{http://arxiv.org/abs/1503.02649}{{\tt arXiv:1503.02649}}.
\bibitem[{{Andrews} et~al.(2018a){Andrews}, {Huang}, {P{\'e}rez}, {Isella},
  {Dullemond}, {Kurtovic}, {Guzm{\'a}n}, {Carpenter}, {Wilner}, {Zhang}, {Zhu},
  {Birnstiel}, {Bai}, {Benisty}, {Hughes}, {{\"O}berg} and
  {Ricci}}]{Andrews2018}
{Andrews}, S.~M. et~al., \bibinfo{year}{2018}a.
\newblock \bibinfo{title}{{The Disk Substructures at High Angular Resolution
  Project (DSHARP). I. Motivation, Sample, Calibration, and Overview}}.
\newblock \bibinfo{journal}{\apjl} \bibinfo{volume}{869}, \bibinfo{pages}{L41}.
\newblock \DOIprefix\doi{10.3847/2041-8213/aaf741},
  \href{http://arxiv.org/abs/1812.04040}{{\tt arXiv:1812.04040}}.
\bibitem[{{Andrews} et~al.(2018b){Andrews}, {Terrell}, {Tripathi}, {Ansdell},
  {Williams} and {Wilner}}]{Andrews2018FluxRadius}
\bibinfo{author}{{Andrews}, S.M.}, \bibinfo{author}{{Terrell}, M.},
  \bibinfo{author}{{Tripathi}, A.}, \bibinfo{author}{{Ansdell}, M.},
  \bibinfo{author}{{Williams}, J.P.}, \bibinfo{author}{{Wilner}, D.J.},
  \bibinfo{year}{2018}b.
\newblock \bibinfo{title}{{Scaling Relations Associated with Millimeter
  Continuum Sizes in Protoplanetary Disks}}.
\newblock \bibinfo{journal}{\apj} \bibinfo{volume}{865}, \bibinfo{pages}{157}.
\newblock \DOIprefix\doi{10.3847/1538-4357/aadd9f},
  \href{http://arxiv.org/abs/1808.10510}{{\tt arXiv:1808.10510}}.
\bibitem[{{Andrews} et~al.(2009){Andrews}, {Wilner}, {Hughes}, {Qi} and
  {Dullemond}}]{Andrews2009}
\bibinfo{author}{{Andrews}, S.M.}, \bibinfo{author}{{Wilner}, D.J.},
  \bibinfo{author}{{Hughes}, A.M.}, \bibinfo{author}{{Qi}, C.},
  \bibinfo{author}{{Dullemond}, C.P.}, \bibinfo{year}{2009}.
\newblock \bibinfo{title}{{Protoplanetary Disk Structures in Ophiuchus}}.
\newblock \bibinfo{journal}{\apj} \bibinfo{volume}{700},
  \bibinfo{pages}{1502--1523}.
\newblock \DOIprefix\doi{10.1088/0004-637X/700/2/1502},
  \href{http://arxiv.org/abs/0906.0730}{{\tt arXiv:0906.0730}}.
\bibitem[{{Andrews} et~al.(2010){Andrews}, {Wilner}, {Hughes}, {Qi} and
  {Dullemond}}]{Andrews2010}
\bibinfo{author}{{Andrews}, S.M.}, \bibinfo{author}{{Wilner}, D.J.},
  \bibinfo{author}{{Hughes}, A.M.}, \bibinfo{author}{{Qi}, C.},
  \bibinfo{author}{{Dullemond}, C.P.}, \bibinfo{year}{2010}.
\newblock \bibinfo{title}{{Protoplanetary Disk Structures in Ophiuchus. II.
  Extension to Fainter Sources}}.
\newblock \bibinfo{journal}{\apj} \bibinfo{volume}{723},
  \bibinfo{pages}{1241--1254}.
\newblock \DOIprefix\doi{10.1088/0004-637X/723/2/1241},
  \href{http://arxiv.org/abs/1007.5070}{{\tt arXiv:1007.5070}}.
\bibitem[{{Ansdell} et~al.(2018){Ansdell}, {Williams}, {Trapman}, {van
  Terwisga}, {Facchini}, {Manara}, {van der Marel}, {Miotello}, {Tazzari},
  {Hogerheijde}, {Guidi}, {Testi} and {van Dishoeck}}]{Ansdell2018}
{Ansdell}, M. et~al., \bibinfo{year}{2018}.
\newblock \bibinfo{title}{{ALMA Survey of Lupus Protoplanetary Disks. II. Gas
  Disk Radii}}.
\newblock \bibinfo{journal}{\apj} \bibinfo{volume}{859}, \bibinfo{pages}{21}.
\newblock \DOIprefix\doi{10.3847/1538-4357/aab890},
  \href{http://arxiv.org/abs/1803.05923}{{\tt arXiv:1803.05923}}.
\bibitem[{{Ataiee} et~al.(2013){Ataiee}, {Pinilla}, {Zsom}, {Dullemond},
  {Dominik} and {Ghanbari}}]{Ataiee2013}
\bibinfo{author}{{Ataiee}, S.}, \bibinfo{author}{{Pinilla}, P.},
  \bibinfo{author}{{Zsom}, A.}, \bibinfo{author}{{Dullemond}, C.P.},
  \bibinfo{author}{{Dominik}, C.}, \bibinfo{author}{{Ghanbari}, J.},
  \bibinfo{year}{2013}.
\newblock \bibinfo{title}{{Asymmetric transition disks: Vorticity or
  eccentricity?}}
\newblock \bibinfo{journal}{\aap} \bibinfo{volume}{553}, \bibinfo{pages}{L3}.
\newblock \DOIprefix\doi{10.1051/0004-6361/201321125},
  \href{http://arxiv.org/abs/1304.1736}{{\tt arXiv:1304.1736}}.
\bibitem[{{Bae} et~al.(2022){Bae}, {Isella}, {Zhu}, {Martin}, {Okuzumi} and
  {Suriano}}]{BaePPVII}
\bibinfo{author}{{Bae}, J.}, \bibinfo{author}{{Isella}, A.},
  \bibinfo{author}{{Zhu}, Z.}, \bibinfo{author}{{Martin}, R.},
  \bibinfo{author}{{Okuzumi}, S.}, \bibinfo{author}{{Suriano}, S.},
  \bibinfo{year}{2022}.
\newblock \bibinfo{title}{{Structured Distributions of Gas and Solids in
  Protoplanetary Disks}}.
\newblock \bibinfo{journal}{arXiv e-prints} ,
  \bibinfo{pages}{arXiv:2210.13314}\href{http://arxiv.org/abs/2210.13314}{{\tt
  arXiv:2210.13314}}.
\bibitem[{{Bae} et~al.(2017){Bae}, {Zhu} and {Hartmann}}]{Bae2017}
\bibinfo{author}{{Bae}, J.}, \bibinfo{author}{{Zhu}, Z.},
  \bibinfo{author}{{Hartmann}, L.}, \bibinfo{year}{2017}.
\newblock \bibinfo{title}{{On the Formation of Multiple Concentric Rings and
  Gaps in Protoplanetary Disks}}.
\newblock \bibinfo{journal}{\apj} \bibinfo{volume}{850}, \bibinfo{pages}{201}.
\newblock \DOIprefix\doi{10.3847/1538-4357/aa9705},
  \href{http://arxiv.org/abs/1706.03066}{{\tt arXiv:1706.03066}}.
\bibitem[{{Balbus} and {Hawley}(1991)}]{BalbusHawley}
\bibinfo{author}{{Balbus}, S.A.}, \bibinfo{author}{{Hawley}, J.F.},
  \bibinfo{year}{1991}.
\newblock \bibinfo{title}{{A Powerful Local Shear Instability in Weakly
  Magnetized Disks. I. Linear Analysis}}.
\newblock \bibinfo{journal}{\apj} \bibinfo{volume}{376}, \bibinfo{pages}{214}.
\newblock \DOIprefix\doi{10.1086/170270}.
\bibitem[{{Barenfeld} et~al.(2017){Barenfeld}, {Carpenter}, {Sargent}, {Isella}
  and {Ricci}}]{Barenfeld2017}
\bibinfo{author}{{Barenfeld}, S.A.}, \bibinfo{author}{{Carpenter}, J.M.},
  \bibinfo{author}{{Sargent}, A.I.}, \bibinfo{author}{{Isella}, A.},
  \bibinfo{author}{{Ricci}, L.}, \bibinfo{year}{2017}.
\newblock \bibinfo{title}{{Measurement of Circumstellar Disk Sizes in the Upper
  Scorpius OB Association with ALMA}}.
\newblock \bibinfo{journal}{\apj} \bibinfo{volume}{851}, \bibinfo{pages}{85}.
\newblock \DOIprefix\doi{10.3847/1538-4357/aa989d},
  \href{http://arxiv.org/abs/1711.04045}{{\tt arXiv:1711.04045}}.
\bibitem[{{Blum} et~al.(2004){Blum}, {Barbosa}, {Damineli}, {Conti} and
  {Ridgway}}]{Blum2004}
\bibinfo{author}{{Blum}, R.D.}, \bibinfo{author}{{Barbosa}, C.L.},
  \bibinfo{author}{{Damineli}, A.}, \bibinfo{author}{{Conti}, P.S.},
  \bibinfo{author}{{Ridgway}, S.}, \bibinfo{year}{2004}.
\newblock \bibinfo{title}{{Accretion Signatures from Massive Young Stellar
  Objects}}.
\newblock \bibinfo{journal}{\apj} \bibinfo{volume}{617},
  \bibinfo{pages}{1167--1176}.
\newblock \DOIprefix\doi{10.1086/425680},
  \href{http://arxiv.org/abs/astro-ph/0409190}{{\tt arXiv:astro-ph/0409190}}.
\bibitem[{{Boehler} et~al.(2013){Boehler}, {Dutrey}, {Guilloteau} and
  {Pi{\'e}tu}}]{Boehler2013}
\bibinfo{author}{{Boehler}, Y.}, \bibinfo{author}{{Dutrey}, A.},
  \bibinfo{author}{{Guilloteau}, S.}, \bibinfo{author}{{Pi{\'e}tu}, V.},
  \bibinfo{year}{2013}.
\newblock \bibinfo{title}{{Probing dust settling in proto-planetary discs with
  ALMA}}.
\newblock \bibinfo{journal}{\mnras} \bibinfo{volume}{431},
  \bibinfo{pages}{1573--1586}.
\newblock \DOIprefix\doi{10.1093/mnras/stt278},
  \href{http://arxiv.org/abs/1303.5906}{{\tt arXiv:1303.5906}}.
\bibitem[{{Boneberg} et~al.(2016){Boneberg}, {Pani{\'c}}, {Haworth}, {Clarke}
  and {Min}}]{Boneberg2016}
\bibinfo{author}{{Boneberg}, D.M.}, \bibinfo{author}{{Pani{\'c}}, O.},
  \bibinfo{author}{{Haworth}, T.J.}, \bibinfo{author}{{Clarke}, C.J.},
  \bibinfo{author}{{Min}, M.}, \bibinfo{year}{2016}.
\newblock \bibinfo{title}{{Determining the mid-plane conditions of
  circumstellar discs using gas and dust modelling: a study of HD 163296}}.
\newblock \bibinfo{journal}{\mnras} \bibinfo{volume}{461},
  \bibinfo{pages}{385--401}.
\newblock \DOIprefix\doi{10.1093/mnras/stw1325},
  \href{http://arxiv.org/abs/1605.07938}{{\tt arXiv:1605.07938}}.
\bibitem[{{Carpenter} et~al.(2022){Carpenter}, {Brogan}, {Iono} and
  {Mroczkowski}}]{CarpenterALMAUpgrade2022}
\bibinfo{author}{{Carpenter}, J.}, \bibinfo{author}{{Brogan}, C.L.},
  \bibinfo{author}{{Iono}, D.}, \bibinfo{author}{{Mroczkowski}, T.},
  \bibinfo{year}{2022}.
\newblock \bibinfo{title}{{The ALMA2030 Wideband Sensitivity Upgrade}}.
\newblock \bibinfo{journal}{arXiv e-prints} ,
  \bibinfo{pages}{arXiv:2211.00195}\href{http://arxiv.org/abs/2211.00195}{{\tt
  arXiv:2211.00195}}.
\bibitem[{{Carr} et~al.(2004){Carr}, {Tokunaga} and {Najita}}]{Carr2004}
\bibinfo{author}{{Carr}, J.S.}, \bibinfo{author}{{Tokunaga}, A.T.},
  \bibinfo{author}{{Najita}, J.}, \bibinfo{year}{2004}.
\newblock \bibinfo{title}{{Hot H$_{2}$O Emission and Evidence for Turbulence in
  the Disk of a Young Star}}.
\newblock \bibinfo{journal}{\apj} \bibinfo{volume}{603},
  \bibinfo{pages}{213--220}.
\newblock \DOIprefix\doi{10.1086/381356},
  \href{http://arxiv.org/abs/astro-ph/0312125}{{\tt arXiv:astro-ph/0312125}}.
\bibitem[{{Casassus} et~al.(2021){Casassus}, {Christiaens}, {C{\'a}rcamo},
  {P{\'e}rez}, {Weber}, {Ercolano}, {van der Marel}, {Pinte}, {Dong},
  {Baruteau}, {Cieza}, {van Dishoeck}, {Jordan}, {Price}, {Absil}, {Arce-Tord},
  {Faramaz}, {Flores} and {Reggiani}}]{Casassus2021}
{Casassus}, S. et~al., \bibinfo{year}{2021}.
\newblock \bibinfo{title}{{A dusty filament and turbulent CO spirals in HD
  135344B - SAO 206462}}.
\newblock \bibinfo{journal}{\mnras} \bibinfo{volume}{507},
  \bibinfo{pages}{3789--3809}.
\newblock \DOIprefix\doi{10.1093/mnras/stab2359},
  \href{http://arxiv.org/abs/2104.08379}{{\tt arXiv:2104.08379}}.
\bibitem[{{Casassus} et~al.(2019){Casassus}, {Marino}, {Lyra}, {Baruteau},
  {Vidal}, {Wootten}, {P{\'e}rez}, {Alarcon}, {Barraza}, {C{\'a}rcamo}, {Dong},
  {Sierra}, {Zhu}, {Ricci}, {Christiaens} and {Cieza}}]{Casassus2019}
{Casassus}, S. et~al., \bibinfo{year}{2019}.
\newblock \bibinfo{title}{{Cm-wavelength observations of MWC 758: resolved dust
  trapping in a vortex}}.
\newblock \bibinfo{journal}{\mnras} \bibinfo{volume}{483},
  \bibinfo{pages}{3278--3287}.
\newblock \DOIprefix\doi{10.1093/mnras/sty3269},
  \href{http://arxiv.org/abs/1805.03023}{{\tt arXiv:1805.03023}}.
\bibitem[{{Chapillon} et~al.(2012){Chapillon}, {Guilloteau}, {Dutrey},
  {Pi{\'e}tu} and {Gu{\'e}lin}}]{Chapillon2012}
\bibinfo{author}{{Chapillon}, E.}, \bibinfo{author}{{Guilloteau}, S.},
  \bibinfo{author}{{Dutrey}, A.}, \bibinfo{author}{{Pi{\'e}tu}, V.},
  \bibinfo{author}{{Gu{\'e}lin}, M.}, \bibinfo{year}{2012}.
\newblock \bibinfo{title}{{Chemistry in disks. VI. CN and HCN in protoplanetary
  disks}}.
\newblock \bibinfo{journal}{\aap} \bibinfo{volume}{537}, \bibinfo{pages}{A60}.
\newblock \DOIprefix\doi{10.1051/0004-6361/201116762},
  \href{http://arxiv.org/abs/1109.5595}{{\tt arXiv:1109.5595}}.
\bibitem[{{Chevance} et~al.(2022){Chevance}, {Krumholz}, {McLeod}, {Ostriker},
  {Rosolowsky} and {Sternberg}}]{Chevance2022}
\bibinfo{author}{{Chevance}, M.}, \bibinfo{author}{{Krumholz}, M.R.},
  \bibinfo{author}{{McLeod}, A.F.}, \bibinfo{author}{{Ostriker}, E.C.},
  \bibinfo{author}{{Rosolowsky}, E.W.}, \bibinfo{author}{{Sternberg}, A.},
  \bibinfo{year}{2022}.
\newblock \bibinfo{title}{{The Life and Times of Giant Molecular Clouds}}.
\newblock \bibinfo{journal}{arXiv e-prints} ,
  \bibinfo{pages}{arXiv:2203.09570}\href{http://arxiv.org/abs/2203.09570}{{\tt
  arXiv:2203.09570}}.
\bibitem[{{Dartois} et~al.(2003){Dartois}, {Dutrey} and
  {Guilloteau}}]{Dartois2003}
\bibinfo{author}{{Dartois}, E.}, \bibinfo{author}{{Dutrey}, A.},
  \bibinfo{author}{{Guilloteau}, S.}, \bibinfo{year}{2003}.
\newblock \bibinfo{title}{{Structure of the DM Tau Outer Disk: Probing the
  vertical kinetic temperature gradient}}.
\newblock \bibinfo{journal}{\aap} \bibinfo{volume}{399},
  \bibinfo{pages}{773--787}.
\newblock \DOIprefix\doi{10.1051/0004-6361:20021638}.
\bibitem[{{de Juan Ovelar} et~al.(2016){de Juan Ovelar}, {Pinilla}, {Min},
  {Dominik} and {Birnstiel}}]{deJuanOvelar2016}
\bibinfo{author}{{de Juan Ovelar}, M.}, \bibinfo{author}{{Pinilla}, P.},
  \bibinfo{author}{{Min}, M.}, \bibinfo{author}{{Dominik}, C.},
  \bibinfo{author}{{Birnstiel}, T.}, \bibinfo{year}{2016}.
\newblock \bibinfo{title}{{Constraining turbulence mixing strength in
  transitional discs with planets using SPHERE and ALMA}}.
\newblock \bibinfo{journal}{\mnras} \bibinfo{volume}{459},
  \bibinfo{pages}{L85--L89}.
\newblock \DOIprefix\doi{10.1093/mnrasl/slw051},
  \href{http://arxiv.org/abs/1603.09357}{{\tt arXiv:1603.09357}}.
\bibitem[{{Doi} and {Kataoka}(2021)}]{DoiKataoka2021}
\bibinfo{author}{{Doi}, K.}, \bibinfo{author}{{Kataoka}, A.},
  \bibinfo{year}{2021}.
\newblock \bibinfo{title}{{Estimate on Dust Scale Height from the ALMA Dust
  Continuum Image of the HD 163296 Protoplanetary Disk}}.
\newblock \bibinfo{journal}{\apj} \bibinfo{volume}{912}, \bibinfo{pages}{164}.
\newblock \DOIprefix\doi{10.3847/1538-4357/abe5a6},
  \href{http://arxiv.org/abs/2102.06209}{{\tt arXiv:2102.06209}}.
\bibitem[{{Dong} et~al.(2017){Dong}, {Li}, {Chiang} and {Li}}]{Dong2017}
\bibinfo{author}{{Dong}, R.}, \bibinfo{author}{{Li}, S.},
  \bibinfo{author}{{Chiang}, E.}, \bibinfo{author}{{Li}, H.},
  \bibinfo{year}{2017}.
\newblock \bibinfo{title}{{Multiple Disk Gaps and Rings Generated by a Single
  Super-Earth}}.
\newblock \bibinfo{journal}{\apj} \bibinfo{volume}{843}, \bibinfo{pages}{127}.
\newblock \DOIprefix\doi{10.3847/1538-4357/aa72f2},
  \href{http://arxiv.org/abs/1705.04687}{{\tt arXiv:1705.04687}}.
\bibitem[{{Dong} et~al.(2018){Dong}, {Li}, {Chiang} and {Li}}]{Dong2018}
\bibinfo{author}{{Dong}, R.}, \bibinfo{author}{{Li}, S.},
  \bibinfo{author}{{Chiang}, E.}, \bibinfo{author}{{Li}, H.},
  \bibinfo{year}{2018}.
\newblock \bibinfo{title}{{Multiple Disk Gaps and Rings Generated by a Single
  Super-Earth. II. Spacings, Depths, and Number of Gaps, with Application to
  Real Systems}}.
\newblock \bibinfo{journal}{\apj} \bibinfo{volume}{866}, \bibinfo{pages}{110}.
\newblock \DOIprefix\doi{10.3847/1538-4357/aadadd},
  \href{http://arxiv.org/abs/1808.06613}{{\tt arXiv:1808.06613}}.
\bibitem[{{Drazkowska} et~al.(2022){Drazkowska}, {Bitsch}, {Lambrechts},
  {Mulders}, {Harsono}, {Vazan}, {Liu}, {Ormel}, {Kretke} and
  {Morbidelli}}]{DrazkowskaPPVII}
{Drazkowska}, J. et~al., \bibinfo{year}{2022}.
\newblock \bibinfo{title}{{Planet Formation Theory in the Era of ALMA and
  Kepler: from Pebbles to Exoplanets}}.
\newblock \bibinfo{journal}{arXiv e-prints} ,
  \bibinfo{pages}{arXiv:2203.09759}\href{http://arxiv.org/abs/2203.09759}{{\tt
  arXiv:2203.09759}}.
\bibitem[{{Dubrulle} et~al.(1995){Dubrulle}, {Morfill} and
  {Sterzik}}]{Dubrulle1995}
\bibinfo{author}{{Dubrulle}, B.}, \bibinfo{author}{{Morfill}, G.},
  \bibinfo{author}{{Sterzik}, M.}, \bibinfo{year}{1995}.
\newblock \bibinfo{title}{{The dust subdisk in the protoplanetary nebula.}}
\newblock \bibinfo{journal}{\icarus} \bibinfo{volume}{114},
  \bibinfo{pages}{237--246}.
\newblock \DOIprefix\doi{10.1006/icar.1995.1058}.
\bibitem[{{Duch{\^e}ne} et~al.(2003){Duch{\^e}ne}, {M{\'e}nard}, {Stapelfeldt}
  and {Duvert}}]{Duchene2003}
\bibinfo{author}{{Duch{\^e}ne}, G.}, \bibinfo{author}{{M{\'e}nard}, F.},
  \bibinfo{author}{{Stapelfeldt}, K.}, \bibinfo{author}{{Duvert}, G.},
  \bibinfo{year}{2003}.
\newblock \bibinfo{title}{{A layered edge-on circumstellar disk around HK Tau
  B}}.
\newblock \bibinfo{journal}{\aap} \bibinfo{volume}{400},
  \bibinfo{pages}{559--565}.
\newblock \DOIprefix\doi{10.1051/0004-6361:20021906},
  \href{http://arxiv.org/abs/astro-ph/0212512}{{\tt arXiv:astro-ph/0212512}}.
\bibitem[{{Dullemond} et~al.(2018){Dullemond}, {Birnstiel}, {Huang},
  {Kurtovic}, {Andrews}, {Guzm{\'a}n}, {P{\'e}rez}, {Isella}, {Zhu}, {Benisty},
  {Wilner}, {Bai}, {Carpenter}, {Zhang} and {Ricci}}]{Dullemond2018}
{Dullemond}, C.~P. et~al., \bibinfo{year}{2018}.
\newblock \bibinfo{title}{{The Disk Substructures at High Angular Resolution
  Project (DSHARP). VI. Dust Trapping in Thin-ringed Protoplanetary Disks}}.
\newblock \bibinfo{journal}{\apjl} \bibinfo{volume}{869}, \bibinfo{pages}{L46}.
\newblock \DOIprefix\doi{10.3847/2041-8213/aaf742},
  \href{http://arxiv.org/abs/1812.04044}{{\tt arXiv:1812.04044}}.
\bibitem[{{Facchini} et~al.(2020){Facchini}, {Benisty}, {Bae}, {Loomis},
  {Perez}, {Ansdell}, {Mayama}, {Pinilla}, {Teague}, {Isella} and
  {Mann}}]{Facchini2020}
{Facchini}, S. et~al., \bibinfo{year}{2020}.
\newblock \bibinfo{title}{{Annular substructures in the transition disks around
  LkCa 15 and J1610}}.
\newblock \bibinfo{journal}{\aap} \bibinfo{volume}{639}, \bibinfo{pages}{A121}.
\newblock \DOIprefix\doi{10.1051/0004-6361/202038027},
  \href{http://arxiv.org/abs/2005.02712}{{\tt arXiv:2005.02712}}.
\bibitem[{{Flaherty}({in prep})}]{FlahertyPrep}
\bibinfo{author}{{Flaherty}, K.}, \bibinfo{year}{{in prep}}.
\newblock \bibinfo{journal}{\apj} .
\bibitem[{{Flaherty} et~al.(2020){Flaherty}, {Hughes}, {Simon}, {Qi}, {Bai},
  {Bulatek}, {Andrews}, {Wilner} and {K{\'o}sp{\'a}l}}]{Flaherty2020}
{Flaherty}, K. et~al., \bibinfo{year}{2020}.
\newblock \bibinfo{title}{{Measuring Turbulent Motion in Planet-forming Disks
  with ALMA: A Detection around DM Tau and Nondetections around MWC 480 and
  V4046 Sgr}}.
\newblock \bibinfo{journal}{\apj} \bibinfo{volume}{895}, \bibinfo{pages}{109}.
\newblock \DOIprefix\doi{10.3847/1538-4357/ab8cc5},
  \href{http://arxiv.org/abs/2004.12176}{{\tt arXiv:2004.12176}}.
\bibitem[{{Flaherty} et~al.(2017){Flaherty}, {Hughes}, {Rose}, {Simon}, {Qi},
  {Andrews}, {K{\'o}sp{\'a}l}, {Wilner}, {Chiang}, {Armitage} and
  {Bai}}]{Flaherty2017}
{Flaherty}, K.~M. et~al., \bibinfo{year}{2017}.
\newblock \bibinfo{title}{{A Three-dimensional View of Turbulence: Constraints
  on Turbulent Motions in the HD 163296 Protoplanetary Disk Using DCO$^{+}$}}.
\newblock \bibinfo{journal}{\apj} \bibinfo{volume}{843}, \bibinfo{pages}{150}.
\newblock \DOIprefix\doi{10.3847/1538-4357/aa79f9},
  \href{http://arxiv.org/abs/1706.04504}{{\tt arXiv:1706.04504}}.
\bibitem[{{Flaherty} et~al.(2015){Flaherty}, {Hughes}, {Rosenfeld}, {Andrews},
  {Chiang}, {Simon}, {Kerzner} and {Wilner}}]{Flaherty2015}
\bibinfo{author}{{Flaherty}, K.M.}, \bibinfo{author}{{Hughes}, A.M.},
  \bibinfo{author}{{Rosenfeld}, K.A.}, \bibinfo{author}{{Andrews}, S.M.},
  \bibinfo{author}{{Chiang}, E.}, \bibinfo{author}{{Simon}, J.B.},
  \bibinfo{author}{{Kerzner}, S.}, \bibinfo{author}{{Wilner}, D.J.},
  \bibinfo{year}{2015}.
\newblock \bibinfo{title}{{Weak Turbulence in the HD 163296 Protoplanetary Disk
  Revealed by ALMA CO Observations}}.
\newblock \bibinfo{journal}{\apj} \bibinfo{volume}{813}, \bibinfo{pages}{99}.
\newblock \DOIprefix\doi{10.1088/0004-637X/813/2/99},
  \href{http://arxiv.org/abs/1510.01375}{{\tt arXiv:1510.01375}}.
\bibitem[{{Flaherty} et~al.(2018){Flaherty}, {Hughes}, {Teague}, {Simon},
  {Andrews} and {Wilner}}]{Flaherty2018}
\bibinfo{author}{{Flaherty}, K.M.}, \bibinfo{author}{{Hughes}, A.M.},
  \bibinfo{author}{{Teague}, R.}, \bibinfo{author}{{Simon}, J.B.},
  \bibinfo{author}{{Andrews}, S.M.}, \bibinfo{author}{{Wilner}, D.J.},
  \bibinfo{year}{2018}.
\newblock \bibinfo{title}{{Turbulence in the TW Hya Disk}}.
\newblock \bibinfo{journal}{\apj} \bibinfo{volume}{856}, \bibinfo{pages}{117}.
\newblock \DOIprefix\doi{10.3847/1538-4357/aab615},
  \href{http://arxiv.org/abs/1803.03842}{{\tt arXiv:1803.03842}}.
\bibitem[{{Frank} et~al.(2002){Frank}, {King} and {Raine}}]{AccretionBook}
\bibinfo{author}{{Frank}, J.}, \bibinfo{author}{{King}, A.},
  \bibinfo{author}{{Raine}, D.J.}, \bibinfo{year}{2002}.
\newblock \bibinfo{title}{{Accretion Power in Astrophysics: Third Edition}}.
\bibitem[{{Fromang} and {Nelson}(2006)}]{FromangNelson2006a}
\bibinfo{author}{{Fromang}, S.}, \bibinfo{author}{{Nelson}, R.P.},
  \bibinfo{year}{2006}.
\newblock \bibinfo{title}{{Global MHD simulations of stratified and turbulent
  protoplanetary discs. I. Model properties}}.
\newblock \bibinfo{journal}{\aap} \bibinfo{volume}{457},
  \bibinfo{pages}{343--358}.
\newblock \DOIprefix\doi{10.1051/0004-6361:20065643},
  \href{http://arxiv.org/abs/astro-ph/0606729}{{\tt arXiv:astro-ph/0606729}}.
\bibitem[{{Gammie}(1996)}]{Gammie1996}
\bibinfo{author}{{Gammie}, C.F.}, \bibinfo{year}{1996}.
\newblock \bibinfo{title}{{Layered Accretion in T Tauri Disks}}.
\newblock \bibinfo{journal}{\apj} \bibinfo{volume}{457}, \bibinfo{pages}{355}.
\newblock \DOIprefix\doi{10.1086/176735}.
\bibitem[{{Godon} and {Livio}(1999)}]{GodonLivio1999}
\bibinfo{author}{{Godon}, P.}, \bibinfo{author}{{Livio}, M.},
  \bibinfo{year}{1999}.
\newblock \bibinfo{title}{{Vortices in Protoplanetary Disks}}.
\newblock \bibinfo{journal}{\apj} \bibinfo{volume}{523},
  \bibinfo{pages}{350--356}.
\newblock \DOIprefix\doi{10.1086/307720},
  \href{http://arxiv.org/abs/astro-ph/9901384}{{\tt arXiv:astro-ph/9901384}}.
\bibitem[{{Gr{\"a}fe} et~al.(2013){Gr{\"a}fe}, {Wolf}, {Guilloteau}, {Dutrey},
  {Stapelfeldt}, {Pontoppidan} and {Sauter}}]{Grafe2013}
\bibinfo{author}{{Gr{\"a}fe}, C.}, \bibinfo{author}{{Wolf}, S.},
  \bibinfo{author}{{Guilloteau}, S.}, \bibinfo{author}{{Dutrey}, A.},
  \bibinfo{author}{{Stapelfeldt}, K.R.}, \bibinfo{author}{{Pontoppidan}, K.M.},
  \bibinfo{author}{{Sauter}, J.}, \bibinfo{year}{2013}.
\newblock \bibinfo{title}{{Vertical settling and radial segregation of large
  dust grains in the circumstellar disk of the Butterfly Star}}.
\newblock \bibinfo{journal}{\aap} \bibinfo{volume}{553}, \bibinfo{pages}{A69}.
\newblock \DOIprefix\doi{10.1051/0004-6361/201220720},
  \href{http://arxiv.org/abs/1303.6499}{{\tt arXiv:1303.6499}}.
\bibitem[{{GRAVITY Collaboration} et~al.(2021){GRAVITY Collaboration},
  {Koutoulaki}, {Garcia Lopez}, {Natta}, {Fedriani}, {Caratti O Garatti},
  {Ray}, {Coffey}, {Brandner}, {Dougados}, {Garcia}, {Klarmann}, {Labadie},
  {Perraut}, {Sanchez-Bermudez}, {Lin}, {Amorim}, {Baub{\"o}ck}, {Benisty},
  {Berger}, {Buron}, {Caselli}, {Cl{\'e}net}, {Coud{\'e} Du Foresto}, {de
  Zeeuw}, {Duvert}, {de Wit}, {Eckart}, {Eisenhauer}, {Filho}, {Gao},
  {Gendron}, {Genzel}, {Gillessen}, {Grellmann}, {Habibi}, {Haubois},
  {Haussmann}, {Henning}, {Hippler}, {Hubert}, {Horrobin}, {Jimenez Rosales},
  {Jocou}, {Kervella}, {Kolb}, {Lacour}, {Le Bouquin}, {L{\'e}na}, {Linz},
  {Ott}, {Paumard}, {Perrin}, {Pfuhl}, {Ram{\'\i}rez-Tannus}, {Rau}, {Rousset},
  {Scheithauer}, {Shangguan}, {Stadler}, {Straub}, {Straubmeier}, {Sturm}, {van
  Dishoeck}, {Vincent}, {von Fellenberg}, {Widmann}, {Wieprecht}, {Wiest},
  {Wiezorrek}, {Yazici} and {Zins}}]{GRAVITYCO2021}
{GRAVITY Collaboration}et~al., \bibinfo{year}{2021}.
\newblock \bibinfo{title}{{The GRAVITY young stellar object survey. IV. The CO
  overtone emission in 51 Oph at sub-au scales}}.
\newblock \bibinfo{journal}{\aap} \bibinfo{volume}{645}, \bibinfo{pages}{A50}.
\newblock \DOIprefix\doi{10.1051/0004-6361/202038000},
  \href{http://arxiv.org/abs/2011.05955}{{\tt arXiv:2011.05955}}.
\bibitem[{{Guidi} et~al.(2022){Guidi}, {Isella}, {Testi}, {Chandler}, {Liu},
  {Schmid}, {Rosotti}, {Meng}, {Jennings}, {Williams}, {Carpenter}, {de
  Gregorio-Monsalvo}, {Li}, {Liu}, {Ortolani}, {Quanz}, {Ricci} and
  {Tazzari}}]{Guidi2022}
{Guidi}, G. et~al., \bibinfo{year}{2022}.
\newblock \bibinfo{title}{{Distribution of solids in the rings of the HD 163296
  disk: a multiwavelength study}}.
\newblock \bibinfo{journal}{\aap} \bibinfo{volume}{664}, \bibinfo{pages}{A137}.
\newblock \DOIprefix\doi{10.1051/0004-6361/202142303},
  \href{http://arxiv.org/abs/2207.01496}{{\tt arXiv:2207.01496}}.
\bibitem[{{Guilloteau} and {Dutrey}(1998)}]{GuilloteauDutrey1998}
\bibinfo{author}{{Guilloteau}, S.}, \bibinfo{author}{{Dutrey}, A.},
  \bibinfo{year}{1998}.
\newblock \bibinfo{title}{{Physical parameters of the Keplerian protoplanetary
  disk of DM Tauri}}.
\newblock \bibinfo{journal}{\aap} \bibinfo{volume}{339},
  \bibinfo{pages}{467--476}.
\bibitem[{{Guilloteau} et~al.(2011){Guilloteau}, {Dutrey}, {Pi{\'e}tu} and
  {Boehler}}]{Guilloteau2011}
\bibinfo{author}{{Guilloteau}, S.}, \bibinfo{author}{{Dutrey}, A.},
  \bibinfo{author}{{Pi{\'e}tu}, V.}, \bibinfo{author}{{Boehler}, Y.},
  \bibinfo{year}{2011}.
\newblock \bibinfo{title}{{A dual-frequency sub-arcsecond study of
  proto-planetary disks at mm wavelengths: first evidence for radial variations
  of the dust properties}}.
\newblock \bibinfo{journal}{\aap} \bibinfo{volume}{529}, \bibinfo{pages}{A105}.
\newblock \DOIprefix\doi{10.1051/0004-6361/201015209},
  \href{http://arxiv.org/abs/1103.1296}{{\tt arXiv:1103.1296}}.
\bibitem[{{Guilloteau} et~al.(2012){Guilloteau}, {Dutrey}, {Wakelam},
  {Hersant}, {Semenov}, {Chapillon}, {Henning} and
  {Pi{\'e}tu}}]{Guilloteau2012}
\bibinfo{author}{{Guilloteau}, S.}, \bibinfo{author}{{Dutrey}, A.},
  \bibinfo{author}{{Wakelam}, V.}, \bibinfo{author}{{Hersant}, F.},
  \bibinfo{author}{{Semenov}, D.}, \bibinfo{author}{{Chapillon}, E.},
  \bibinfo{author}{{Henning}, T.}, \bibinfo{author}{{Pi{\'e}tu}, V.},
  \bibinfo{year}{2012}.
\newblock \bibinfo{title}{{Chemistry in disks. VIII. The CS molecule as an
  analytic tracer of turbulence in disks}}.
\newblock \bibinfo{journal}{\aap} \bibinfo{volume}{548}, \bibinfo{pages}{A70}.
\newblock \DOIprefix\doi{10.1051/0004-6361/201220331},
  \href{http://arxiv.org/abs/1211.4969}{{\tt arXiv:1211.4969}}.
\bibitem[{{Hammer} et~al.(2021){Hammer}, {Lin}, {Kratter} and
  {Pinilla}}]{Hammer2021}
\bibinfo{author}{{Hammer}, M.}, \bibinfo{author}{{Lin}, M.K.},
  \bibinfo{author}{{Kratter}, K.M.}, \bibinfo{author}{{Pinilla}, P.},
  \bibinfo{year}{2021}.
\newblock \bibinfo{title}{{Which planets trigger longer lived vortices:
  low-mass or high-mass?}}
\newblock \bibinfo{journal}{\mnras} \bibinfo{volume}{504},
  \bibinfo{pages}{3963--3985}.
\newblock \DOIprefix\doi{10.1093/mnras/stab1079},
  \href{http://arxiv.org/abs/2104.02782}{{\tt arXiv:2104.02782}}.
\bibitem[{{Harrison} et~al.(2021){Harrison}, {Looney}, {Stephens}, {Li},
  {Teague}, {Crutcher}, {Yang}, {Cox}, {Fern{\'a}ndez-L{\'o}pez} and
  {Shinnaga}}]{Harrison2021}
{Harrison}, R.~E. et~al., \bibinfo{year}{2021}.
\newblock \bibinfo{title}{{ALMA CN Zeeman Observations of AS 209: Limits on
  Magnetic Field Strength and Magnetically Driven Accretion Rate}}.
\newblock \bibinfo{journal}{\apj} \bibinfo{volume}{908}, \bibinfo{pages}{141}.
\newblock \DOIprefix\doi{10.3847/1538-4357/abd94e},
  \href{http://arxiv.org/abs/2101.01846}{{\tt arXiv:2101.01846}}.
\bibitem[{{Hartmann} et~al.(1998){Hartmann}, {Calvet}, {Gullbring} and
  {D'Alessio}}]{Hartmann1998}
\bibinfo{author}{{Hartmann}, L.}, \bibinfo{author}{{Calvet}, N.},
  \bibinfo{author}{{Gullbring}, E.}, \bibinfo{author}{{D'Alessio}, P.},
  \bibinfo{year}{1998}.
\newblock \bibinfo{title}{{Accretion and the Evolution of T Tauri Disks}}.
\newblock \bibinfo{journal}{\apj} \bibinfo{volume}{495},
  \bibinfo{pages}{385--400}.
\newblock \DOIprefix\doi{10.1086/305277}.
\bibitem[{{Hartmann} et~al.(2004){Hartmann}, {Hinkle} and
  {Calvet}}]{Hartmann2004}
\bibinfo{author}{{Hartmann}, L.}, \bibinfo{author}{{Hinkle}, K.},
  \bibinfo{author}{{Calvet}, N.}, \bibinfo{year}{2004}.
\newblock \bibinfo{title}{{High-Resolution Near-Infrared Spectroscopy of FU
  Orionis Objects}}.
\newblock \bibinfo{journal}{\apj} \bibinfo{volume}{609},
  \bibinfo{pages}{906--916}.
\newblock \DOIprefix\doi{10.1086/421317}.
\bibitem[{{Hendler} et~al.(2020){Hendler}, {Pascucci}, {Pinilla}, {Tazzari},
  {Carpenter}, {Malhotra} and {Testi}}]{Hendler2020}
\bibinfo{author}{{Hendler}, N.}, \bibinfo{author}{{Pascucci}, I.},
  \bibinfo{author}{{Pinilla}, P.}, \bibinfo{author}{{Tazzari}, M.},
  \bibinfo{author}{{Carpenter}, J.}, \bibinfo{author}{{Malhotra}, R.},
  \bibinfo{author}{{Testi}, L.}, \bibinfo{year}{2020}.
\newblock \bibinfo{title}{{The Evolution of Dust Disk Sizes from a Homogeneous
  Analysis of 1-10 Myr old Stars}}.
\newblock \bibinfo{journal}{\apj} \bibinfo{volume}{895}, \bibinfo{pages}{126}.
\newblock \DOIprefix\doi{10.3847/1538-4357/ab70ba},
  \href{http://arxiv.org/abs/2001.02666}{{\tt arXiv:2001.02666}}.
\bibitem[{{Huang} et~al.(2018){Huang}, {Andrews}, {Dullemond}, {Isella},
  {P{\'e}rez}, {Guzm{\'a}n}, {{\"O}berg}, {Zhu}, {Zhang}, {Bai}, {Benisty},
  {Birnstiel}, {Carpenter}, {Hughes}, {Ricci}, {Weaver} and
  {Wilner}}]{Huang2018}
{Huang}, J. et~al., \bibinfo{year}{2018}.
\newblock \bibinfo{title}{{The Disk Substructures at High Angular Resolution
  Project (DSHARP). II. Characteristics of Annular Substructures}}.
\newblock \bibinfo{journal}{\apjl} \bibinfo{volume}{869}, \bibinfo{pages}{L42}.
\newblock \DOIprefix\doi{10.3847/2041-8213/aaf740},
  \href{http://arxiv.org/abs/1812.04041}{{\tt arXiv:1812.04041}}.
\bibitem[{{Hughes} et~al.(2011){Hughes}, {Wilner}, {Andrews}, {Qi} and
  {Hogerheijde}}]{Hughes2011}
\bibinfo{author}{{Hughes}, A.M.}, \bibinfo{author}{{Wilner}, D.J.},
  \bibinfo{author}{{Andrews}, S.M.}, \bibinfo{author}{{Qi}, C.},
  \bibinfo{author}{{Hogerheijde}, M.R.}, \bibinfo{year}{2011}.
\newblock \bibinfo{title}{{Empirical Constraints on Turbulence in
  Protoplanetary Accretion Disks}}.
\newblock \bibinfo{journal}{\apj} \bibinfo{volume}{727}, \bibinfo{pages}{85}.
\newblock \DOIprefix\doi{10.1088/0004-637X/727/2/85},
  \href{http://arxiv.org/abs/1011.3826}{{\tt arXiv:1011.3826}}.
\bibitem[{{Ilee} et~al.(2014){Ilee}, {Fairlamb}, {Oudmaijer},
  {Mendigut{\'\i}a}, {van den Ancker}, {Kraus} and {Wheelwright}}]{Ilee2014}
\bibinfo{author}{{Ilee}, J.D.}, \bibinfo{author}{{Fairlamb}, J.},
  \bibinfo{author}{{Oudmaijer}, R.D.}, \bibinfo{author}{{Mendigut{\'\i}a}, I.},
  \bibinfo{author}{{van den Ancker}, M.E.}, \bibinfo{author}{{Kraus}, S.},
  \bibinfo{author}{{Wheelwright}, H.E.}, \bibinfo{year}{2014}.
\newblock \bibinfo{title}{{Investigating the inner discs of Herbig Ae/Be stars
  with CO bandhead and Br{\ensuremath{\gamma}} emission}}.
\newblock \bibinfo{journal}{\mnras} \bibinfo{volume}{445},
  \bibinfo{pages}{3723--3736}.
\newblock \DOIprefix\doi{10.1093/mnras/stu1942},
  \href{http://arxiv.org/abs/1409.4897}{{\tt arXiv:1409.4897}}.
\bibitem[{{Ilee} et~al.(2013){Ilee}, {Wheelwright}, {Oudmaijer}, {de Wit},
  {Maud}, {Hoare}, {Lumsden}, {Moore}, {Urquhart} and {Mottram}}]{Ilee2013}
{Ilee}, J.~D. et~al., \bibinfo{year}{2013}.
\newblock \bibinfo{title}{{CO bandhead emission of massive young stellar
  objects: determining disc properties}}.
\newblock \bibinfo{journal}{\mnras} \bibinfo{volume}{429},
  \bibinfo{pages}{2960--2973}.
\newblock \DOIprefix\doi{10.1093/mnras/sts537},
  \href{http://arxiv.org/abs/1212.0554}{{\tt arXiv:1212.0554}}.
\bibitem[{{Isella} et~al.(2009){Isella}, {Carpenter} and
  {Sargent}}]{Isella2009}
\bibinfo{author}{{Isella}, A.}, \bibinfo{author}{{Carpenter}, J.M.},
  \bibinfo{author}{{Sargent}, A.I.}, \bibinfo{year}{2009}.
\newblock \bibinfo{title}{{Structure and Evolution of Pre-main-sequence
  Circumstellar Disks}}.
\newblock \bibinfo{journal}{\apj} \bibinfo{volume}{701},
  \bibinfo{pages}{260--282}.
\newblock \DOIprefix\doi{10.1088/0004-637X/701/1/260},
  \href{http://arxiv.org/abs/0906.2227}{{\tt arXiv:0906.2227}}.
\bibitem[{{Izquierdo} et~al.({submitted}){Izquierdo}, {Testi}, {Facchini},
  {Rosotti}, {van Dishoeck}, {Wolfer} and {Paneque-Carreno}}]{IzquierdoMAPS}
\bibinfo{author}{{Izquierdo}, A.F.}, \bibinfo{author}{{Testi}, L.},
  \bibinfo{author}{{Facchini}, S.}, \bibinfo{author}{{Rosotti}, G.P.},
  \bibinfo{author}{{van Dishoeck}, E.F.}, \bibinfo{author}{{Wolfer}, L.},
  \bibinfo{author}{{Paneque-Carreno}, T.}, \bibinfo{year}{{submitted}}.
\newblock \bibinfo{title}{{The Disc Miner II: Revealing gas substructure and
  kinematic signatures of planet-disc interaction through line profile
  analysis}}.
\newblock \bibinfo{journal}{\aap} .
\bibitem[{{Jiang} et~al.(2022){Jiang}, {Zhu} and {Ormel}}]{Jiang2022}
\bibinfo{author}{{Jiang}, H.}, \bibinfo{author}{{Zhu}, W.},
  \bibinfo{author}{{Ormel}, C.W.}, \bibinfo{year}{2022}.
\newblock \bibinfo{title}{{No Significant Correlation between Line-emission and
  Continuum Substructures in the Molecules with ALMA at Planet-forming Scales
  Program}}.
\newblock \bibinfo{journal}{\apjl} \bibinfo{volume}{924}, \bibinfo{pages}{L31}.
\newblock \DOIprefix\doi{10.3847/2041-8213/ac46fe},
  \href{http://arxiv.org/abs/2112.13859}{{\tt arXiv:2112.13859}}.
\bibitem[{{Johansen} and {Lambrechts}(2017)}]{JohansenLambrechtsReview}
\bibinfo{author}{{Johansen}, A.}, \bibinfo{author}{{Lambrechts}, M.},
  \bibinfo{year}{2017}.
\newblock \bibinfo{title}{{Forming Planets via Pebble Accretion}}.
\newblock \bibinfo{journal}{Annual Review of Earth and Planetary Sciences}
  \bibinfo{volume}{45}, \bibinfo{pages}{359--387}.
\newblock \DOIprefix\doi{10.1146/annurev-earth-063016-020226}.
\bibitem[{{Jones} et~al.(2012){Jones}, {Pringle} and {Alexander}}]{Jones2012}
\bibinfo{author}{{Jones}, M.G.}, \bibinfo{author}{{Pringle}, J.E.},
  \bibinfo{author}{{Alexander}, R.D.}, \bibinfo{year}{2012}.
\newblock \bibinfo{title}{{The relationship between accretion disc age and
  stellar age and its consequences for protostellar discs}}.
\newblock \bibinfo{journal}{\mnras} \bibinfo{volume}{419},
  \bibinfo{pages}{925--935}.
\newblock \DOIprefix\doi{10.1111/j.1365-2966.2011.19730.x},
  \href{http://arxiv.org/abs/1109.0276}{{\tt arXiv:1109.0276}}.
\bibitem[{{Krijt} et~al.(2020){Krijt}, {Bosman}, {Zhang}, {Schwarz}, {Ciesla}
  and {Bergin}}]{Krijt2020}
\bibinfo{author}{{Krijt}, S.}, \bibinfo{author}{{Bosman}, A.D.},
  \bibinfo{author}{{Zhang}, K.}, \bibinfo{author}{{Schwarz}, K.R.},
  \bibinfo{author}{{Ciesla}, F.J.}, \bibinfo{author}{{Bergin}, E.A.},
  \bibinfo{year}{2020}.
\newblock \bibinfo{title}{{CO Depletion in Protoplanetary Disks: A Unified
  Picture Combining Physical Sequestration and Chemical Processing}}.
\newblock \bibinfo{journal}{\apj} \bibinfo{volume}{899}, \bibinfo{pages}{134}.
\newblock \DOIprefix\doi{10.3847/1538-4357/aba75d},
  \href{http://arxiv.org/abs/2007.09517}{{\tt arXiv:2007.09517}}.
\bibitem[{{Larson}(1981)}]{Larson1981}
\bibinfo{author}{{Larson}, R.B.}, \bibinfo{year}{1981}.
\newblock \bibinfo{title}{{Turbulence and star formation in molecular clouds.}}
\newblock \bibinfo{journal}{\mnras} \bibinfo{volume}{194},
  \bibinfo{pages}{809--826}.
\newblock \DOIprefix\doi{10.1093/mnras/194.4.809}.
\bibitem[{{Law} et~al.(2021){Law}, {Loomis}, {Teague}, {{\"O}berg}, {Czekala},
  {Andrews}, {Huang}, {Aikawa}, {Alarc{\'o}n}, {Bae}, {Bergin}, {Bergner},
  {Boehler}, {Booth}, {Bosman}, {Calahan}, {Cataldi}, {Cleeves}, {Furuya},
  {Guzm{\'a}n}, {Ilee}, {Le Gal}, {Liu}, {Long}, {M{\'e}nard}, {Nomura}, {Qi},
  {Schwarz}, {Sierra}, {Tsukagoshi}, {Yamato}, {van't Hoff}, {Walsh}, {Wilner}
  and {Zhang}}]{Law2021Radial}
{Law}, C.~J. et~al., \bibinfo{year}{2021}.
\newblock \bibinfo{title}{{Molecules with ALMA at Planet-forming Scales (MAPS).
  III. Characteristics of Radial Chemical Substructures}}.
\newblock \bibinfo{journal}{\apjs} \bibinfo{volume}{257}, \bibinfo{pages}{3}.
\newblock \DOIprefix\doi{10.3847/1538-4365/ac1434},
  \href{http://arxiv.org/abs/2109.06210}{{\tt arXiv:2109.06210}}.
\bibitem[{{Lesur} et~al.(2022){Lesur}, {Ercolano}, {Flock}, {Lin}, {Yang},
  {Barranco}, {Benitez-Llambay}, {Goodman}, {Johansen}, {Klahr}, {Laibe},
  {Lyra}, {Marcus}, {Nelson}, {Squire}, {Simon}, {Turner}, {Umurhan} and
  {Youdin}}]{LesurPPVII}
{Lesur}, G. et~al., \bibinfo{year}{2022}.
\newblock \bibinfo{title}{{Hydro-, Magnetohydro-, and Dust-Gas Dynamics of
  Protoplanetary Disks}}.
\newblock \bibinfo{journal}{arXiv e-prints} ,
  \bibinfo{pages}{arXiv:2203.09821}\href{http://arxiv.org/abs/2203.09821}{{\tt
  arXiv:2203.09821}}.
\bibitem[{{Lin} et~al.(2021){Lin}, {Lee}, {Li}, {Tobin} and
  {Turner}}]{Lin2021HH212}
\bibinfo{author}{{Lin}, Z.Y.D.}, \bibinfo{author}{{Lee}, C.F.},
  \bibinfo{author}{{Li}, Z.Y.}, \bibinfo{author}{{Tobin}, J.J.},
  \bibinfo{author}{{Turner}, N.J.}, \bibinfo{year}{2021}.
\newblock \bibinfo{title}{{Inferring (sub)millimetre dust opacities and
  temperature structure in edge-on protostellar discs from resolved
  multiwavelength continuum observations: the case of the HH 212 disc}}.
\newblock \bibinfo{journal}{\mnras} \bibinfo{volume}{501},
  \bibinfo{pages}{1316--1335}.
\newblock \DOIprefix\doi{10.1093/mnras/staa3685},
  \href{http://arxiv.org/abs/2008.08627}{{\tt arXiv:2008.08627}}.
\bibitem[{{Liu} et~al.(2018){Liu}, {Jin}, {Li}, {Isella} and
  {Li}}]{Liu2018HD163296}
\bibinfo{author}{{Liu}, S.F.}, \bibinfo{author}{{Jin}, S.},
  \bibinfo{author}{{Li}, S.}, \bibinfo{author}{{Isella}, A.},
  \bibinfo{author}{{Li}, H.}, \bibinfo{year}{2018}.
\newblock \bibinfo{title}{{New Constraints on Turbulence and Embedded Planet
  Mass in the HD 163296 Disk from Planet-Disk Hydrodynamic Simulations}}.
\newblock \bibinfo{journal}{\apj} \bibinfo{volume}{857}, \bibinfo{pages}{87}.
\newblock \DOIprefix\doi{10.3847/1538-4357/aab718},
  \href{http://arxiv.org/abs/1803.05437}{{\tt arXiv:1803.05437}}.
\bibitem[{{Liu} et~al.(2022){Liu}, {Bertrang}, {Flock}, {Rosotti}, {van
  Dishoeck}, {Boehler}, {Facchini}, {Cui}, {Wolf} and {Fang}}]{Liu2022HD163296}
{Liu}, Y. et~al., \bibinfo{year}{2022}.
\newblock \bibinfo{title}{{Millimeter Gap Contrast as a Probe for Turbulence
  Level in Protoplanetary Disks}}.
\newblock \bibinfo{journal}{arXiv e-prints} ,
  \bibinfo{pages}{arXiv:2208.09230}\href{http://arxiv.org/abs/2208.09230}{{\tt
  arXiv:2208.09230}}.
\bibitem[{{Lodato} et~al.(2017){Lodato}, {Scardoni}, {Manara} and
  {Testi}}]{Lodato2017}
\bibinfo{author}{{Lodato}, G.}, \bibinfo{author}{{Scardoni}, C.E.},
  \bibinfo{author}{{Manara}, C.F.}, \bibinfo{author}{{Testi}, L.},
  \bibinfo{year}{2017}.
\newblock \bibinfo{title}{{Protoplanetary disc `isochrones' and the evolution
  of discs in the M˙-M$_{d}$ plane}}.
\newblock \bibinfo{journal}{\mnras} \bibinfo{volume}{472},
  \bibinfo{pages}{4700--4706}.
\newblock \DOIprefix\doi{10.1093/mnras/stx2273},
  \href{http://arxiv.org/abs/1708.09467}{{\tt arXiv:1708.09467}}.
\bibitem[{{Long} et~al.(2022){Long}, {Andrews}, {Rosotti}, {Harsono},
  {Pinilla}, {Wilner}, {{\"O}berg}, {Teague}, {Trapman} and
  {Tabone}}]{Long2022}
{Long}, F. et~al., \bibinfo{year}{2022}.
\newblock \bibinfo{title}{{Gas Disk Sizes from CO Line Observations: A Test of
  Angular Momentum Evolution}}.
\newblock \bibinfo{journal}{\apj} \bibinfo{volume}{931}, \bibinfo{pages}{6}.
\newblock \DOIprefix\doi{10.3847/1538-4357/ac634e},
  \href{http://arxiv.org/abs/2203.16735}{{\tt arXiv:2203.16735}}.
\bibitem[{{Long} et~al.(2018){Long}, {Pinilla}, {Herczeg}, {Harsono},
  {Dipierro}, {Pascucci}, {Hendler}, {Tazzari}, {Ragusa}, {Salyk}, {Edwards},
  {Lodato}, {van de Plas}, {Johnstone}, {Liu}, {Boehler}, {Cabrit}, {Manara},
  {Menard}, {Mulders}, {Nisini}, {Fischer}, {Rigliaco}, {Banzatti}, {Avenhaus}
  and {Gully-Santiago}}]{Long2018}
{Long}, F. et~al., \bibinfo{year}{2018}.
\newblock \bibinfo{title}{{Gaps and Rings in an ALMA Survey of Disks in the
  Taurus Star-forming Region}}.
\newblock \bibinfo{journal}{\apj} \bibinfo{volume}{869}, \bibinfo{pages}{17}.
\newblock \DOIprefix\doi{10.3847/1538-4357/aae8e1},
  \href{http://arxiv.org/abs/1810.06044}{{\tt arXiv:1810.06044}}.
\bibitem[{{Lynden-Bell} and {Pringle}(1974)}]{LyndenBellPringle1974}
\bibinfo{author}{{Lynden-Bell}, D.}, \bibinfo{author}{{Pringle}, J.E.},
  \bibinfo{year}{1974}.
\newblock \bibinfo{title}{{The evolution of viscous discs and the origin of the
  nebular variables.}}
\newblock \bibinfo{journal}{\mnras} \bibinfo{volume}{168},
  \bibinfo{pages}{603--637}.
\newblock \DOIprefix\doi{10.1093/mnras/168.3.603}.
\bibitem[{{Manara} et~al.(2022){Manara}, {Ansdell}, {Rosotti}, {Hughes},
  {Armitage}, {Lodato} and {Williams}}]{ManaraPPVII}
\bibinfo{author}{{Manara}, C.F.}, \bibinfo{author}{{Ansdell}, M.},
  \bibinfo{author}{{Rosotti}, G.P.}, \bibinfo{author}{{Hughes}, A.M.},
  \bibinfo{author}{{Armitage}, P.J.}, \bibinfo{author}{{Lodato}, G.},
  \bibinfo{author}{{Williams}, J.P.}, \bibinfo{year}{2022}.
\newblock \bibinfo{title}{{Demographics of young stars and their protoplanetary
  disks: lessons learned on disk evolution and its connection to planet
  formation}}.
\newblock \bibinfo{journal}{arXiv e-prints} ,
  \bibinfo{pages}{arXiv:2203.09930}\href{http://arxiv.org/abs/2203.09930}{{\tt
  arXiv:2203.09930}}.
\bibitem[{{Manara} et~al.(2020){Manara}, {Natta}, {Rosotti}, {Alcal{\'a}},
  {Nisini}, {Lodato}, {Testi}, {Pascucci}, {Hillenbrand}, {Carpenter},
  {Scholz}, {Fedele}, {Frasca}, {Mulders}, {Rigliaco}, {Scardoni} and
  {Zari}}]{Manara2020}
{Manara}, C.~F. et~al., \bibinfo{year}{2020}.
\newblock \bibinfo{title}{{X-shooter survey of disk accretion in Upper
  Scorpius. I. Very high accretion rates at age > 5 Myr}}.
\newblock \bibinfo{journal}{\aap} \bibinfo{volume}{639}, \bibinfo{pages}{A58}.
\newblock \DOIprefix\doi{10.1051/0004-6361/202037949},
  \href{http://arxiv.org/abs/2004.14232}{{\tt arXiv:2004.14232}}.
\bibitem[{{Manara} et~al.(2016){Manara}, {Rosotti}, {Testi}, {Natta},
  {Alcal{\'a}}, {Williams}, {Ansdell}, {Miotello}, {van der Marel}, {Tazzari},
  {Carpenter}, {Guidi}, {Mathews}, {Oliveira}, {Prusti} and {van
  Dishoeck}}]{Manara2016}
{Manara}, C.~F. et~al., \bibinfo{year}{2016}.
\newblock \bibinfo{title}{{Evidence for a correlation between mass accretion
  rates onto young stars and the mass of their protoplanetary disks}}.
\newblock \bibinfo{journal}{\aap} \bibinfo{volume}{591}, \bibinfo{pages}{L3}.
\newblock \DOIprefix\doi{10.1051/0004-6361/201628549},
  \href{http://arxiv.org/abs/1605.03050}{{\tt arXiv:1605.03050}}.
\bibitem[{{Meru} et~al.(2019){Meru}, {Rosotti}, {Booth}, {Nazari} and
  {Clarke}}]{Meru2019}
\bibinfo{author}{{Meru}, F.}, \bibinfo{author}{{Rosotti}, G.P.},
  \bibinfo{author}{{Booth}, R.A.}, \bibinfo{author}{{Nazari}, P.},
  \bibinfo{author}{{Clarke}, C.J.}, \bibinfo{year}{2019}.
\newblock \bibinfo{title}{{Is the ring inside or outside the planet?: the
  effect of planet migration on dust rings}}.
\newblock \bibinfo{journal}{\mnras} \bibinfo{volume}{482},
  \bibinfo{pages}{3678--3695}.
\newblock \DOIprefix\doi{10.1093/mnras/sty2847},
  \href{http://arxiv.org/abs/1810.06573}{{\tt arXiv:1810.06573}}.
\bibitem[{{Michel} et~al.(2022){Michel}, {Sadavoy}, {Sheehan}, {Looney} and
  {Cox}}]{Michel2022VLA1623W}
\bibinfo{author}{{Michel}, A.}, \bibinfo{author}{{Sadavoy}, S.I.},
  \bibinfo{author}{{Sheehan}, P.D.}, \bibinfo{author}{{Looney}, L.W.},
  \bibinfo{author}{{Cox}, E.G.}, \bibinfo{year}{2022}.
\newblock \bibinfo{title}{{A Millimeter-multiwavelength Continuum Study of VLA
  1623 West}}.
\newblock \bibinfo{journal}{\apj} \bibinfo{volume}{937}, \bibinfo{pages}{104}.
\newblock \DOIprefix\doi{10.3847/1538-4357/ac905c},
  \href{http://arxiv.org/abs/2209.06781}{{\tt arXiv:2209.06781}}.
\bibitem[{{Min} et~al.(2011){Min}, {Dullemond}, {Kama} and
  {Dominik}}]{MinDullemond2011}
\bibinfo{author}{{Min}, M.}, \bibinfo{author}{{Dullemond}, C.P.},
  \bibinfo{author}{{Kama}, M.}, \bibinfo{author}{{Dominik}, C.},
  \bibinfo{year}{2011}.
\newblock \bibinfo{title}{{The thermal structure and the location of the snow
  line in the protosolar nebula: Axisymmetric models with full 3-D radiative
  transfer}}.
\newblock \bibinfo{journal}{\icarus} \bibinfo{volume}{212},
  \bibinfo{pages}{416--426}.
\newblock \DOIprefix\doi{10.1016/j.icarus.2010.12.002},
  \href{http://arxiv.org/abs/1012.0727}{{\tt arXiv:1012.0727}}.
\bibitem[{{Miotello} et~al.(2022){Miotello}, {Kamp}, {Birnstiel}, {Cleeves} and
  {Kataoka}}]{MiotelloPPVII}
\bibinfo{author}{{Miotello}, A.}, \bibinfo{author}{{Kamp}, I.},
  \bibinfo{author}{{Birnstiel}, T.}, \bibinfo{author}{{Cleeves}, L.I.},
  \bibinfo{author}{{Kataoka}, A.}, \bibinfo{year}{2022}.
\newblock \bibinfo{title}{{Setting the Stage for Planet Formation: Measurements
  and Implications of the Fundamental Disk Properties}}.
\newblock \bibinfo{journal}{arXiv e-prints} ,
  \bibinfo{pages}{arXiv:2203.09818}\href{http://arxiv.org/abs/2203.09818}{{\tt
  arXiv:2203.09818}}.
\bibitem[{{Miranda} and {Rafikov}(2019)}]{MirandaRafikov2019}
\bibinfo{author}{{Miranda}, R.}, \bibinfo{author}{{Rafikov}, R.R.},
  \bibinfo{year}{2019}.
\newblock \bibinfo{title}{{On the Planetary Interpretation of Multiple Gaps and
  Rings in Protoplanetary Disks Seen By ALMA}}.
\newblock \bibinfo{journal}{\apjl} \bibinfo{volume}{878}, \bibinfo{pages}{L9}.
\newblock \DOIprefix\doi{10.3847/2041-8213/ab22a7},
  \href{http://arxiv.org/abs/1905.08259}{{\tt arXiv:1905.08259}}.
\bibitem[{{Montesinos} et~al.(2020){Montesinos}, {Garrido-Deutelmoser},
  {Olofsson}, {Giuppone}, {Cuadra}, {Bayo}, {Sucerquia} and
  {Cuello}}]{Montesinos2020}
\bibinfo{author}{{Montesinos}, M.}, \bibinfo{author}{{Garrido-Deutelmoser},
  J.}, \bibinfo{author}{{Olofsson}, J.}, \bibinfo{author}{{Giuppone}, C.A.},
  \bibinfo{author}{{Cuadra}, J.}, \bibinfo{author}{{Bayo}, A.},
  \bibinfo{author}{{Sucerquia}, M.}, \bibinfo{author}{{Cuello}, N.},
  \bibinfo{year}{2020}.
\newblock \bibinfo{title}{{Dust trapping around Lagrangian points in
  protoplanetary disks}}.
\newblock \bibinfo{journal}{\aap} \bibinfo{volume}{642}, \bibinfo{pages}{A224}.
\newblock \DOIprefix\doi{10.1051/0004-6361/202038758},
  \href{http://arxiv.org/abs/2009.10768}{{\tt arXiv:2009.10768}}.
\bibitem[{{Mulders} and {Dominik}(2012)}]{MuldersDominik2012}
\bibinfo{author}{{Mulders}, G.D.}, \bibinfo{author}{{Dominik}, C.},
  \bibinfo{year}{2012}.
\newblock \bibinfo{title}{{Probing the turbulent mixing strength in
  protoplanetary disks across the stellar mass range: no significant
  variations}}.
\newblock \bibinfo{journal}{\aap} \bibinfo{volume}{539}, \bibinfo{pages}{A9}.
\newblock \DOIprefix\doi{10.1051/0004-6361/201118127},
  \href{http://arxiv.org/abs/1201.1453}{{\tt arXiv:1201.1453}}.
\bibitem[{{Mulders} et~al.(2017){Mulders}, {Pascucci}, {Manara}, {Testi},
  {Herczeg}, {Henning}, {Mohanty} and {Lodato}}]{Mulders2017}
\bibinfo{author}{{Mulders}, G.D.}, \bibinfo{author}{{Pascucci}, I.},
  \bibinfo{author}{{Manara}, C.F.}, \bibinfo{author}{{Testi}, L.},
  \bibinfo{author}{{Herczeg}, G.J.}, \bibinfo{author}{{Henning}, T.},
  \bibinfo{author}{{Mohanty}, S.}, \bibinfo{author}{{Lodato}, G.},
  \bibinfo{year}{2017}.
\newblock \bibinfo{title}{{Constraints from Dust Mass and Mass Accretion Rate
  Measurements on Angular Momentum Transport in Protoplanetary Disks}}.
\newblock \bibinfo{journal}{\apj} \bibinfo{volume}{847}, \bibinfo{pages}{31}.
\newblock \DOIprefix\doi{10.3847/1538-4357/aa8906},
  \href{http://arxiv.org/abs/1708.09464}{{\tt arXiv:1708.09464}}.
\bibitem[{{Najita} et~al.(1996){Najita}, {Carr}, {Glassgold}, {Shu} and
  {Tokunaga}}]{Najita1996}
\bibinfo{author}{{Najita}, J.}, \bibinfo{author}{{Carr}, J.S.},
  \bibinfo{author}{{Glassgold}, A.E.}, \bibinfo{author}{{Shu}, F.H.},
  \bibinfo{author}{{Tokunaga}, A.T.}, \bibinfo{year}{1996}.
\newblock \bibinfo{title}{{Kinematic Diagnostics of Disks around Young Stars:
  CO Overtone Emission from WL 16 and 1548C27}}.
\newblock \bibinfo{journal}{\apj} \bibinfo{volume}{462}, \bibinfo{pages}{919}.
\newblock \DOIprefix\doi{10.1086/177205},
  \href{http://arxiv.org/abs/astro-ph/9512109}{{\tt arXiv:astro-ph/9512109}}.
\bibitem[{{Najita} and {Bergin}(2018)}]{NajitaBergin2018}
\bibinfo{author}{{Najita}, J.R.}, \bibinfo{author}{{Bergin}, E.A.},
  \bibinfo{year}{2018}.
\newblock \bibinfo{title}{{Protoplanetary Disk Sizes and Angular Momentum
  Transport}}.
\newblock \bibinfo{journal}{\apj} \bibinfo{volume}{864}, \bibinfo{pages}{168}.
\newblock \DOIprefix\doi{10.3847/1538-4357/aad80c},
  \href{http://arxiv.org/abs/1808.05618}{{\tt arXiv:1808.05618}}.
\bibitem[{{Najita} et~al.(2009){Najita}, {Doppmann}, {Carr}, {Graham} and
  {Eisner}}]{Najita2009}
\bibinfo{author}{{Najita}, J.R.}, \bibinfo{author}{{Doppmann}, G.W.},
  \bibinfo{author}{{Carr}, J.S.}, \bibinfo{author}{{Graham}, J.R.},
  \bibinfo{author}{{Eisner}, J.A.}, \bibinfo{year}{2009}.
\newblock \bibinfo{title}{{High-Resolution K-Band Spectroscopy of MWC 480 and
  V1331 Cyg}}.
\newblock \bibinfo{journal}{\apj} \bibinfo{volume}{691},
  \bibinfo{pages}{738--748}.
\newblock \DOIprefix\doi{10.1088/0004-637X/691/1/738},
  \href{http://arxiv.org/abs/0809.4267}{{\tt arXiv:0809.4267}}.
\bibitem[{{Nazari} et~al.(2019){Nazari}, {Booth}, {Clarke}, {Rosotti},
  {Tazzari}, {Juhasz} and {Meru}}]{Nazari2019}
\bibinfo{author}{{Nazari}, P.}, \bibinfo{author}{{Booth}, R.A.},
  \bibinfo{author}{{Clarke}, C.J.}, \bibinfo{author}{{Rosotti}, G.P.},
  \bibinfo{author}{{Tazzari}, M.}, \bibinfo{author}{{Juhasz}, A.},
  \bibinfo{author}{{Meru}, F.}, \bibinfo{year}{2019}.
\newblock \bibinfo{title}{{Revealing signatures of planets migrating in
  protoplanetary discs with ALMA multiwavelength observations}}.
\newblock \bibinfo{journal}{\mnras} \bibinfo{volume}{485},
  \bibinfo{pages}{5914--5923}.
\newblock \DOIprefix\doi{10.1093/mnras/stz836},
  \href{http://arxiv.org/abs/1903.03114}{{\tt arXiv:1903.03114}}.
\bibitem[{{Ohashi} and {Kataoka}(2019)}]{OhashiKataoka2019}
\bibinfo{author}{{Ohashi}, S.}, \bibinfo{author}{{Kataoka}, A.},
  \bibinfo{year}{2019}.
\newblock \bibinfo{title}{{Radial Variations in Grain Sizes and Dust Scale
  Heights in the Protoplanetary Disk around HD 163296 Revealed by ALMA
  Polarization Observations}}.
\newblock \bibinfo{journal}{\apj} \bibinfo{volume}{886}, \bibinfo{pages}{103}.
\newblock \DOIprefix\doi{10.3847/1538-4357/ab5107},
  \href{http://arxiv.org/abs/1910.12868}{{\tt arXiv:1910.12868}}.
\bibitem[{{Paardekooper} et~al.(2022){Paardekooper}, {Dong}, {Duffell}, {Fung},
  {Masset}, {Ogilvie} and {Tanaka}}]{PaardekooperPPVII}
\bibinfo{author}{{Paardekooper}, S.J.}, \bibinfo{author}{{Dong}, R.},
  \bibinfo{author}{{Duffell}, P.}, \bibinfo{author}{{Fung}, J.},
  \bibinfo{author}{{Masset}, F.S.}, \bibinfo{author}{{Ogilvie}, G.},
  \bibinfo{author}{{Tanaka}, H.}, \bibinfo{year}{2022}.
\newblock \bibinfo{title}{{Planet-Disk Interactions}}.
\newblock \bibinfo{journal}{arXiv e-prints} ,
  \bibinfo{pages}{arXiv:2203.09595}\href{http://arxiv.org/abs/2203.09595}{{\tt
  arXiv:2203.09595}}.
\bibitem[{{P{\'e}rez} et~al.(2019){P{\'e}rez}, {Casassus}, {Baruteau}, {Dong},
  {Hales} and {Cieza}}]{Perez2019}
\bibinfo{author}{{P{\'e}rez}, S.}, \bibinfo{author}{{Casassus}, S.},
  \bibinfo{author}{{Baruteau}, C.}, \bibinfo{author}{{Dong}, R.},
  \bibinfo{author}{{Hales}, A.}, \bibinfo{author}{{Cieza}, L.},
  \bibinfo{year}{2019}.
\newblock \bibinfo{title}{{Dust Unveils the Formation of a Mini-Neptune Planet
  in a Protoplanetary Ring}}.
\newblock \bibinfo{journal}{\aj} \bibinfo{volume}{158}, \bibinfo{pages}{15}.
\newblock \DOIprefix\doi{10.3847/1538-3881/ab1f88},
  \href{http://arxiv.org/abs/1902.05143}{{\tt arXiv:1902.05143}}.
\bibitem[{{Pi{\'e}tu} et~al.(2007){Pi{\'e}tu}, {Dutrey} and
  {Guilloteau}}]{Pietu2007}
\bibinfo{author}{{Pi{\'e}tu}, V.}, \bibinfo{author}{{Dutrey}, A.},
  \bibinfo{author}{{Guilloteau}, S.}, \bibinfo{year}{2007}.
\newblock \bibinfo{title}{{Probing the structure of protoplanetary disks: a
  comparative study of DM Tau, LkCa 15, and MWC 480}}.
\newblock \bibinfo{journal}{\aap} \bibinfo{volume}{467},
  \bibinfo{pages}{163--178}.
\newblock \DOIprefix\doi{10.1051/0004-6361:20066537},
  \href{http://arxiv.org/abs/astro-ph/0701425}{{\tt arXiv:astro-ph/0701425}}.
\bibitem[{{Pinte} et~al.(2016){Pinte}, {Dent}, {M{\'e}nard}, {Hales}, {Hill},
  {Cortes} and {de Gregorio-Monsalvo}}]{Pinte2016}
\bibinfo{author}{{Pinte}, C.}, \bibinfo{author}{{Dent}, W.R.F.},
  \bibinfo{author}{{M{\'e}nard}, F.}, \bibinfo{author}{{Hales}, A.},
  \bibinfo{author}{{Hill}, T.}, \bibinfo{author}{{Cortes}, P.},
  \bibinfo{author}{{de Gregorio-Monsalvo}, I.}, \bibinfo{year}{2016}.
\newblock \bibinfo{title}{{Dust and Gas in the Disk of HL Tauri: Surface
  Density, Dust Settling, and Dust-to-gas Ratio}}.
\newblock \bibinfo{journal}{\apj} \bibinfo{volume}{816}, \bibinfo{pages}{25}.
\newblock \DOIprefix\doi{10.3847/0004-637X/816/1/25},
  \href{http://arxiv.org/abs/1508.00584}{{\tt arXiv:1508.00584}}.
\bibitem[{{Pinte} et~al.(2007){Pinte}, {Fouchet}, {M{\'e}nard}, {Gonzalez} and
  {Duch{\^e}ne}}]{Pinte2007}
\bibinfo{author}{{Pinte}, C.}, \bibinfo{author}{{Fouchet}, L.},
  \bibinfo{author}{{M{\'e}nard}, F.}, \bibinfo{author}{{Gonzalez}, J.F.},
  \bibinfo{author}{{Duch{\^e}ne}, G.}, \bibinfo{year}{2007}.
\newblock \bibinfo{title}{{On the stratified dust distribution of the GG Tauri
  circumbinary ring}}.
\newblock \bibinfo{journal}{\aap} \bibinfo{volume}{469},
  \bibinfo{pages}{963--971}.
\newblock \DOIprefix\doi{10.1051/0004-6361:20077137},
  \href{http://arxiv.org/abs/0704.2747}{{\tt arXiv:0704.2747}}.
\bibitem[{{Pinte} et~al.(2022){Pinte}, {Teague}, {Flaherty}, {Hall}, {Facchini}
  and {Casassus}}]{PintePPVII}
\bibinfo{author}{{Pinte}, C.}, \bibinfo{author}{{Teague}, R.},
  \bibinfo{author}{{Flaherty}, K.}, \bibinfo{author}{{Hall}, C.},
  \bibinfo{author}{{Facchini}, S.}, \bibinfo{author}{{Casassus}, S.},
  \bibinfo{year}{2022}.
\newblock \bibinfo{title}{{Kinematic Structures in Planet-Forming Disks}}.
\newblock \bibinfo{journal}{arXiv e-prints} ,
  \bibinfo{pages}{arXiv:2203.09528}\href{http://arxiv.org/abs/2203.09528}{{\tt
  arXiv:2203.09528}}.
\bibitem[{{Pizzati} et~al.({in prep}){Pizzati}, {Rosotti} and
  {Tabone}}]{Pizzatiprep}
\bibinfo{author}{{Pizzati}, E.}, \bibinfo{author}{{Rosotti}, G.P.},
  \bibinfo{author}{{Tabone}, B.}, \bibinfo{year}{{in prep}}.
\newblock \bibinfo{journal}{\mnras} .
\bibitem[{{Pringle}(1981)}]{Pringle1981}
\bibinfo{author}{{Pringle}, J.E.}, \bibinfo{year}{1981}.
\newblock \bibinfo{title}{{Accretion discs in astrophysics}}.
\newblock \bibinfo{journal}{\araa} \bibinfo{volume}{19},
  \bibinfo{pages}{137--162}.
\newblock \DOIprefix\doi{10.1146/annurev.aa.19.090181.001033}.
\bibitem[{{Qi} et~al.(2004){Qi}, {Ho}, {Wilner}, {Takakuwa}, {Hirano},
  {Ohashi}, {Bourke}, {Zhang}, {Blake}, {Hogerheijde}, {Saito}, {Choi} and
  {Yang}}]{Qi2004}
{Qi}, C. et~al., \bibinfo{year}{2004}.
\newblock \bibinfo{title}{{Imaging the Disk around TW Hydrae with the
  Submillimeter Array}}.
\newblock \bibinfo{journal}{\apjl} \bibinfo{volume}{616},
  \bibinfo{pages}{L11--L14}.
\newblock \DOIprefix\doi{10.1086/421063},
  \href{http://arxiv.org/abs/astro-ph/0403412}{{\tt arXiv:astro-ph/0403412}}.
\bibitem[{{Rafikov}(2017)}]{Rafikov2017}
\bibinfo{author}{{Rafikov}, R.R.}, \bibinfo{year}{2017}.
\newblock \bibinfo{title}{{Protoplanetary Disks as (Possibly) Viscous Disks}}.
\newblock \bibinfo{journal}{\apj} \bibinfo{volume}{837}, \bibinfo{pages}{163}.
\newblock \DOIprefix\doi{10.3847/1538-4357/aa6249},
  \href{http://arxiv.org/abs/1701.02352}{{\tt arXiv:1701.02352}}.
\bibitem[{{Reg{\'a}ly} et~al.(2017){Reg{\'a}ly}, {Juh{\'a}sz} and
  {Neh{\'e}z}}]{Regaly2017}
\bibinfo{author}{{Reg{\'a}ly}, Z.}, \bibinfo{author}{{Juh{\'a}sz}, A.},
  \bibinfo{author}{{Neh{\'e}z}, D.}, \bibinfo{year}{2017}.
\newblock \bibinfo{title}{{Interpreting Brightness Asymmetries in Transition
  Disks: Vortex at Dead Zone or Planet-carved Gap Edges?}}
\newblock \bibinfo{journal}{\apj} \bibinfo{volume}{851}, \bibinfo{pages}{89}.
\newblock \DOIprefix\doi{10.3847/1538-4357/aa9a3f},
  \href{http://arxiv.org/abs/1711.03548}{{\tt arXiv:1711.03548}}.
\bibitem[{{Ribas} et~al.(2020){Ribas}, {Espaillat}, {Mac{\'\i}as} and
  {Sarro}}]{Ribas2020}
\bibinfo{author}{{Ribas}, {\'A}.}, \bibinfo{author}{{Espaillat}, C.C.},
  \bibinfo{author}{{Mac{\'\i}as}, E.}, \bibinfo{author}{{Sarro}, L.M.},
  \bibinfo{year}{2020}.
\newblock \bibinfo{title}{{Modeling protoplanetary disk SEDs with artificial
  neural networks. Revisiting the viscous disk model and updated disk masses}}.
\newblock \bibinfo{journal}{\aap} \bibinfo{volume}{642}, \bibinfo{pages}{A171}.
\newblock \DOIprefix\doi{10.1051/0004-6361/202038352},
  \href{http://arxiv.org/abs/2009.03323}{{\tt arXiv:2009.03323}}.
\bibitem[{{Rometsch} et~al.(2021){Rometsch}, {Ziampras}, {Kley} and
  {B{\'e}thune}}]{Rometsch2021}
\bibinfo{author}{{Rometsch}, T.}, \bibinfo{author}{{Ziampras}, A.},
  \bibinfo{author}{{Kley}, W.}, \bibinfo{author}{{B{\'e}thune}, W.},
  \bibinfo{year}{2021}.
\newblock \bibinfo{title}{{Survival of planet-induced vortices in 2D disks}}.
\newblock \bibinfo{journal}{\aap} \bibinfo{volume}{656}, \bibinfo{pages}{A130}.
\newblock \DOIprefix\doi{10.1051/0004-6361/202142105},
  \href{http://arxiv.org/abs/2110.00589}{{\tt arXiv:2110.00589}}.
\bibitem[{{Rosotti} et~al.(2019a){Rosotti}, {Booth}, {Tazzari}, {Clarke},
  {Lodato} and {Testi}}]{Rosotti2019letter}
\bibinfo{author}{{Rosotti}, G.P.}, \bibinfo{author}{{Booth}, R.A.},
  \bibinfo{author}{{Tazzari}, M.}, \bibinfo{author}{{Clarke}, C.},
  \bibinfo{author}{{Lodato}, G.}, \bibinfo{author}{{Testi}, L.},
  \bibinfo{year}{2019}a.
\newblock \bibinfo{title}{{On the millimetre continuum flux-radius correlation
  of proto-planetary discs}}.
\newblock \bibinfo{journal}{\mnras} \bibinfo{volume}{486},
  \bibinfo{pages}{L63--L68}.
\newblock \DOIprefix\doi{10.1093/mnrasl/slz064},
  \href{http://arxiv.org/abs/1905.00021}{{\tt arXiv:1905.00021}}.
\bibitem[{{Rosotti} et~al.(2017){Rosotti}, {Clarke}, {Manara} and
  {Facchini}}]{Rosotti2017}
\bibinfo{author}{{Rosotti}, G.P.}, \bibinfo{author}{{Clarke}, C.J.},
  \bibinfo{author}{{Manara}, C.F.}, \bibinfo{author}{{Facchini}, S.},
  \bibinfo{year}{2017}.
\newblock \bibinfo{title}{{Constraining protoplanetary disc evolution using
  accretion rate and disc mass measurements: the usefulness of the
  dimensionless accretion parameter}}.
\newblock \bibinfo{journal}{\mnras} \bibinfo{volume}{468},
  \bibinfo{pages}{1631--1638}.
\newblock \DOIprefix\doi{10.1093/mnras/stx595},
  \href{http://arxiv.org/abs/1703.02974}{{\tt arXiv:1703.02974}}.
\bibitem[{{Rosotti} et~al.(2019b){Rosotti}, {Tazzari}, {Booth}, {Testi},
  {Lodato} and {Clarke}}]{Rosotti2019Paper}
\bibinfo{author}{{Rosotti}, G.P.}, \bibinfo{author}{{Tazzari}, M.},
  \bibinfo{author}{{Booth}, R.A.}, \bibinfo{author}{{Testi}, L.},
  \bibinfo{author}{{Lodato}, G.}, \bibinfo{author}{{Clarke}, C.},
  \bibinfo{year}{2019}b.
\newblock \bibinfo{title}{{The time evolution of dusty protoplanetary disc
  radii: observed and physical radii differ}}.
\newblock \bibinfo{journal}{\mnras} \bibinfo{volume}{486},
  \bibinfo{pages}{4829--4844}.
\newblock \DOIprefix\doi{10.1093/mnras/stz1190},
  \href{http://arxiv.org/abs/1905.00019}{{\tt arXiv:1905.00019}}.
\bibitem[{{Rosotti} et~al.(2020){Rosotti}, {Teague}, {Dullemond}, {Booth} and
  {Clarke}}]{Rosotti2020}
\bibinfo{author}{{Rosotti}, G.P.}, \bibinfo{author}{{Teague}, R.},
  \bibinfo{author}{{Dullemond}, C.}, \bibinfo{author}{{Booth}, R.A.},
  \bibinfo{author}{{Clarke}, C.J.}, \bibinfo{year}{2020}.
\newblock \bibinfo{title}{{The efficiency of dust trapping in ringed
  protoplanetary discs}}.
\newblock \bibinfo{journal}{\mnras} \bibinfo{volume}{495},
  \bibinfo{pages}{173--181}.
\newblock \DOIprefix\doi{10.1093/mnras/staa1170},
  \href{http://arxiv.org/abs/2004.11394}{{\tt arXiv:2004.11394}}.
\bibitem[{{Sanchis} et~al.(2021){Sanchis}, {Testi}, {Natta}, {Facchini},
  {Manara}, {Miotello}, {Ercolano}, {Henning}, {Preibisch}, {Carpenter}, {de
  Gregorio-Monsalvo}, {Jayawardhana}, {Lopez}, {Mu{\v{z}}i{\'c}}, {Pascucci},
  {Santamar{\'\i}a-Miranda}, {van Terwisga} and {Williams}}]{Sanchis2021}
{Sanchis}, E. et~al., \bibinfo{year}{2021}.
\newblock \bibinfo{title}{{Measuring the ratio of the gas and dust emission
  radii of protoplanetary disks in the Lupus star-forming region}}.
\newblock \bibinfo{journal}{\aap} \bibinfo{volume}{649}, \bibinfo{pages}{A19}.
\newblock \DOIprefix\doi{10.1051/0004-6361/202039733},
  \href{http://arxiv.org/abs/2101.11307}{{\tt arXiv:2101.11307}}.
\bibitem[{{Sellek} et~al.(2020){Sellek}, {Booth} and
  {Clarke}}]{Sellek2020Correlation}
\bibinfo{author}{{Sellek}, A.D.}, \bibinfo{author}{{Booth}, R.A.},
  \bibinfo{author}{{Clarke}, C.J.}, \bibinfo{year}{2020}.
\newblock \bibinfo{title}{{A dusty origin for the correlation between
  protoplanetary disc accretion rates and dust masses}}.
\newblock \bibinfo{journal}{\mnras} \bibinfo{volume}{498},
  \bibinfo{pages}{2845--2863}.
\newblock \DOIprefix\doi{10.1093/mnras/staa2519},
  \href{http://arxiv.org/abs/2008.07530}{{\tt arXiv:2008.07530}}.
\bibitem[{{Semenov} and {Wiebe}(2011)}]{SemenovWiebe2011}
\bibinfo{author}{{Semenov}, D.}, \bibinfo{author}{{Wiebe}, D.},
  \bibinfo{year}{2011}.
\newblock \bibinfo{title}{{Chemical Evolution of Turbulent Protoplanetary Disks
  and the Solar Nebula}}.
\newblock \bibinfo{journal}{\apjs} \bibinfo{volume}{196}, \bibinfo{pages}{25}.
\newblock \DOIprefix\doi{10.1088/0067-0049/196/2/25},
  \href{http://arxiv.org/abs/1104.4358}{{\tt arXiv:1104.4358}}.
\bibitem[{{Shakura} and {Sunyaev}(1973)}]{ShakuraSunyaev1973}
\bibinfo{author}{{Shakura}, N.I.}, \bibinfo{author}{{Sunyaev}, R.A.},
  \bibinfo{year}{1973}.
\newblock \bibinfo{title}{{Black holes in binary systems. Observational
  appearance.}}
\newblock \bibinfo{journal}{\aap} \bibinfo{volume}{24},
  \bibinfo{pages}{337--355}.
\bibitem[{{Sheehan} et~al.(2022){Sheehan}, {Tobin}, {Li}, {van't Hoff},
  {J{\o}rgensen}, {Kwon}, {Looney}, {Ohashi}, {Takakuwa}, {Williams}, {Aso},
  {Gavino}, {Gregorio-Monsalvo}, {Han}, {Lee}, {Plunkett}, {Sharma}, {Aikawa},
  {Lai}, {Lee}, {Lin}, {Saigo}, {Tomida} and {Yen}}]{Sheehan2022}
{Sheehan}, P.~D. et~al., \bibinfo{year}{2022}.
\newblock \bibinfo{title}{{A VLA View of the Flared, Asymmetric Disk around the
  Class 0 Protostar L1527 IRS}}.
\newblock \bibinfo{journal}{\apj} \bibinfo{volume}{934}, \bibinfo{pages}{95}.
\newblock \DOIprefix\doi{10.3847/1538-4357/ac7a3b},
  \href{http://arxiv.org/abs/2206.13548}{{\tt arXiv:2206.13548}}.
\bibitem[{{Sierra} et~al.(2019){Sierra}, {Lizano}, {Mac{\'\i}as},
  {Carrasco-Gonz{\'a}lez}, {Osorio} and {Flock}}]{Sierra2019}
\bibinfo{author}{{Sierra}, A.}, \bibinfo{author}{{Lizano}, S.},
  \bibinfo{author}{{Mac{\'\i}as}, E.},
  \bibinfo{author}{{Carrasco-Gonz{\'a}lez}, C.}, \bibinfo{author}{{Osorio},
  M.}, \bibinfo{author}{{Flock}, M.}, \bibinfo{year}{2019}.
\newblock \bibinfo{title}{{An Analytical Model of Radial Dust Trapping in
  Protoplanetary Disks}}.
\newblock \bibinfo{journal}{\apj} \bibinfo{volume}{876}, \bibinfo{pages}{7}.
\newblock \DOIprefix\doi{10.3847/1538-4357/ab1265},
  \href{http://arxiv.org/abs/1903.08769}{{\tt arXiv:1903.08769}}.
\bibitem[{{Sierra} et~al.(2021){Sierra}, {P{\'e}rez}, {Zhang}, {Law},
  {Guzm{\'a}n}, {Qi}, {Bosman}, {{\"O}berg}, {Andrews}, {Long}, {Teague},
  {Booth}, {Walsh}, {Wilner}, {M{\'e}nard}, {Cataldi}, {Czekala}, {Bae},
  {Huang}, {Bergner}, {Ilee}, {Benisty}, {Le Gal}, {Loomis}, {Tsukagoshi},
  {Liu}, {Yamato} and {Aikawa}}]{Sierra2021}
{Sierra}, A. et~al., \bibinfo{year}{2021}.
\newblock \bibinfo{title}{{Molecules with ALMA at Planet-forming Scales (MAPS).
  XIV. Revealing Disk Substructures in Multiwavelength Continuum Emission}}.
\newblock \bibinfo{journal}{\apjs} \bibinfo{volume}{257}, \bibinfo{pages}{14}.
\newblock \DOIprefix\doi{10.3847/1538-4365/ac1431},
  \href{http://arxiv.org/abs/2109.06433}{{\tt arXiv:2109.06433}}.
\bibitem[{{Simon} et~al.(2015){Simon}, {Hughes}, {Flaherty}, {Bai} and
  {Armitage}}]{Simon2015}
\bibinfo{author}{{Simon}, J.B.}, \bibinfo{author}{{Hughes}, A.M.},
  \bibinfo{author}{{Flaherty}, K.M.}, \bibinfo{author}{{Bai}, X.N.},
  \bibinfo{author}{{Armitage}, P.J.}, \bibinfo{year}{2015}.
\newblock \bibinfo{title}{{Signatures of MRI-driven Turbulence in
  Protoplanetary Disks: Predictions for ALMA Observations}}.
\newblock \bibinfo{journal}{\apj} \bibinfo{volume}{808}, \bibinfo{pages}{180}.
\newblock \DOIprefix\doi{10.1088/0004-637X/808/2/180},
  \href{http://arxiv.org/abs/1501.02808}{{\tt arXiv:1501.02808}}.
\bibitem[{{Somigliana} et~al.(2020){Somigliana}, {Toci}, {Lodato}, {Rosotti}
  and {Manara}}]{Somigliana2020}
\bibinfo{author}{{Somigliana}, A.}, \bibinfo{author}{{Toci}, C.},
  \bibinfo{author}{{Lodato}, G.}, \bibinfo{author}{{Rosotti}, G.},
  \bibinfo{author}{{Manara}, C.F.}, \bibinfo{year}{2020}.
\newblock \bibinfo{title}{{Effects of photoevaporation on protoplanetary disc
  `isochrones'}}.
\newblock \bibinfo{journal}{\mnras} \bibinfo{volume}{492},
  \bibinfo{pages}{1120--1126}.
\newblock \DOIprefix\doi{10.1093/mnras/stz3481},
  \href{http://arxiv.org/abs/1912.05623}{{\tt arXiv:1912.05623}}.
\bibitem[{{Tabone} et~al.(2022){Tabone}, {Rosotti}, {Lodato}, {Armitage},
  {Cridland} and {van Dishoeck}}]{Tabone2022Letter}
\bibinfo{author}{{Tabone}, B.}, \bibinfo{author}{{Rosotti}, G.P.},
  \bibinfo{author}{{Lodato}, G.}, \bibinfo{author}{{Armitage}, P.J.},
  \bibinfo{author}{{Cridland}, A.J.}, \bibinfo{author}{{van Dishoeck}, E.F.},
  \bibinfo{year}{2022}.
\newblock \bibinfo{title}{{MHD disc winds can reproduce fast disc dispersal and
  the correlation between accretion rate and disc mass in Lupus}}.
\newblock \bibinfo{journal}{\mnras} \bibinfo{volume}{512},
  \bibinfo{pages}{L74--L79}.
\newblock \DOIprefix\doi{10.1093/mnrasl/slab124},
  \href{http://arxiv.org/abs/2111.14473}{{\tt arXiv:2111.14473}}.
\bibitem[{{Tazzari} et~al.(2021){Tazzari}, {Clarke}, {Testi}, {Williams},
  {Facchini}, {Manara}, {Natta} and {Rosotti}}]{Tazzari2021}
\bibinfo{author}{{Tazzari}, M.}, \bibinfo{author}{{Clarke}, C.J.},
  \bibinfo{author}{{Testi}, L.}, \bibinfo{author}{{Williams}, J.P.},
  \bibinfo{author}{{Facchini}, S.}, \bibinfo{author}{{Manara}, C.F.},
  \bibinfo{author}{{Natta}, A.}, \bibinfo{author}{{Rosotti}, G.},
  \bibinfo{year}{2021}.
\newblock \bibinfo{title}{{Multiwavelength continuum sizes of protoplanetary
  discs: scaling relations and implications for grain growth and radial
  drift}}.
\newblock \bibinfo{journal}{\mnras} \bibinfo{volume}{506},
  \bibinfo{pages}{2804--2823}.
\newblock \DOIprefix\doi{10.1093/mnras/stab1808},
  \href{http://arxiv.org/abs/2010.02249}{{\tt arXiv:2010.02249}}.
\bibitem[{{Teague} et~al.(2018a){Teague}, {Bae}, {Bergin}, {Birnstiel} and
  {Foreman-Mackey}}]{Teague2018}
\bibinfo{author}{{Teague}, R.}, \bibinfo{author}{{Bae}, J.},
  \bibinfo{author}{{Bergin}, E.A.}, \bibinfo{author}{{Birnstiel}, T.},
  \bibinfo{author}{{Foreman-Mackey}, D.}, \bibinfo{year}{2018}a.
\newblock \bibinfo{title}{{A Kinematical Detection of Two Embedded Jupiter-mass
  Planets in HD 163296}}.
\newblock \bibinfo{journal}{\apjl} \bibinfo{volume}{860}, \bibinfo{pages}{L12}.
\newblock \DOIprefix\doi{10.3847/2041-8213/aac6d7},
  \href{http://arxiv.org/abs/1805.10290}{{\tt arXiv:1805.10290}}.
\bibitem[{{Teague} et~al.(2016){Teague}, {Guilloteau}, {Semenov}, {Henning},
  {Dutrey}, {Pi{\'e}tu}, {Birnstiel}, {Chapillon}, {Hollenbach} and
  {Gorti}}]{Teague2016}
{Teague}, R. et~al., \bibinfo{year}{2016}.
\newblock \bibinfo{title}{{Measuring turbulence in TW Hydrae with ALMA: methods
  and limitations}}.
\newblock \bibinfo{journal}{\aap} \bibinfo{volume}{592}, \bibinfo{pages}{A49}.
\newblock \DOIprefix\doi{10.1051/0004-6361/201628550},
  \href{http://arxiv.org/abs/1606.00005}{{\tt arXiv:1606.00005}}.
\bibitem[{{Teague} et~al.(2018b){Teague}, {Henning}, {Guilloteau}, {Bergin},
  {Semenov}, {Dutrey}, {Flock}, {Gorti} and {Birnstiel}}]{Teague2018CS}
{Teague}, R. et~al., \bibinfo{year}{2018}b.
\newblock \bibinfo{title}{{Temperature, Mass, and Turbulence: A Spatially
  Resolved Multiband Non-LTE Analysis of CS in TW Hya}}.
\newblock \bibinfo{journal}{\apj} \bibinfo{volume}{864}, \bibinfo{pages}{133}.
\newblock \DOIprefix\doi{10.3847/1538-4357/aad80e},
  \href{http://arxiv.org/abs/1808.01768}{{\tt arXiv:1808.01768}}.
\bibitem[{{Tobin} et~al.(2020){Tobin}, {Sheehan}, {Megeath},
  {D{\'\i}az-Rodr{\'\i}guez}, {Offner}, {Murillo}, {van 't Hoff}, {van
  Dishoeck}, {Osorio}, {Anglada}, {Furlan}, {Stutz}, {Reynolds}, {Karnath},
  {Fischer}, {Persson}, {Looney}, {Li}, {Stephens}, {Chandler}, {Cox},
  {Dunham}, {Tychoniec}, {Kama}, {Kratter}, {Kounkel}, {Mazur}, {Maud},
  {Patel}, {Perez}, {Sadavoy}, {Segura-Cox}, {Sharma}, {Stephenson}, {Watson}
  and {Wyrowski}}]{Tobin2020}
{Tobin}, J.~J. et~al., \bibinfo{year}{2020}.
\newblock \bibinfo{title}{{The VLA/ALMA Nascent Disk and Multiplicity (VANDAM)
  Survey of Orion Protostars. II. A Statistical Characterization of Class 0 and
  Class I Protostellar Disks}}.
\newblock \bibinfo{journal}{\apj} \bibinfo{volume}{890}, \bibinfo{pages}{130}.
\newblock \DOIprefix\doi{10.3847/1538-4357/ab6f64},
  \href{http://arxiv.org/abs/2001.04468}{{\tt arXiv:2001.04468}}.
\bibitem[{{Toci} et~al.(2021){Toci}, {Rosotti}, {Lodato}, {Testi} and
  {Trapman}}]{Toci2021}
\bibinfo{author}{{Toci}, C.}, \bibinfo{author}{{Rosotti}, G.},
  \bibinfo{author}{{Lodato}, G.}, \bibinfo{author}{{Testi}, L.},
  \bibinfo{author}{{Trapman}, L.}, \bibinfo{year}{2021}.
\newblock \bibinfo{title}{{On the secular evolution of the ratio between gas
  and dust radii in protoplanetary discs}}.
\newblock \bibinfo{journal}{\mnras} \bibinfo{volume}{507},
  \bibinfo{pages}{818--833}.
\newblock \DOIprefix\doi{10.1093/mnras/stab2112},
  \href{http://arxiv.org/abs/2107.09914}{{\tt arXiv:2107.09914}}.
\bibitem[{{Trapman} et~al.(2020){Trapman}, {Rosotti}, {Bosman}, {Hogerheijde}
  and {van Dishoeck}}]{Trapman2020}
\bibinfo{author}{{Trapman}, L.}, \bibinfo{author}{{Rosotti}, G.},
  \bibinfo{author}{{Bosman}, A.D.}, \bibinfo{author}{{Hogerheijde}, M.R.},
  \bibinfo{author}{{van Dishoeck}, E.F.}, \bibinfo{year}{2020}.
\newblock \bibinfo{title}{{Observed sizes of planet-forming disks trace viscous
  spreading}}.
\newblock \bibinfo{journal}{\aap} \bibinfo{volume}{640}, \bibinfo{pages}{A5}.
\newblock \DOIprefix\doi{10.1051/0004-6361/202037673},
  \href{http://arxiv.org/abs/2005.11330}{{\tt arXiv:2005.11330}}.
\bibitem[{{Tripathi} et~al.(2017){Tripathi}, {Andrews}, {Birnstiel} and
  {Wilner}}]{Tripathi2017}
\bibinfo{author}{{Tripathi}, A.}, \bibinfo{author}{{Andrews}, S.M.},
  \bibinfo{author}{{Birnstiel}, T.}, \bibinfo{author}{{Wilner}, D.J.},
  \bibinfo{year}{2017}.
\newblock \bibinfo{title}{{A millimeter Continuum Size-Luminosity Relationship
  for Protoplanetary Disks}}.
\newblock \bibinfo{journal}{\apj} \bibinfo{volume}{845}, \bibinfo{pages}{44}.
\newblock \DOIprefix\doi{10.3847/1538-4357/aa7c62},
  \href{http://arxiv.org/abs/1706.08977}{{\tt arXiv:1706.08977}}.
\bibitem[{{Ueda} et~al.(2021){Ueda}, {Kataoka}, {Zhang}, {Zhu},
  {Carrasco-Gonz{\'a}lez} and {Sierra}}]{Ueda2020}
\bibinfo{author}{{Ueda}, T.}, \bibinfo{author}{{Kataoka}, A.},
  \bibinfo{author}{{Zhang}, S.}, \bibinfo{author}{{Zhu}, Z.},
  \bibinfo{author}{{Carrasco-Gonz{\'a}lez}, C.}, \bibinfo{author}{{Sierra},
  A.}, \bibinfo{year}{2021}.
\newblock \bibinfo{title}{{Impact of Differential Dust Settling on the SED and
  Polarization: Application to the Inner Region of the HL Tau Disk}}.
\newblock \bibinfo{journal}{\apj} \bibinfo{volume}{913}, \bibinfo{pages}{117}.
\newblock \DOIprefix\doi{10.3847/1538-4357/abf7b8},
  \href{http://arxiv.org/abs/2104.05927}{{\tt arXiv:2104.05927}}.
\bibitem[{{van der Marel} et~al.(2019){van der Marel}, {Dong}, {di Francesco},
  {Williams} and {Tobin}}]{vanderMarel2019}
\bibinfo{author}{{van der Marel}, N.}, \bibinfo{author}{{Dong}, R.},
  \bibinfo{author}{{di Francesco}, J.}, \bibinfo{author}{{Williams}, J.P.},
  \bibinfo{author}{{Tobin}, J.}, \bibinfo{year}{2019}.
\newblock \bibinfo{title}{{Protoplanetary Disk Rings and Gaps across Ages and
  Luminosities}}.
\newblock \bibinfo{journal}{\apj} \bibinfo{volume}{872}, \bibinfo{pages}{112}.
\newblock \DOIprefix\doi{10.3847/1538-4357/aafd31},
  \href{http://arxiv.org/abs/1901.03680}{{\tt arXiv:1901.03680}}.
\bibitem[{{Villenave} et~al.(2020){Villenave}, {M{\'e}nard}, {Dent},
  {Duch{\^e}ne}, {Stapelfeldt}, {Benisty}, {Boehler}, {van der Plas}, {Pinte},
  {Telkamp}, {Wolff}, {Flores}, {Lesur}, {Louvet}, {Riols}, {Dougados},
  {Williams} and {Padgett}}]{Villenave2020}
{Villenave}, M. et~al., \bibinfo{year}{2020}.
\newblock \bibinfo{title}{{Observations of edge-on protoplanetary disks with
  ALMA. I. Results from continuum data}}.
\newblock \bibinfo{journal}{\aap} \bibinfo{volume}{642}, \bibinfo{pages}{A164}.
\newblock \DOIprefix\doi{10.1051/0004-6361/202038087},
  \href{http://arxiv.org/abs/2008.06518}{{\tt arXiv:2008.06518}}.
\bibitem[{{Villenave} et~al.({submitted}){Villenave}, {Podio}, {Duchene},
  {Stapefeldt}, {Melis}, {Carrasco-Gonzalez}, {Le Gouellec}, {Menard}, {de
  Simone}, {Chandler}, {Garufi}, {Pinte}, {Bianchi} and
  {Codella}}]{Villenave2023}
{Villenave}, M. et~al., \bibinfo{year}{{submitted}}.
\newblock \bibinfo{title}{{Modest dust settling in the IRAS04302+2247 Class I
  protoplanetary disk}}.
\newblock \bibinfo{journal}{\apj} .
\bibitem[{{Villenave} et~al.(2022){Villenave}, {Stapelfeldt}, {Duch{\^e}ne},
  {M{\'e}nard}, {Lambrechts}, {Sierra}, {Flores}, {Dent}, {Wolff}, {Ribas},
  {Benisty}, {Cuello} and {Pinte}}]{Villenave2022}
{Villenave}, M. et~al., \bibinfo{year}{2022}.
\newblock \bibinfo{title}{{A Highly Settled Disk around Oph163131}}.
\newblock \bibinfo{journal}{\apj} \bibinfo{volume}{930}, \bibinfo{pages}{11}.
\newblock \DOIprefix\doi{10.3847/1538-4357/ac5fae},
  \href{http://arxiv.org/abs/2204.00640}{{\tt arXiv:2204.00640}}.
\bibitem[{{Vlemmings} et~al.(2019){Vlemmings}, {Lankhaar}, {Cazzoletti},
  {Ceccobello}, {Dall'Olio}, {van Dishoeck}, {Facchini}, {Humphreys},
  {Persson}, {Testi} and {Williams}}]{Vlemmings2019}
{Vlemmings}, W.~H.~T. et~al., \bibinfo{year}{2019}.
\newblock \bibinfo{title}{{Stringent limits on the magnetic field strength in
  the disc of TW Hya. ALMA observations of CN polarisation}}.
\newblock \bibinfo{journal}{\aap} \bibinfo{volume}{624}, \bibinfo{pages}{L7}.
\newblock \DOIprefix\doi{10.1051/0004-6361/201935459},
  \href{http://arxiv.org/abs/1904.01632}{{\tt arXiv:1904.01632}}.
\bibitem[{{Weber} et~al.(2019){Weber}, {P{\'e}rez}, {Ben{\'\i}tez-Llambay},
  {Gressel}, {Casassus} and {Krapp}}]{Weber2019}
\bibinfo{author}{{Weber}, P.}, \bibinfo{author}{{P{\'e}rez}, S.},
  \bibinfo{author}{{Ben{\'\i}tez-Llambay}, P.}, \bibinfo{author}{{Gressel},
  O.}, \bibinfo{author}{{Casassus}, S.}, \bibinfo{author}{{Krapp}, L.},
  \bibinfo{year}{2019}.
\newblock \bibinfo{title}{{Predicting the Observational Signature of Migrating
  Neptune-sized Planets in Low-viscosity Disks}}.
\newblock \bibinfo{journal}{\apj} \bibinfo{volume}{884}, \bibinfo{pages}{178}.
\newblock \DOIprefix\doi{10.3847/1538-4357/ab412f},
  \href{http://arxiv.org/abs/1909.01661}{{\tt arXiv:1909.01661}}.
\bibitem[{{Wheelwright} et~al.(2010){Wheelwright}, {Oudmaijer}, {de Wit},
  {Hoare}, {Lumsden} and {Urquhart}}]{Wheelwright2010}
\bibinfo{author}{{Wheelwright}, H.E.}, \bibinfo{author}{{Oudmaijer}, R.D.},
  \bibinfo{author}{{de Wit}, W.J.}, \bibinfo{author}{{Hoare}, M.G.},
  \bibinfo{author}{{Lumsden}, S.L.}, \bibinfo{author}{{Urquhart}, J.S.},
  \bibinfo{year}{2010}.
\newblock \bibinfo{title}{{Probing discs around massive young stellar objects
  with CO first overtone emission{\textdagger}}}.
\newblock \bibinfo{journal}{\mnras} \bibinfo{volume}{408},
  \bibinfo{pages}{1840--1850}.
\newblock \DOIprefix\doi{10.1111/j.1365-2966.2010.17250.x},
  \href{http://arxiv.org/abs/1007.3289}{{\tt arXiv:1007.3289}}.
\bibitem[{{Whipple}(1972)}]{Whipple1972}
\bibinfo{author}{{Whipple}, F.L.}, \bibinfo{year}{1972}.
\newblock \bibinfo{title}{{On certain aerodynamic processes for asteroids and
  comets}}, in: \bibinfo{editor}{{Elvius}, A.} (Ed.), \bibinfo{booktitle}{From
  Plasma to Planet}, p. \bibinfo{pages}{211}.
\bibitem[{{Wolff} et~al.(2021){Wolff}, {Duch{\^e}ne}, {Stapelfeldt},
  {M{\'e}nard}, {Flores}, {Padgett}, {Pinte}, {Villenave}, {van der Plas} and
  {Perrin}}]{Wolff2021}
{Wolff}, S.~G. et~al., \bibinfo{year}{2021}.
\newblock \bibinfo{title}{{The Anatomy of an Unusual Edge-on Protoplanetary
  Disk. I. Dust Settling in a Cold Disk}}.
\newblock \bibinfo{journal}{\aj} \bibinfo{volume}{161}, \bibinfo{pages}{238}.
\newblock \DOIprefix\doi{10.3847/1538-3881/abeb1d},
  \href{http://arxiv.org/abs/2103.02665}{{\tt arXiv:2103.02665}}.
\bibitem[{{Youdin} and {Lithwick}(2007)}]{YoudinLithwick2007}
\bibinfo{author}{{Youdin}, A.N.}, \bibinfo{author}{{Lithwick}, Y.},
  \bibinfo{year}{2007}.
\newblock \bibinfo{title}{{Particle stirring in turbulent gas disks: Including
  orbital oscillations}}.
\newblock \bibinfo{journal}{\icarus} \bibinfo{volume}{192},
  \bibinfo{pages}{588--604}.
\newblock \DOIprefix\doi{10.1016/j.icarus.2007.07.012},
  \href{http://arxiv.org/abs/0707.2975}{{\tt arXiv:0707.2975}}.
\bibitem[{{Zagaria} et~al.(2022a){Zagaria}, {Clarke}, {Rosotti} and
  {Manara}}]{Zagaria2022UpperSco}
\bibinfo{author}{{Zagaria}, F.}, \bibinfo{author}{{Clarke}, C.J.},
  \bibinfo{author}{{Rosotti}, G.P.}, \bibinfo{author}{{Manara}, C.F.},
  \bibinfo{year}{2022}a.
\newblock \bibinfo{title}{{Stellar multiplicity affects the correlation between
  protoplanetary disc masses and accretion rates: binaries explain high
  accretors in Upper Sco}}.
\newblock \bibinfo{journal}{\mnras} \bibinfo{volume}{512},
  \bibinfo{pages}{3538--3550}.
\newblock \DOIprefix\doi{10.1093/mnras/stac621},
  \href{http://arxiv.org/abs/2203.01986}{{\tt arXiv:2203.01986}}.
\bibitem[{{Zagaria} et~al.(2022b){Zagaria}, {Rosotti}, {Clarke} and
  {Tabone}}]{Zagaria2022}
\bibinfo{author}{{Zagaria}, F.}, \bibinfo{author}{{Rosotti}, G.P.},
  \bibinfo{author}{{Clarke}, C.J.}, \bibinfo{author}{{Tabone}, B.},
  \bibinfo{year}{2022}b.
\newblock \bibinfo{title}{{Modelling the secular evolution of protoplanetary
  disc dust sizes - a comparison between the viscous and magnetic wind case}}.
\newblock \bibinfo{journal}{\mnras} \bibinfo{volume}{514},
  \bibinfo{pages}{1088--1106}.
\newblock \DOIprefix\doi{10.1093/mnras/stac1461},
  \href{http://arxiv.org/abs/2205.10931}{{\tt arXiv:2205.10931}}.
\bibitem[{{Zhang} et~al.(2021){Zhang}, {Booth}, {Law}, {Bosman}, {Schwarz},
  {Bergin}, {{\"O}berg}, {Andrews}, {Guzm{\'a}n}, {Walsh}, {Qi}, {van't Hoff},
  {Long}, {Wilner}, {Huang}, {Czekala}, {Ilee}, {Cataldi}, {Bergner}, {Aikawa},
  {Teague}, {Bae}, {Loomis}, {Calahan}, {Alarc{\'o}n}, {M{\'e}nard}, {Le Gal},
  {Sierra}, {Yamato}, {Nomura}, {Tsukagoshi}, {P{\'e}rez}, {Trapman}, {Liu} and
  {Furuya}}]{Zhang2021MAPS}
{Zhang}, K. et~al., \bibinfo{year}{2021}.
\newblock \bibinfo{title}{{Molecules with ALMA at Planet-forming Scales (MAPS).
  V. CO Gas Distributions}}.
\newblock \bibinfo{journal}{\apjs} \bibinfo{volume}{257}, \bibinfo{pages}{5}.
\newblock \DOIprefix\doi{10.3847/1538-4365/ac1580},
  \href{http://arxiv.org/abs/2109.06233}{{\tt arXiv:2109.06233}}.
\bibitem[{{Zhang} and {Zhu}(2020)}]{Zhang2020}
\bibinfo{author}{{Zhang}, S.}, \bibinfo{author}{{Zhu}, Z.},
  \bibinfo{year}{2020}.
\newblock \bibinfo{title}{{The effects of disc self-gravity and radiative
  cooling on the formation of gaps and spirals by young planets}}.
\newblock \bibinfo{journal}{\mnras} \bibinfo{volume}{493},
  \bibinfo{pages}{2287--2305}.
\newblock \DOIprefix\doi{10.1093/mnras/staa404},
  \href{http://arxiv.org/abs/1911.01530}{{\tt arXiv:1911.01530}}.
\bibitem[{{Zhang} et~al.(2018){Zhang}, {Zhu}, {Huang}, {Guzm{\'a}n}, {Andrews},
  {Birnstiel}, {Dullemond}, {Carpenter}, {Isella}, {P{\'e}rez}, {Benisty},
  {Wilner}, {Baruteau}, {Bai} and {Ricci}}]{Zhang2018DSHARP}
{Zhang}, S. et~al., \bibinfo{year}{2018}.
\newblock \bibinfo{title}{{The Disk Substructures at High Angular Resolution
  Project (DSHARP). VII. The Planet-Disk Interactions Interpretation}}.
\newblock \bibinfo{journal}{\apjl} \bibinfo{volume}{869}, \bibinfo{pages}{L47}.
\newblock \DOIprefix\doi{10.3847/2041-8213/aaf744},
  \href{http://arxiv.org/abs/1812.04045}{{\tt arXiv:1812.04045}}.
\bibitem[{{Ziampras} et~al.(2020){Ziampras}, {Kley} and
  {Dullemond}}]{Ziampras2020}
\bibinfo{author}{{Ziampras}, A.}, \bibinfo{author}{{Kley}, W.},
  \bibinfo{author}{{Dullemond}, C.P.}, \bibinfo{year}{2020}.
\newblock \bibinfo{title}{{Importance of radiative effects in gap opening by
  planets in protoplanetary disks}}.
\newblock \bibinfo{journal}{\aap} \bibinfo{volume}{637}, \bibinfo{pages}{A50}.
\newblock \DOIprefix\doi{10.1051/0004-6361/201937048},
  \href{http://arxiv.org/abs/2003.02298}{{\tt arXiv:2003.02298}}.
\bibitem[{{Zormpas} et~al.(2022){Zormpas}, {Birnstiel}, {Rosotti} and
  {Andrews}}]{Zormpas2022}
\bibinfo{author}{{Zormpas}, A.}, \bibinfo{author}{{Birnstiel}, T.},
  \bibinfo{author}{{Rosotti}, G.P.}, \bibinfo{author}{{Andrews}, S.M.},
  \bibinfo{year}{2022}.
\newblock \bibinfo{title}{{A large population study of protoplanetary disks.
  Explaining the millimeter size-luminosity relation with or without
  substructure}}.
\newblock \bibinfo{journal}{\aap} \bibinfo{volume}{661}, \bibinfo{pages}{A66}.
\newblock \DOIprefix\doi{10.1051/0004-6361/202142046},
  \href{http://arxiv.org/abs/2202.01241}{{\tt arXiv:2202.01241}}.
\bibitem[{{Zuckerman} and {Evans}(1974)}]{ZuckermanEvans1974}
\bibinfo{author}{{Zuckerman}, B.}, \bibinfo{author}{{Evans}, N.~J., I.},
  \bibinfo{year}{1974}.
\newblock \bibinfo{title}{{Models of Massive Molecular Clouds}}.
\newblock \bibinfo{journal}{\apjl} \bibinfo{volume}{192},
  \bibinfo{pages}{L149}.
\newblock \DOIprefix\doi{10.1086/181613}.

\end{thebibliography}





\end{document}